\documentstyle[aas2pp4]{article}
\tighten

\newcounter{savefig}
\newcommand{\alphfig}{\setcounter{savefig}{\value{figure}}
     \stepcounter{savefig}\setcounter{figure}{0}
     \renewcommand{\thefigure}{{\arabic{savefig}{\it \alph{figure}}}}}
\newcommand{\alphfigtwo}{\setcounter{figure}{0}
 \renewcommand{\thefigure}{{\arabic{savefig}{\it \alph{figure}\alph{figure}}}}}
\newcommand{\resetfig}{\setcounter{figure}{\value{savefig}}
     \renewcommand{\thefigure}{\arabic{figure}}}

\def\myplottwo#1{\plotfiddle{#1}{2.2in}{0}{75.}{75.}{-160}{-390}}  
\def\myplotthree#1{\plotfiddle{#1}{2.8in}{0}{75.}{75.}{-160}{-338}}
\def\myplotfour#1{\plotfiddle{#1}{4.0in}{0}{75.}{75.}{-160}{-280}} 
\def\myplotfive#1{\plotfiddle{#1}{4.5in}{0}{75.}{75.}{-160}{-221}} 
\def\myplotsix#1{\plotfiddle{#1}{5.3in}{0}{75.}{75.}{-160}{-168}}  
\def\myplotseven#1{\plotfiddle{#1}{6.2in}{0}{75.}{75.}{-160}{-110}}
\def\myploteight#1{\plotfiddle{#1}{7.0in}{0}{75.}{75.}{-160}{-50}} 
 
\def\CIVdblt{{\rm C}\kern 0.1em{\sc iv}~$\lambda\lambda 1548, 1550$}
\def\MgIIdblt{{\rm Mg}\kern 0.1em{\sc ii}~$\lambda\lambda 2976, 2803$}
\def\NVdblt{{\rm N}\kern 0.1em{\sc v}~$\lambda\lambda 1238, 1242$}  
\def\OVIdblt{{\rm O}\kern 0.1em{\sc vi}~$\lambda\lambda 1031, 1037$} 
\def\SiIVdblt{{\rm Si}\kern 0.1em{\sc iv}~$\lambda\lambda1393, 1402$}  

\def\AlII{\hbox{{\rm Al}\kern 0.1em{\sc ii}}}
\def\AlIII{{\hbox{\rm Al}\kern 0.1em{\sc iii}}}
\def\CaII{\hbox{{\rm Ca}\kern 0.1em{\sc ii}}}
\def\CrII{\hbox{{\rm Cr}\kern 0.1em{\sc ii}}}
\def\CII{\hbox{{\rm C}\kern 0.1em{\sc ii}}}
\def\CIII{\hbox{{\rm C}\kern 0.1em{\sc iii}}}
\def\CIV{\hbox{{\rm C}\kern 0.1em{\sc iv}}}
\def\CV{\hbox{{\rm C}\kern 0.1em{\sc v}}}
\def\HI{\hbox{{\rm H}\kern 0.1em{\sc i}}}
\def\HII{\hbox{{\rm H}\kern 0.1em{\sc ii}}}
\def\Lya{\hbox{{\rm Ly}\kern 0.1em$\alpha$}}
\def\Lyb{\hbox{{\rm Ly}\kern 0.1em$\beta$}}
\def\Lyg{\hbox{{\rm Ly}\kern 0.1em$\gamma$}}
\def\Lyfive{\hbox{{\rm Ly}\kern 0.1em$5$}}
\def\Lysix{\hbox{{\rm Ly}\kern 0.1em$6$}}
\def\Lyseven{\hbox{{\rm Ly}\kern 0.1em$7$}}
\def\Lyeight{\hbox{{\rm Ly}\kern 0.1em$8$}}
\def\Lynine{\hbox{{\rm Ly}\kern 0.1em$9$}}
\def\Lyten{\hbox{{\rm Ly}\kern 0.1em$10$}}
\def\HeI{\hbox{{\rm He}\kern 0.1em{\sc i}}}
\def\HeII{\hbox{{\rm He}\kern 0.1em{\sc ii}}}
\def\FeI{\hbox{{\rm Fe}\kern 0.1em{\sc i}}}
\def\FeII{\hbox{{\rm Fe}\kern 0.1em{\sc ii}}}
\def\FeIII{\hbox{{\rm Fe}\kern 0.1em{\sc iii}}}
\def\MnII{\hbox{{\rm Mn}\kern 0.1em{\sc ii}}}
\def\MgI{\hbox{{\rm Mg}\kern 0.1em{\sc i}}}
\def\MgII{\hbox{{\rm Mg}\kern 0.1em{\sc ii}}}
\def\MgIII{\hbox{{\rm Mg}\kern 0.1em{\sc iii}}}
\def\MgIV{\hbox{{\rm Mg}\kern 0.1em{\sc iv}}}
\def\NaI{\hbox{{\rm Na}\kern 0.1em{\sc i}}}
\def\NV{\hbox{{\rm N}\kern 0.1em{\sc v}}}
\def\NII{\hbox{{\rm N}\kern 0.1em{\sc ii}}}
\def\NIII{\hbox{{\rm N}\kern 0.1em{\sc iii}}}
\def\OVI{\hbox{{\rm O}\kern 0.1em{\sc vi}}}
\def\OII{\hbox{[{\rm O}\kern 0.1em{\sc ii}]}}
\def\SiII{\hbox{{\rm Si}\kern 0.1em{\sc ii}}}
\def\SiIII{\hbox{{\rm Si}\kern 0.1em{\sc iii}}}
\def\SiIV{\hbox{{\rm Si}\kern 0.1em{\sc iv}}}
\def\SII{\hbox{{\rm S}\kern 0.1em{\sc ii}}}
\def\SIII{\hbox{{\rm S}\kern 0.1em{\sc iii}}}
\def\SIV{\hbox{{\rm S}\kern 0.1em{\sc iv}}}
\def\TiII{\hbox{{\rm Ti}\kern 0.1em{\sc ii}}}
\def\ZnII{\hbox{{\rm Zn}\kern 0.1em{\sc ii}}}

\def\kms{\hbox{km~s$^{-1}$}}       
\def\cm2{\hbox{cm$^{-2}$}}
\def\etal{et~al.\ }

\lefthead{Churchill et~al.}
\righthead{Weak {\MgII} Absorbers}

\begin{document}
 
\title{The Population of Weak Mg II Absorbers I. \\
A Survey of 26 QSO HIRES/Keck Spectra\altaffilmark{1,2}}
 
\thispagestyle{empty}
 
\author{Christopher W. Churchill\altaffilmark{3}, 
        Jane R. Rigby, 
        Jane C. Charlton\altaffilmark{4}}
\affil{Department of Astronomy and Astrophysics \\ 
      The Pennsylvania State University, University Park PA 16802 \\ 
      {\it cwc, jrigby, charlton@astro.psu.edu}}
 
\and
\author{\vglue -0.3in
Steven S. Vogt\altaffilmark{3,5}}
\affil{Board of Astronomy and Astrophysics \\
       University of California, Santa Cruz CA 95064 \\
       {\it vogt@ucolick.org}}

\altaffiltext{1}{Based in part on observations obtained at the
W.~M. Keck Observatory, which is jointly operated by the University of
California and the California Institute of Technology.}
\altaffiltext{2}{Based in part on observations obtained with the
NASA/ESA {\it Hubble Space Telescope}, which is operated by the STScI
for the Association of Universities for Research in Astronomy, Inc.,
under NASA contract NAS5--26555.}
\altaffiltext{3}{Visiting Astronomer at the W.~M. Keck Observatory}
\altaffiltext{4}{Center for Gravitational Physics and Geometry,
Pennsylvania State University}
\altaffiltext{5}{UCO/Lick Observatories, University of California}

\begin{abstract}
\begingroup
\normalsize
We present a search for ``weak'' {\MgII} absorbers [those with $W_{\rm
r}(2796) < 0.3$~{\AA}] in the HIRES/Keck spectra of 26 QSOs.
We found 30, of which 23 are newly discovered.
The spectra are 80\% complete to $W_{\rm r}(2796) = 0.02$~{\AA} and
have a cumulative redshift path of $\sim 17.2$ for the redshift range
$0.4 \leq z \leq 1.4$.
The number of absorbers per unit redshift, $dN/dz$, is seen to
increase as the equivalent width threshold is decreased; we obtained
$dN/dz = 1.74\pm0.10$ for our $0.02 \leq W_{\rm r}(2796) <
0.3$~{\AA} sample.
The equivalent width distribution follows a power law, $N(W)\propto
W^{-\delta}$, with $\delta \sim 1.0$; there is no turnover down to
$W_{\rm r}(2796) = 0.02$~{\AA} at $\left< z \right> = 0.9$.
Weak absorbers comprise at least 65\% of the total {\MgII}
absorption population, which outnumbers Lyman limit systems (LLS) by a
factor of $3.8\pm1.1$; the majority of weak {\MgII} absorbers must
arise in sub--LLS environments.
Tentatively, we predict that $\sim 5$\% of the {\Lya} forest clouds with
$W_{\rm r}({\Lya}) \geq 0.1$~{\AA} will have detectable {\MgII}
absorption to $W^{\rm min}_{\rm r}(2796) = 0.02$~{\AA} and that this
is primarily a high--metallicity selection effect ($[Z/Z_{\odot}]
\geq -1$).
This implies that {\MgII} absorbing structures figure prominently
as tracers of sub--LLS environments where gas has been processed by
stars.
We compare the number density of  $W_{\rm r}(2796) \geq
0.02$~{\AA} absorbers with that of both high and low surface
brightness galaxies and find a fiducial absorber size of $35h^{-1}$ to
$63h^{-1}$~kpc, depending upon the assumed galaxy population and
their absorption properties.
The individual absorbing ``clouds'' have  $W_{\rm r}(2796) \leq
0.15$~{\AA} and their narrow (often unresolved) line widths imply
temperatures of $\sim 25,000$~K.
We measured $W_{\rm r}(1548)$ from {\CIV} in FOS/{\it HST\/}
archival spectra and, based upon comparisons with {\FeII}, found
a range of ionization conditions (low, high, and multi--phase) in
absorbers selected by weak {\MgII}.
\endgroup
\end{abstract}

 
 
\section{Introduction}
\label{sec:intro}

We present a survey for ``weak'' {\MgIIdblt} absorption in the
spectra of QSOs obtained with the HIRES spectrograph (\cite{vogtspie})
on the Keck~I telescope.
There have been several {\MgII} surveys over the previous decade
(\cite{ltw87}, hereafter LTW; \cite{tytler87}, hereafter TBSYK;
\cite{caulet89}; \cite{pb90}, hereafter PB90; \cite{ssb88}, hereafter
SSB;  \cite{ss92}, hereafter SS92).
These surveys were complete to a rest--frame {\MgII} $\lambda
2796$ equivalent width, $W_{\rm r}(2796)$, of $0.3$~{\AA} and above.
The more comprehensive work of SS92\nocite{ss92} yielded solid
statistics on the equivalent width distribution, redshift number
density, large scale velocity clustering, and evolution over the
redshift range $0.3 \leq z \leq 2.2$, for which the {\MgII} doublet
can be observed with ground based telescopes.

The shape of the  $W_{\rm r}(2796)$ distribution function at smaller
equivalent widths has important implications for our understanding of
cosmic chemical evolution and its connection to star producing
environments.
At $0.3 \leq z \leq 1.0$,  {\MgII} absorption with $W_{\rm r}(2796) \geq
0.3$~{\AA} {\MgII} has been found to arise within 
$\sim 40h^{-1}$~kpc\footnote{Throughout this paper we use $q_0=0.05$
and express physical sizes in terms of $h=H_{0}/100$.} 
of normal galaxies (\cite{bb91}; \cite{sdp94}).
These galaxies exhibit a wide range of colors from late--type spirals
to the reddest ellipticals (though more luminous galaxies are redder),
and have $L_{B}$ and $L_{K}$ luminosity functions consistent with the
local luminosity function (types later than Sd are absent).
It has been concluded that a wide range of morphological types are
contributing to the {\MgII} absorbing gas cross section, except that
isolated low--mass ($L_{K} \leq 0.06L_{K}^{\ast}$) ``faint blue
galaxies'' are not (\cite{sdp94}).
A rapid cut off in the equivalent width distribution would imply that
this observed sample of galaxies provides a complete picture of the
star forming environments that give rise to {\MgII} absorbing gas.
Currently, that picture is one in which the galaxy population selected
by {\MgII} absorption is stable, showing very little cosmological
evolution from $z\sim 1$ (but see \cite{lilly95}).

If, on the other hand, the number of {\MgII} absorbers per unit
redshift continues to rise for decreasing $W_{\rm r}(2796)$,
it would be implied that surveys complete to $0.3$~{\AA} have unveiled
only a small portion of the population of metal--line systems selected
by {\MgII} absorption.
Churchill \& Le~Brun (1998\nocite{cl98}) have discussed this
possibility based upon the discovery of two near--solar to
super--solar metallicity {\MgII} absorbers in the {\Lya} forest of
PKS~$0454+039$. 
As such, our current picture of the relationship between {\MgII}
absorbing gas and star forming environments may require some
modification.
Perhaps the observed $L_{K} \sim 0.06L_{K}^{\ast}$ cut off in the
luminosity function of {\MgII} absorption--selected galaxies or the
cut off at $\sim 40h^{-1}$~kpc of {\MgII} absorption around these
galaxies does not apply for $W_{\rm r}(2796) < 0.3$~{\AA}.
Alternatively, perhaps another population of star forming objects that
preferentially give rise to weaker {\MgII} absorption is implied.
Or, perhaps both a slight modification to the current picture of
{\MgII} absorbing galaxies {\it and\/} the incorporation of another
population of objects would be implied.
Measuring the statistical absorption properties of the weakest {\MgII}
absorbers is a first step toward verifying or casting new light on
such speculations.

For $W_{\rm r}(2796) < 0.3$~{\AA}, PB90\nocite{pb90} and
SS92\nocite{ss92} inferred a cut off in the {\MgII} equivalent width
distribution, the number of absorbers per unit redshift per unit
equivalent width. 
However, the completeness of their data below $0.3$~{\AA} dropped
rapidly.
Their conclusions have necessarily been based upon a comparison
of the number of {\MgII} absorbers detected with $W_{\rm
r}(2796) < 0.3$~{\AA}, corrected for completeness, to the number of
these absorbers {\it predicted\/} by extrapolating the equivalent
width distribution.
Measurements of the equivalent width distribution, however, were
somewhat inconclusive; the data were adequately described either by a
power law distribution with slope $\delta = 2.0$
(TBSYK\nocite{tytler87}) or $\delta = 1.6$ (SS92\nocite{ss92}),
or by an exponential with a characteristic equivalent width of $W_{\rm
r}(2796) = 0.66$~{\AA} (SS92\nocite{ss92}; LTW\nocite{ltw87}).
Womble (1995\nocite{womble95}) and Tripp, Lu, \& Savage
(1997\nocite{tripp97}) have tentatively concluded that the 
equivalent width distribution continues to rise below 0.3~{\AA}.

The HIRES/Keck spectra obtained for the thesis of Churchill
(1997a\nocite{thesis}) provide duplicate redshift coverage of 26 of the
103 QSO sight lines studied by SS92\nocite{ss92}.
The spectra are 80\% 
complete to $W_{\rm r}^{\rm min}(2796) = 0.02$~{\AA}, and provide an
opportunity to directly investigate if in fact there is a paucity of
{\MgII} absorbers with $W_{\rm r}(2796) < 0.3$~{\AA} and to measure
the shape of the distribution of equivalent widths well below
0.3~{\AA}.
Archival FOS spectra from the {\it Hubble Space Telescope\/} ({\it
HST\/}) cover the {\CIVdblt} doublet, which is a useful indicator of
the ionization conditions.
For systems with redshifts greater than $\sim 1.2$, ground--based
spectra (taken from the literature) cover {\CIV}.

The acquisition of the data and their reduction and analysis are
described in \S\ref{sec:data}.  
In \S\ref{sec:sample}, we present our adopted sample of weak {\MgII}
systems.
In \S\ref{sec:statistics} we compute the redshift path density,
determine the shape of the equivalent width distribution, examine
the clustering, and discuss the general absorption properties of weak
{\MgII} absorbers.
We present the results from a FOS/{\it HST\/} survey and literature
search for associated {\CIV} and discuss the ionization conditions and
constraints on the absorption cross sections in \S\ref{sec:discuss}.
Our main results are summarized in \S\ref{sec:conclude}.


\section{Observations and Data Analysis}
\label{sec:data}

\subsection{The Sample of QSO Spectra}

A full description of the HIRES data acquisition is given in Churchill
(1997a\nocite{thesis}) and in Churchill, Vogt, \& Charlton (1998,
hereafter CVC98\nocite{cvc98}).
In short, 26 QSO spectra were obtained with the HIRES spectrometer
(\cite{vogtspie}) on the Keck~I telescope. 
The QSOs and their emission redshifts, the UT date of the
observations, the summed exposure times, and the approximate
wavelength coverage are presented in Table~\ref{tab:obsjournal}.
The spectra comprise a biased sample of QSO sight lines with
``strong'' {\MgII} absorbers .
However, the QSO lines of sight are unbiased for {\MgII}
systems with $W_{\rm r}(2796) < 0.3$~{\AA}, since nothing is
known {\it a priori\/} about the presence of these ``weak'' {\MgII}
systems\footnote{From this point forward, we refer to ``weak'' systems
as those having $W_{\rm r}(2796) < 0.3$~{\AA} and call $W_{\rm
r}(2796) \geq 0.3$~{\AA} systems ``strong''.}.
It is possible that weak systems are preferentially found 
in sight lines along which strong systems are present, in which case
these QSO spectra would not comprise an unbiased sample.
We return to this point in \S\ref{sec:clustering}.
Because of the echelle format, each spectrum has breaks in wavelength
coverage redward of 5100~{\AA}.
The gaps in redshift coverage begin at $z\sim 0.83$ and become more
pronounced toward higher redshifts (see Figure~\ref{fig:ewlimits}).
The FOS/{\it HST\/} data have either been collected from the {\it
HST\/} archive (in collaboration with S. Kirhakos, B. Jannuzi, and
D. Schneider) or are from the QSO Absorption Line Key Project
(Bahcall \etal 1996\nocite{kp7}; Jannuzi \etal 1998\nocite{kp13}).

\subsection{The HIRES/Keck Data}
\label{sec:datared}

The raw data frames were bias corrected, flat fielded, cosmic ray
cleaned, combined, and scattered light corrected, using the standard
IRAF\footnote{IRAF is distributed by the National Optical Astronomy
Observatories, which are operated by AURA, Inc., under contract to the
National Science Foundation.} packages following the techniques
described in Churchill (1995\nocite{cwc95}).
The individual spectra were extracted using the optimal extraction
algorithms provided in the {\it Apextract\/} package.
The wavelength calibration was done interactively using the {\it
ecidentify\/} task of the {\it Echelle\/} package.
We have not resampled the data by linearizing the wavelength as a
function of pixel along the dispersion direction.
The continuum fits, based upon the formalism of Sembach \&
Savage (1992\nocite{sembach92}), were obtained by minimizing $\chi
^{2}$ between the flux values and a smooth function, usually a
Legendre polynomial.
For the detection of weak unresolved features, it is critical that the
measured uncertainty in each resolution element accurately reflects
the continuum noise.
Therefore, for each echelle order, we enforced a unity $\chi_{\nu}^{2}$
between the smooth fitted continuum model and the spectrum by scaling
the uncertainty spectrum (output by {\it apextract\/}) by a single
``optimizing'' multiplicative factor, $f$.
To obtain $f$, we root solved the function $1 - \chi_{\nu}^{2}$,
where $\chi_{\nu}^{2} = V_{\rm fit}/fV_{\rm \sigma}$, and $V_{\rm
fit}$ and $V_{\sigma}$ are the variance in the continuum fit and the
uncertainty spectrum, respectively.
The process required an automated iterative convergence algorithm in
order to objectively mask absorption features from the fit.

\alphfig

\begin{figure*}[t]
\plotfiddle{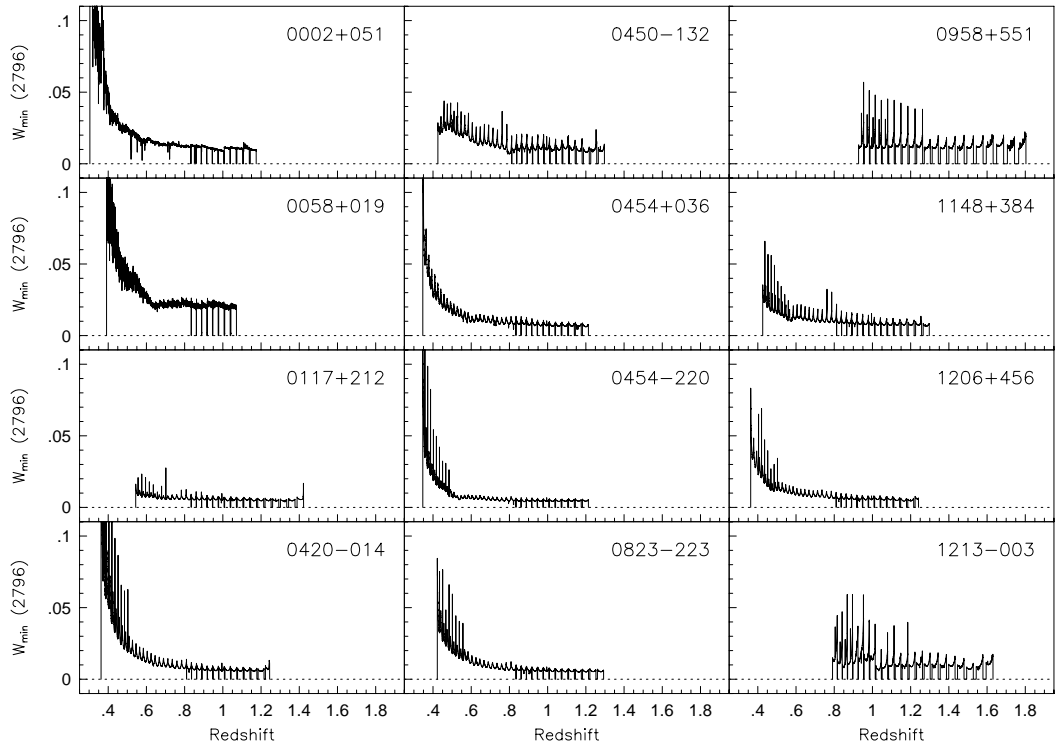}{3.5in}{0.}{60.}{60.}{-250}{0}
\vglue -0.5in
\protect\caption
{The survey sensitivity, given as the $5\sigma$ $W_{\rm r}
^{\rm min} (2796)$, is shown as a function of redshift. Breaks in the
echelle spectral coverage, where there is no possibility of detecting
a {\MgII} doublet in a given redshift range, are arbitrarily assigned
$W_{\rm r} ^{\rm min} (2796) = 0$. \label{fig:ewlimits}}
\end{figure*}

\begin{figure*}[t]
\plotfiddle{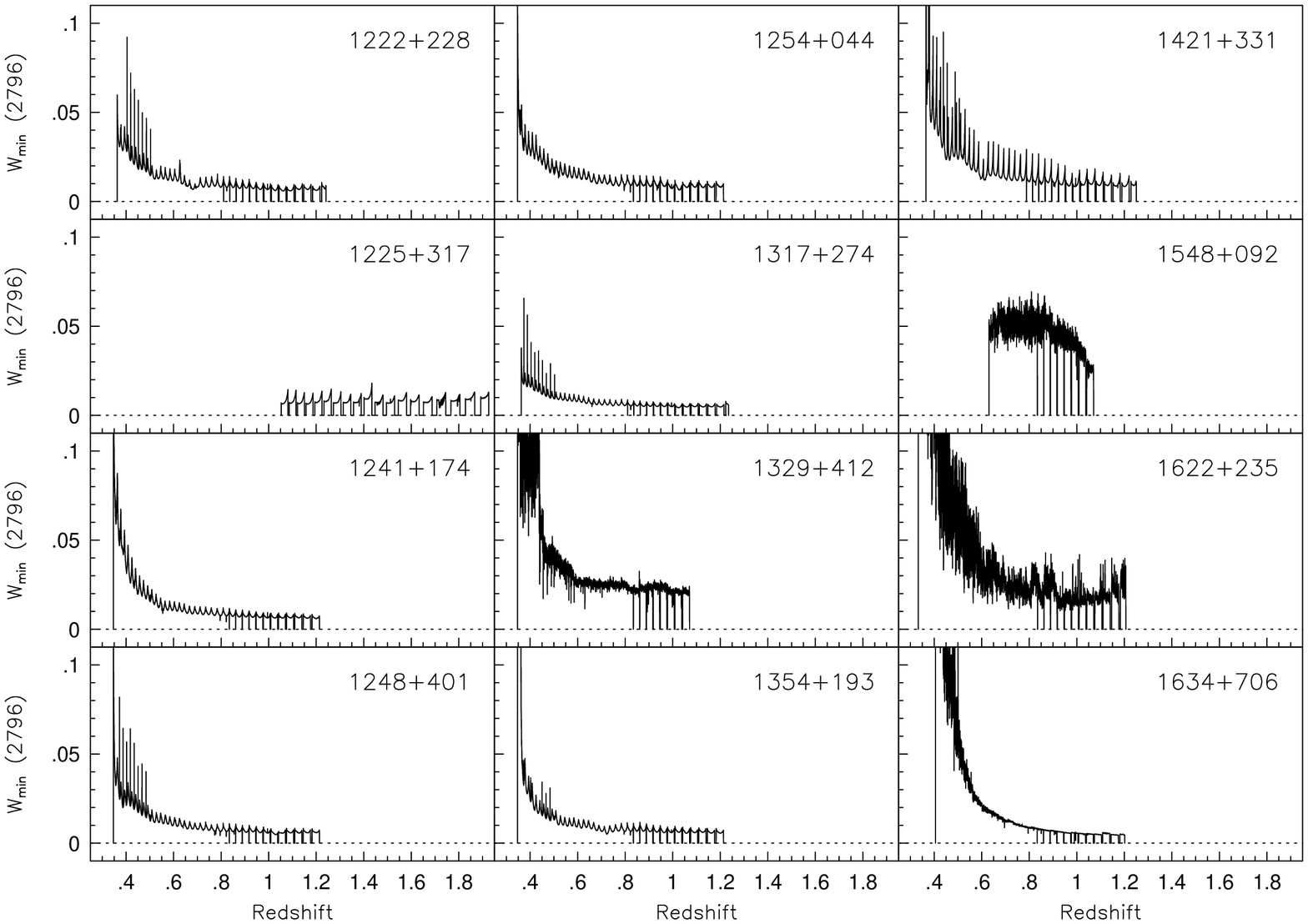}{3.5in}{0.}{60.}{60.}{-250}{0}
\vglue -0.5in
\protect\caption
{Same as for figure \ref{fig:ewlimits}}
\end{figure*}

\begin{figure}[th]
\plotfiddle{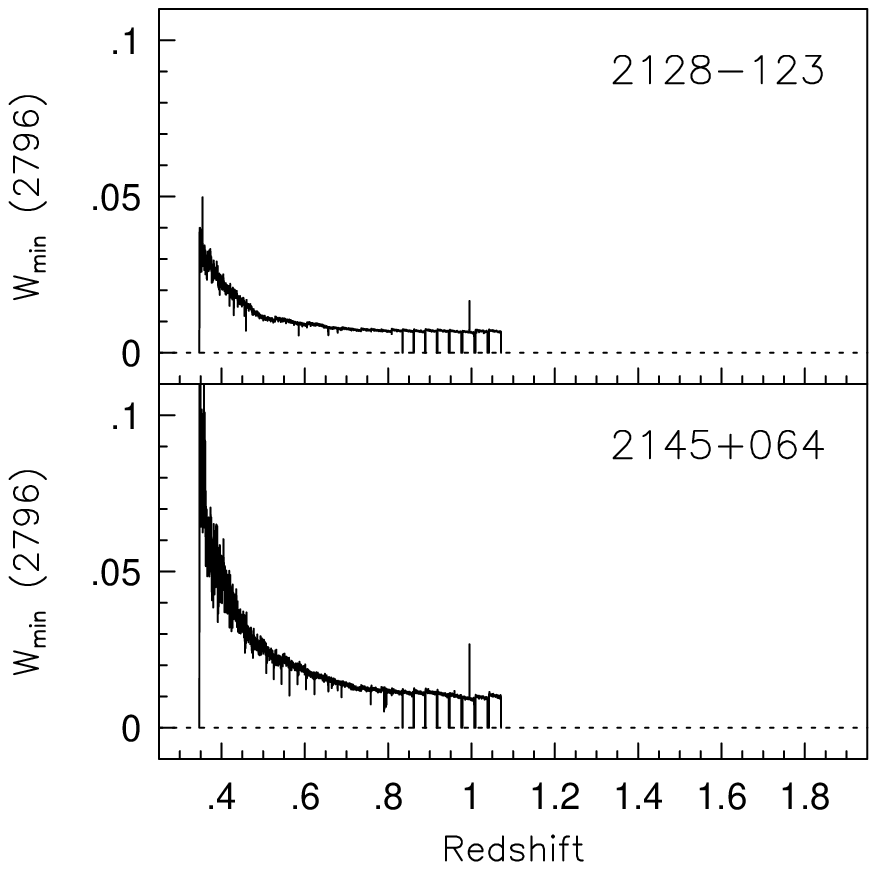}{3.5in}{0.}{60.}{60.}{-118}{-98}
\vglue -0.93in
\protect\caption
{Same as for figure \ref{fig:ewlimits} \label{fig:ewlimitsend}}
\end{figure}
 
For the objective identification of unresolved absorption (and
emission) features, we have used the formalism presented by Schneider
\etal (1993\nocite{schneider93}).
The resulting equivalent width uncertainty spectra provide the
observed equivalent width detection threshold as a function of wavelength.
The $5\sigma$ rest--frame equivalent width limits of the $\lambda
2796$ transition are shown in Figure~\ref{fig:ewlimits} as a function
of redshift.
Where there is no redshift coverage (interorder gaps), we have
arbitrarily set the limiting equivalent width to zero.

Only a single exposure of Q$0002+051$ was obtained so that removing
cosmic rays, especially from the sky, was problematic.
As a consequence, the zero level of the spectrum is uncertain by
$\sim 10$\% 
and the measured equivalent widths may be biased by this probable
zero--point offset, which varied from echelle order to echelle order.
This uncertainty has not been included in the error estimate of the 
quoted equivalent widths.
In Q$1548+092$, the {\Lya} forest compromises all transitions of metal
line systems that lie blueward of $\sim 4560$~{\AA}.  
Thus, we have not searched for {\MgII} doublets in this spectrum below
$z\sim 0.62$.

\subsection{The FOS/{\it HST\/} data}

We searched for {\CIVdblt} doublets at the redshifts of weak {\MgII}
absorbers in FOS/{\it HST\/} spectra.
These spectra were originally collected together for a companion paper
(Churchill \etal 1998, \cite{paper2}).
The archival data have been retrieved and reduced in collaboration
with S. Kirhakos, B. Jannuzi, and D. Schneider using the techniques
and software of the {\it HST\/} QSO Absorption Line Key Project.
The remaining spectra, which were originally obtained for the {\it
HST\/} Key Project, have been kindly provided by our collaborators in
fully reduced form. 
In Table~\ref{tab:obsjournal}, we reference the available spectra and
the grating that covered {\CIV}.
For details of the data reduction see Schneider \etal
(1993\nocite{schneider93}),  Bahcall \etal (1996\nocite{kp7}), and
Jannuzi \etal (1998\nocite{kp13}).

\subsection{Doublet Searching}
\label{sec:searching}

The search for {\MgII} doublets involved the following steps.
First, a complete list of $5\sigma$ features were objectively defined
in each HIRES spectrum.
To locate candidate {\MgII} doublets, we then tested the features 
one by one, starting at the smallest wavelength feature and moving
toward larger wavelengths.
We assumed each feature was a candidate $\lambda 2796$ line with
observed central wavelength $\lambda _{27}$.
Then, the expected location of the $\lambda 2803$ feature was computed
from $\lambda _{\rm 28} =  2803.531(\lambda _{\rm 27}/2796.352)$.
Centered about $\lambda _{\rm 28}$, an equivalent width and its
uncertainty were measured in an aperture with the same full width at
the continuum as that of the candidate $\lambda 2796$ feature.
These quantities were measured using the formalism of Sembach
\& Savage (1992\nocite{sembach92}).
The pair was designated as a candidate doublet when the $\lambda 2803$
detection significance was roughly equal to or greater than half that
of the  $\lambda 2796$ feature (given by the ratio of the $f\lambda$)
and when the doublet ratio was consistent with $1 \leq {\rm DR} \leq
2$ within the $1\sigma$ uncertainties.

We also employed a quantitative measure of the chance that the
candidate $\lambda 2803$ transition could be a random feature or a
transition from another system\footnote{We have performed an
exhaustive cross--checking of line identifications from {\CIV} and
{\MgII} redshifts reported in the literature and from those discovered
in our spectra.  Details of the line identifications and the list of
searched transitions and redshifts will be presented in
CVC98\nocite{cvc98}.} along the QSO sight line.
A ``false alarm'' probability is computed by scanning
the spectrum pixel by pixel (also with an aperture given by the full
width at the continuum of the candidate $\lambda 2796$ feature).
The scan is performed about $50$~{\AA} to both sides of the candidate
$\lambda 2803$ feature. 
This corresponds to a total redshift window of $\Delta z \sim 0.02$,
or $\pm 3000$~{\kms} about the feature.
The false alarm probability is simply the fraction of pixels with
detected features (both emission and absorption) having a significance
level greater than or equal to the candidate $\lambda 2803$ feature.
The detection levels of the $\lambda 2803$ transitions
for all included doublets were greater then $4.5\sigma$.
Most bonafide {\MgII} doublets have false alarm probabilities of
$P_{\rm fa} \leq 10^{-6}$.
The largest false alarm probability, for S24 in Q$1213-003$ at
$z=1.1277$, was $P _{\rm fa} \sim 0.009$.
We are quite certain that our adopted sample is not contaminated by
false {\MgII} systems.

In addition to the above constraints, a nearest neighbor velocity
separation of greater than 500~{\kms} was enforced.
The velocity filter was applied so that small equivalent width, high
velocity components in strong systems would not be included in the
sample.
The 500~{\kms} separation criterion resulted in our
dropping only a single potential weak {\MgII} absorber from our sample
in Q$1206+456$ (see ``system A'' in \cite{q1206}).

The search for {\CIV} in the FOS/{\it HST\/} spectra was performed in
an identical fashion as the {\MgII} doublets, except that the
objective line list was relaxed to $3\sigma$ detections\footnote{We
acknowledge A. Dobrzycki, who kindly assisted us in a preliminary
search for {\CIV} doublets in the FOS/{\it HST\/} spectra of Dobrzycki
\etal (1998)}.
We visually inspected each spectrum to determine if a candidate {\CIV}
doublet (at the redshift of a weak {\MgII} system) was real or a blend
of {\Lya} lines.
If real, we measured $W_{\rm r}(1548)$ using the same techniques used
for the HIRES data. 
If not, we measure the $3\sigma$ equivalent width limit.
We verified our results with the literature, when possible, but all
quoted {\CIV} equivalent widths and limits are based upon our
measurements in order to maintain uniformity.


\section{The Doublet Sample}
\label{sec:sample}

Our adopted sample includes 30 systems and is presented in
Table~\ref{tab:systems}.
Tabulated are the system number, the absorption redshift, the QSO
spectrum in which it was detected, the velocity width in {\kms} of the
$\lambda 2796$ transition, the rest--frame equivalent width of the
$\lambda 2796$ transition, and the {\MgII} doublet ratio.
The last column contains the cumulative redshift path, $Z(W_{\rm
r},{\rm DR})$, which is the total redshift path over which the tabulated
system could have been detected in this survey (see
Eq.~\ref{eq:redpath}).
The velocity widths, $\omega _{v}$, are measured directly from
the flux values according to
\begin{equation}
\omega _{v}^{2} = 
   \int _{v_{-}} ^{v_{+}} \tau _{a} (v) (\Delta v)^{2} dv \bigg/ 
    \int _{v_{-}} ^{v_{+}} \tau _{a} (v) dv ,
\end{equation}
where $\tau _{a}(v) = \ln [I_{\rm c}(v)/I(v)]$ is the apparent optical
depth (\cite{savage91}), $I_{\rm c}(v)$ is the continuum flux at
velocity $v$, $I(v)$ is the measured flux at $v$, $\Delta v = 
v - \left< v \right>$, and $\left< v \right>$ is the velocity centroid
of the absorption profile.
The $\omega _{v}$ are mathematically equivalent to the Gaussian width
of a normal distribution.

\begin{figure}[hbt]
\myplottwo{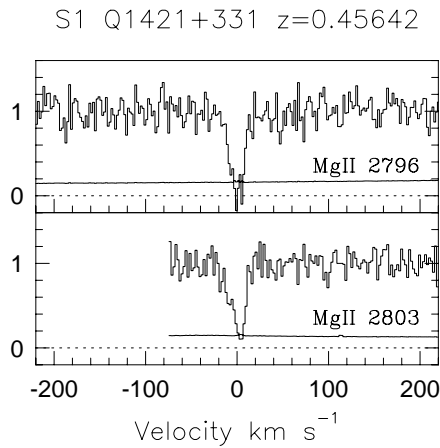}
\protect\caption
{The detected transitions of the weak {\MgII} systems. All
detections are $5\sigma$ unless annotated otherwise.  Absorption
features marked with ``$\ast$'' are not associated with the system
being presented.  In systems with multiple components, ticks mark the
components that are detected at the $3\sigma$ level. \label{fig:s1}}
\end{figure}

In Table~\ref{tab:ews}, we present the properties of each system,
including the transition identity, the observed wavelength, and the
rest--frame equivalent width or its $3\sigma$ upper limit.
The equivalent widths and their uncertainties are computed using the
methods of Sembach \& Savage (1992\nocite{sembach92}). 
For all {\MgII} systems, we searched the HIRES spectrum for other
transitions with the significance level relaxed to $3\sigma$.
Because the {\FeII} and {\MgI} $\lambda 2853$ transitions are the
strongest and most commonly found in {\MgII} absorbers, we have
presented their limits.
Other transitions are included in Table~\ref{tab:ews} when they have
been detected.
The data are shown in Figure~\ref{fig:s1}.
Features marked with a ``$\ast$'' are either members of other systems
or are unidentified.

\subsection{Discussion of Systems}
\label{sec:qsonotes}

\subsubsection{\rm S1  (Q$1421+331$;  $z_{\rm abs}=0.45642$)}

There is no previous report of S1.
No {\FeII} transitions were captured by the CCD.
{\MgI} was not detected.
A FOS/{\it HST\/} spectrum of this QSO was not available.
There is no galaxy candidate (C. Steidel, private communication).

\subsubsection{\rm S2  (Q$1329+412$; $z_{\rm abs}=0.500786$)}

S2 was reported as a ``probable'' {\MgII} doublet by
SSB\nocite{ssb88}.  
This system is associated with a blueish low mass galaxy with $\simeq
0.05L^{\ast}_{K}$ at an impact parameter of $\simeq 5h^{-1}$~kpc
(C. Steidel, private communication). 
The signal--to--noise ratio of the HIRES spectrum is fairly low, and
only the {\MgII} doublet was detected.
{\CIV} was not detected in the FOS/{\it HST\/} spectrum.

\subsubsection{\rm S3  (Q$1354+193$; $z_{\rm abs}=0.52149$)}

There is no previous report of S3.
Though {\FeII} and {\MgI} were captured by the CCD, only the {\MgII}
doublet was detected.
{\CIV} was not detected in the FOS/{\it HST\/} spectrum (also see
\cite{kp13}).
This QSO field has about five galaxies at $z\sim 0.5$, which are part
of a foreground cluster (C. Steidel, private communication).

\resetfig
\alphfig
 
\begin{figure}[ht]
\myplottwo{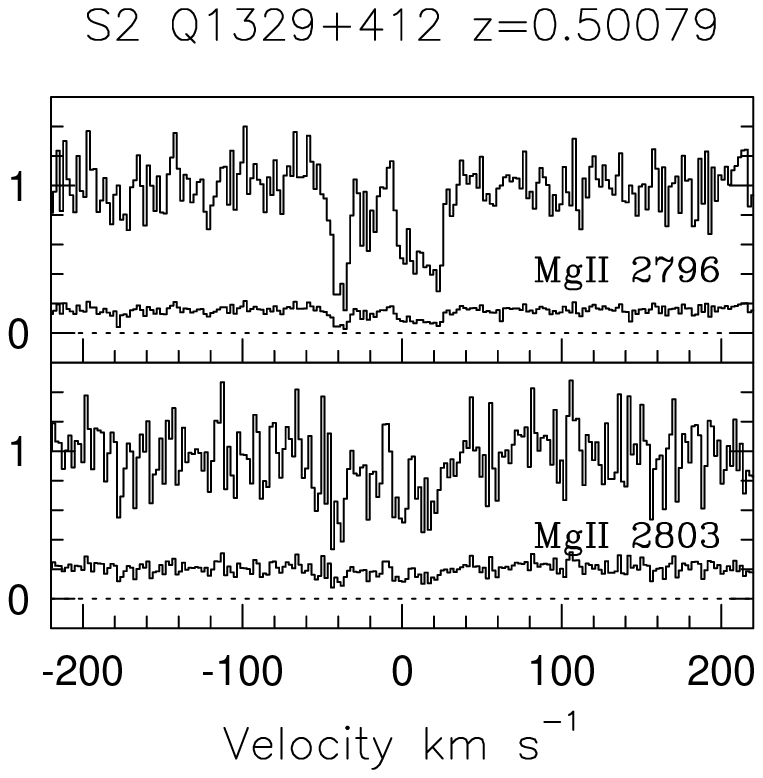}
\protect\caption
{Same as for figure \ref{fig:s1}}
\label{fig:s2}
\end{figure}

\begin{figure}[ht]
\myplottwo{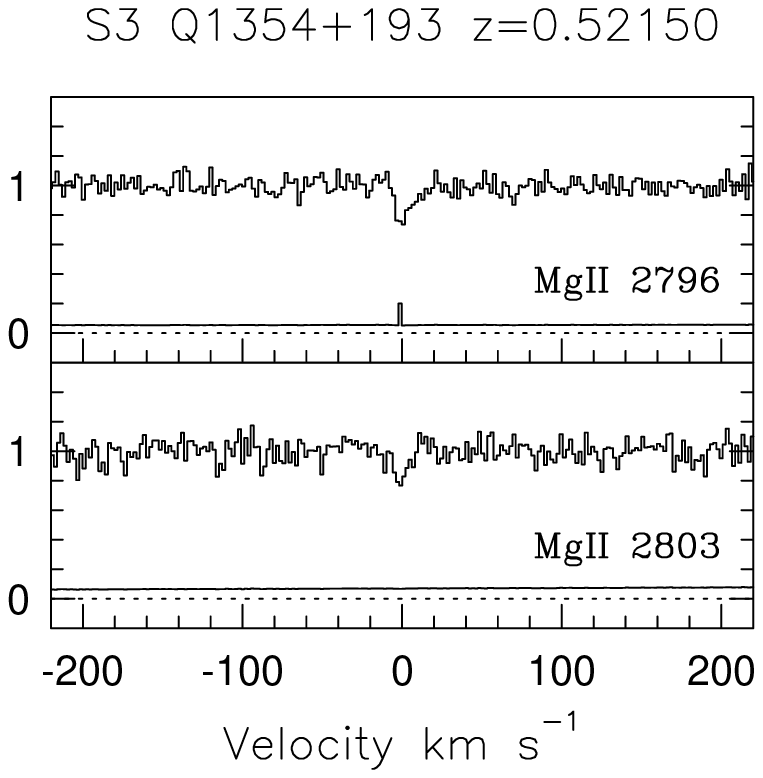}
\protect\caption
{Same as for figure \ref{fig:s1}}
\label{fig:s3}
\end{figure}

\begin{figure}[ht]
\myplottwo{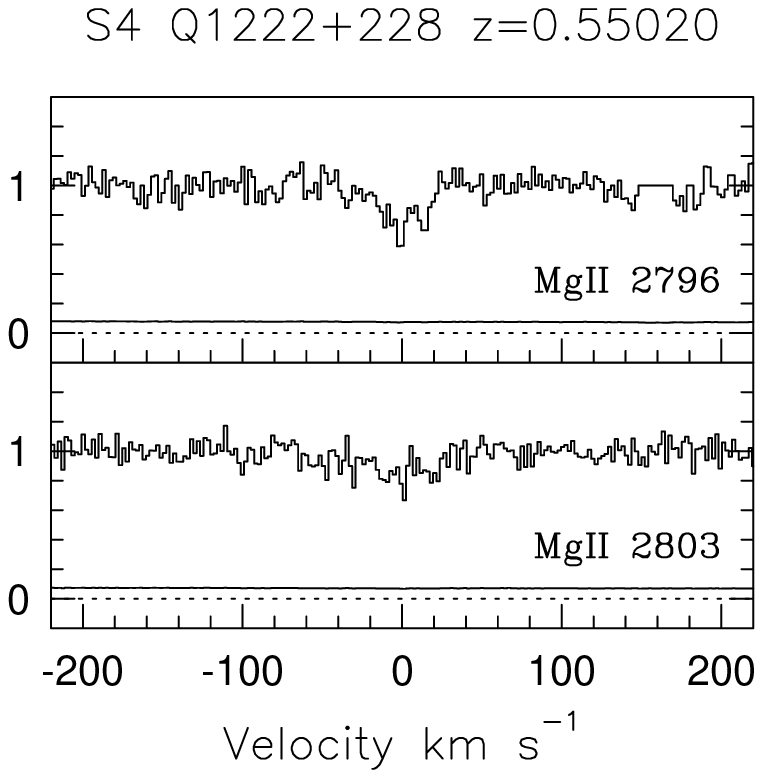}
\protect\caption
{Same as for figure \ref{fig:s1}}
\label{fig:s4}
\end{figure}
 
\begin{figure}[th]
\myplottwo{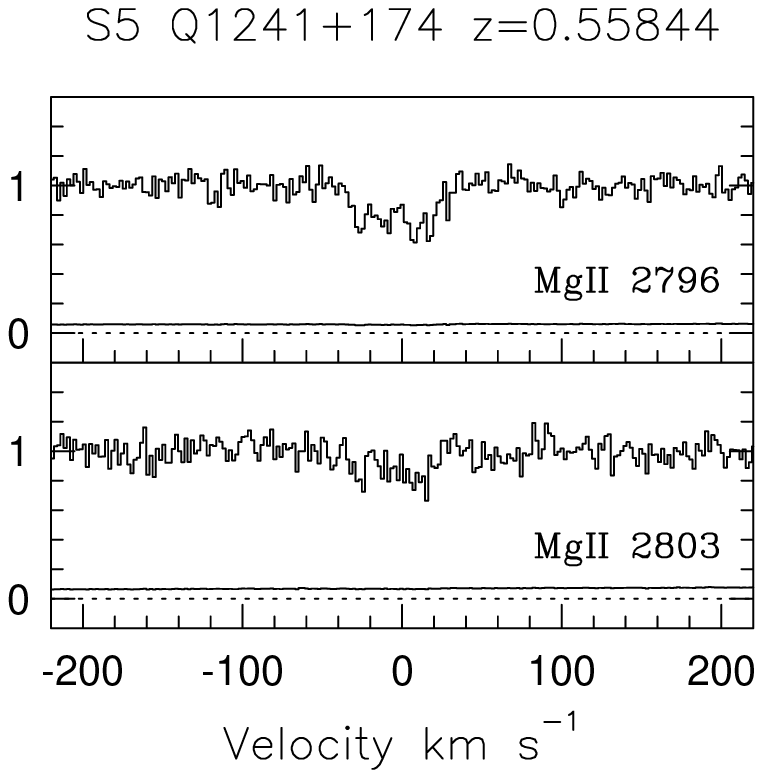}
\protect\caption
{Same as for figure \ref{fig:s1}}
\label{fig:s5}
\end{figure}

\begin{figure}[ht]
\myplottwo{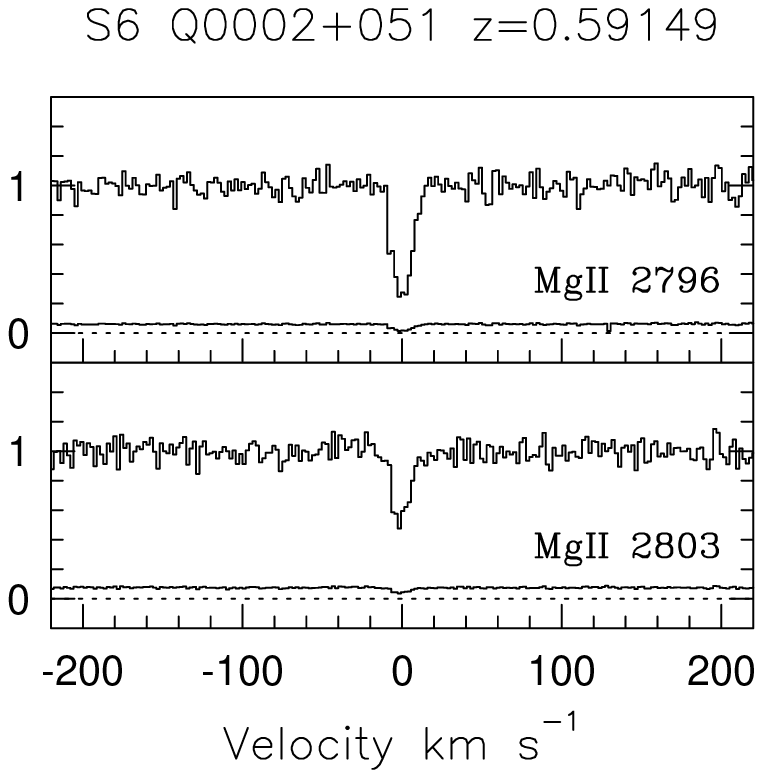}
\protect\caption
{Same as for figure \ref{fig:s1}}
\label{fig:s6}
\end{figure}

\clearpage

\begin{figure}[hb]
\myplotfive{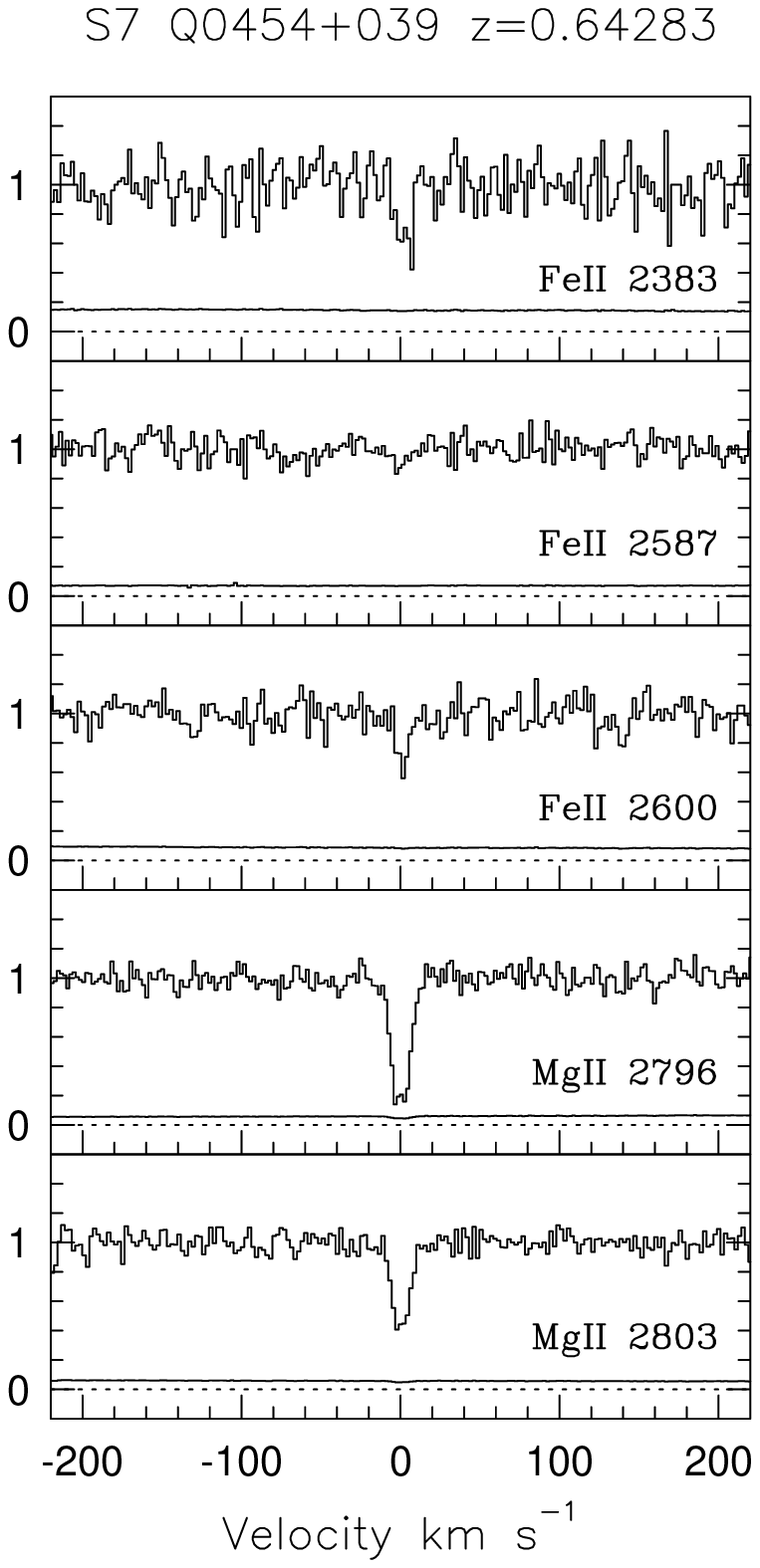}
\protect\caption
{Same as for figure \ref{fig:s1}}
\label{fig:s7}
\end{figure}

\begin{figure}[ht]
\myplottwo{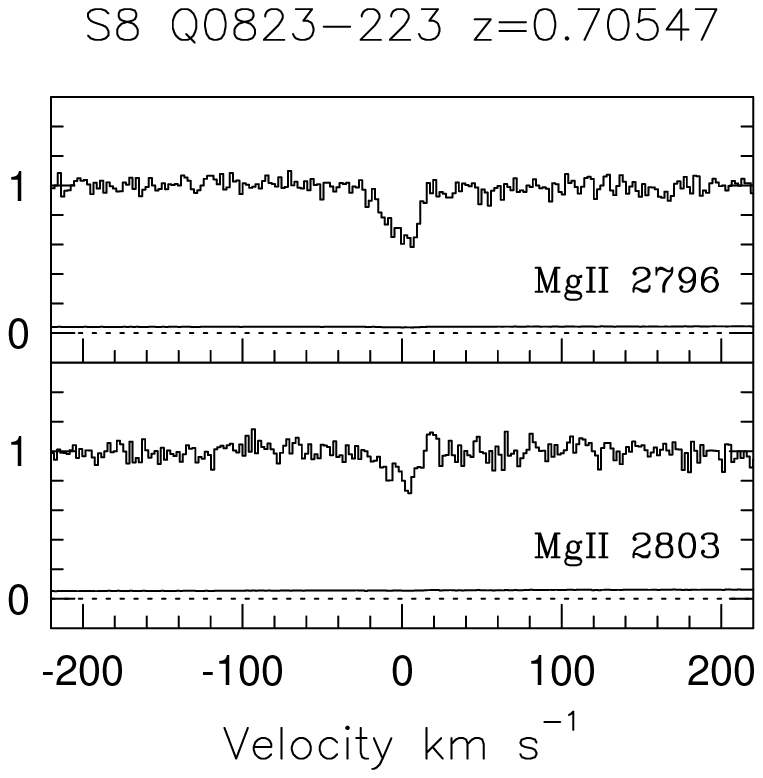}
\protect\caption
{Same as for figure \ref{fig:s1}}
\label{fig:s8}
\end{figure}

\begin{figure}[ht]
\myplotfour{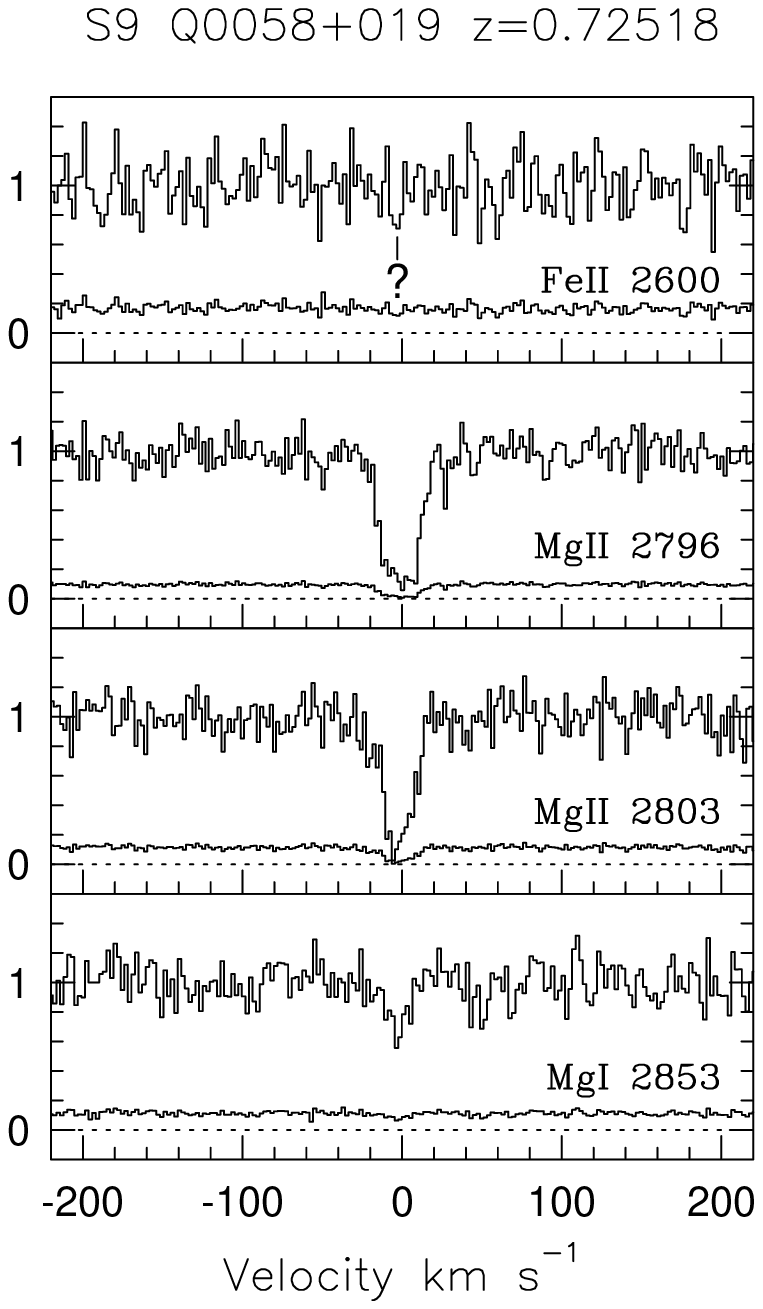}
\protect\caption
{Same as for figure \ref{fig:s1}}
\label{fig:s9}
\end{figure}

\begin{figure}[ht]
\myplotfive{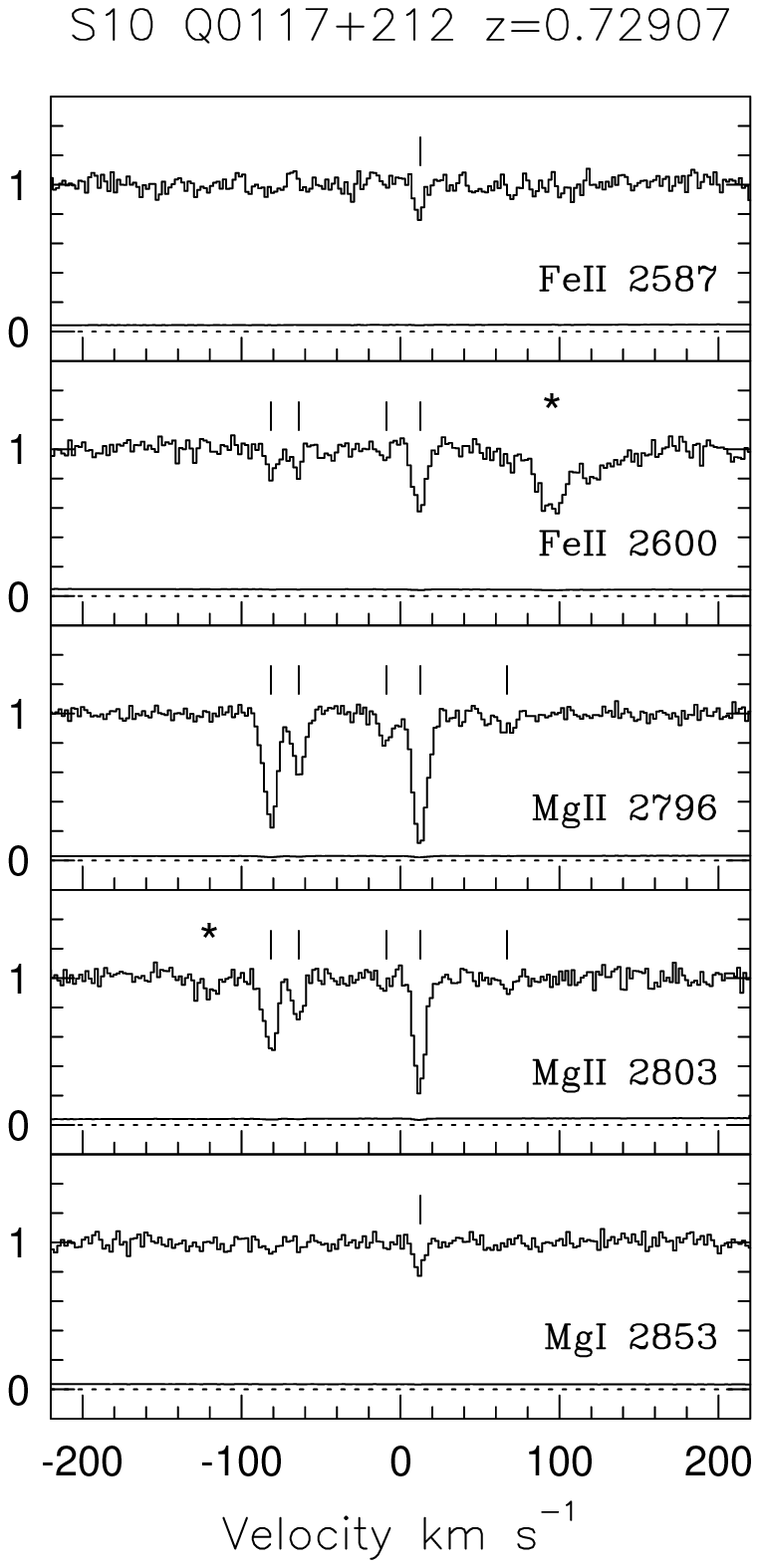}
\protect\caption
{Same as for figure \ref{fig:s1}}
\label{fig:s10}
\end{figure}

\begin{figure}[ht]
\myplotfive{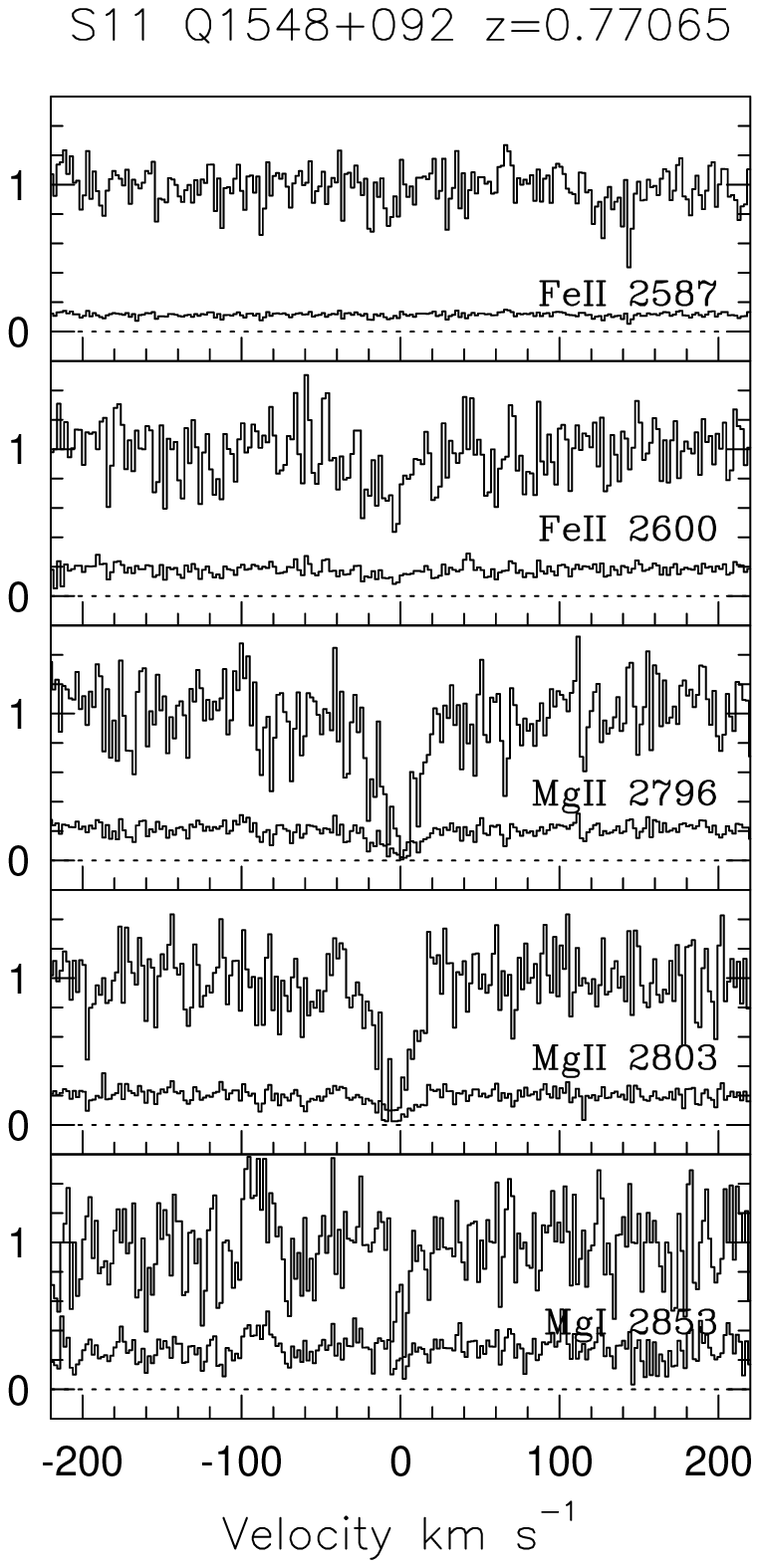}
\protect\caption
{Same as for figure \ref{fig:s1}}
\label{fig:s11}
\end{figure}

\begin{figure}[ht]
\myplottwo{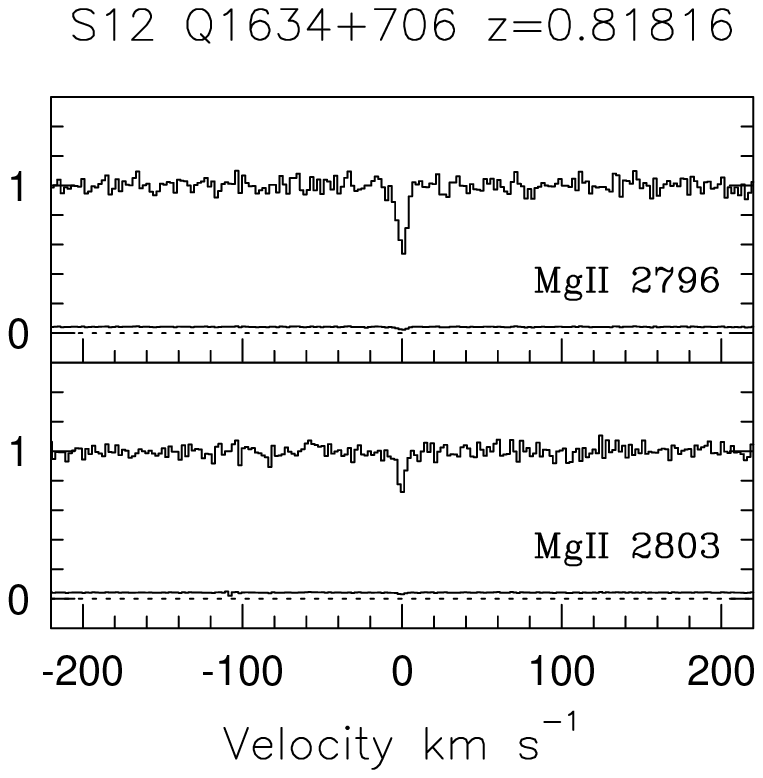}
\protect\caption
{Same as for figure \ref{fig:s1}}
\label{fig:s12}
\end{figure}
 
\begin{figure}[ht]
\myploteight{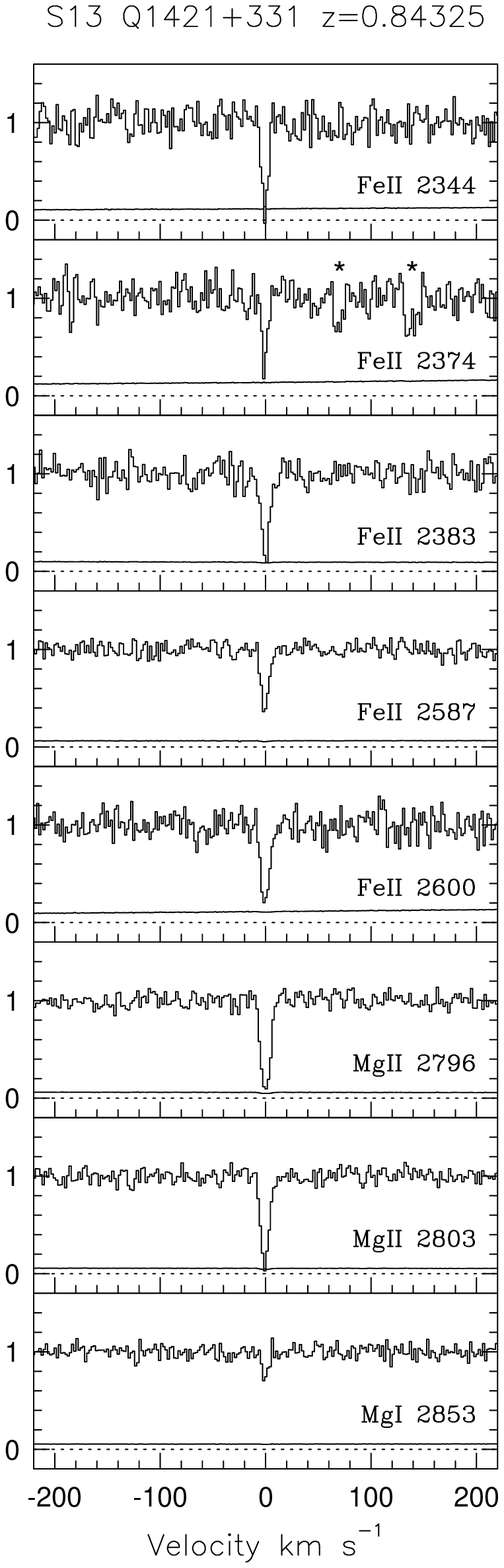}
\protect\caption
{Same as for figure \ref{fig:s1}}
\label{fig:s13}
\end{figure}

\begin{figure}[ht]
\myplotsix{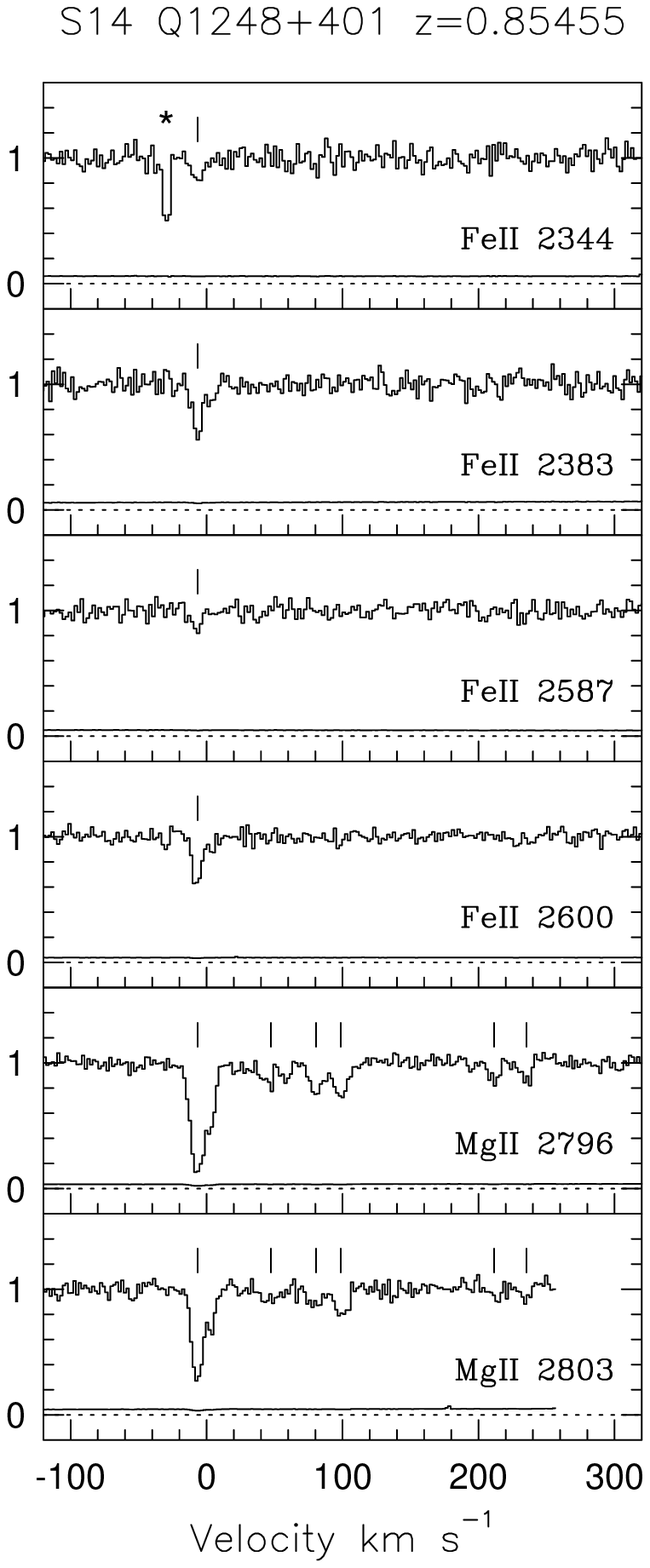}
\protect\caption
{Same as for figure \ref{fig:s1}}
\label{fig:s14}
\end{figure}

\begin{figure}[ht]
\myplottwo{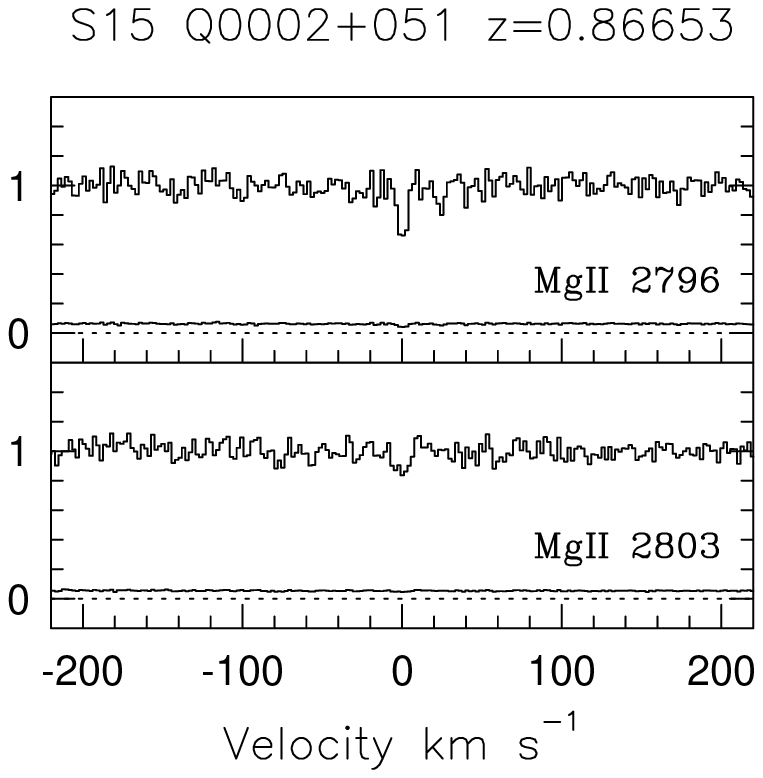}
\protect\caption
{Same as for figure \ref{fig:s1}}
\label{fig:s15}
\end{figure}
 
\begin{figure}[ht]
\myplottwo{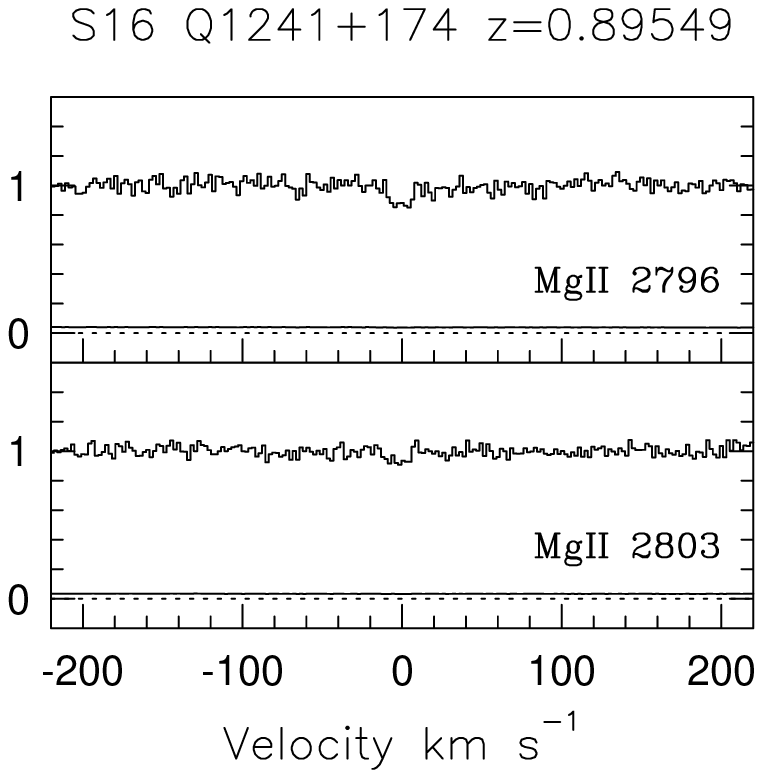}
\protect\caption
{Same as for figure \ref{fig:s1}}
\label{fig:s16}
\end{figure}
 
\begin{figure}[ht]
\myplottwo{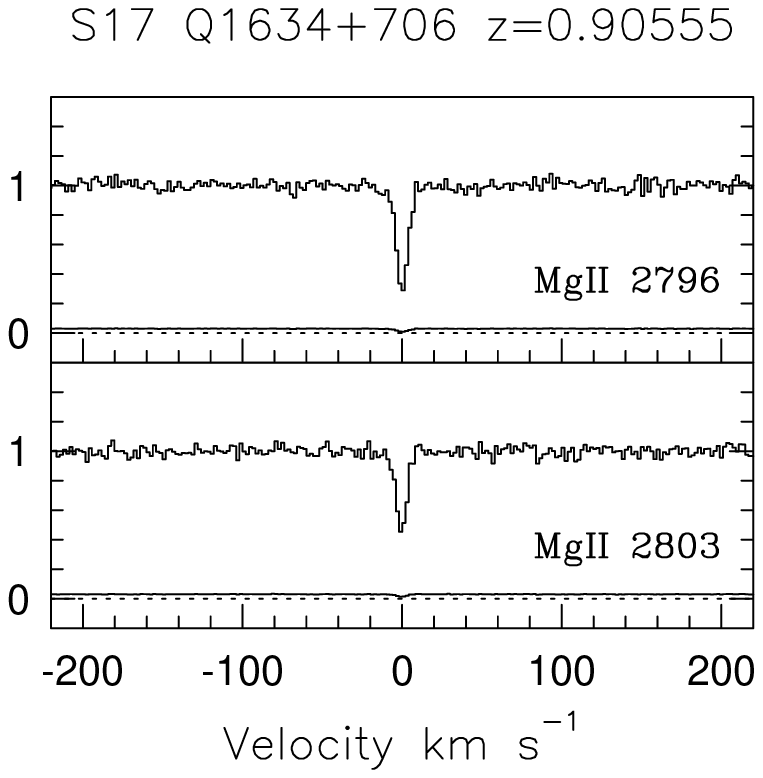}
\protect\caption
{Same as for figure \ref{fig:s1}}
\label{fig:s17}
\end{figure}
 
\begin{figure}[ht]
\myplotthree{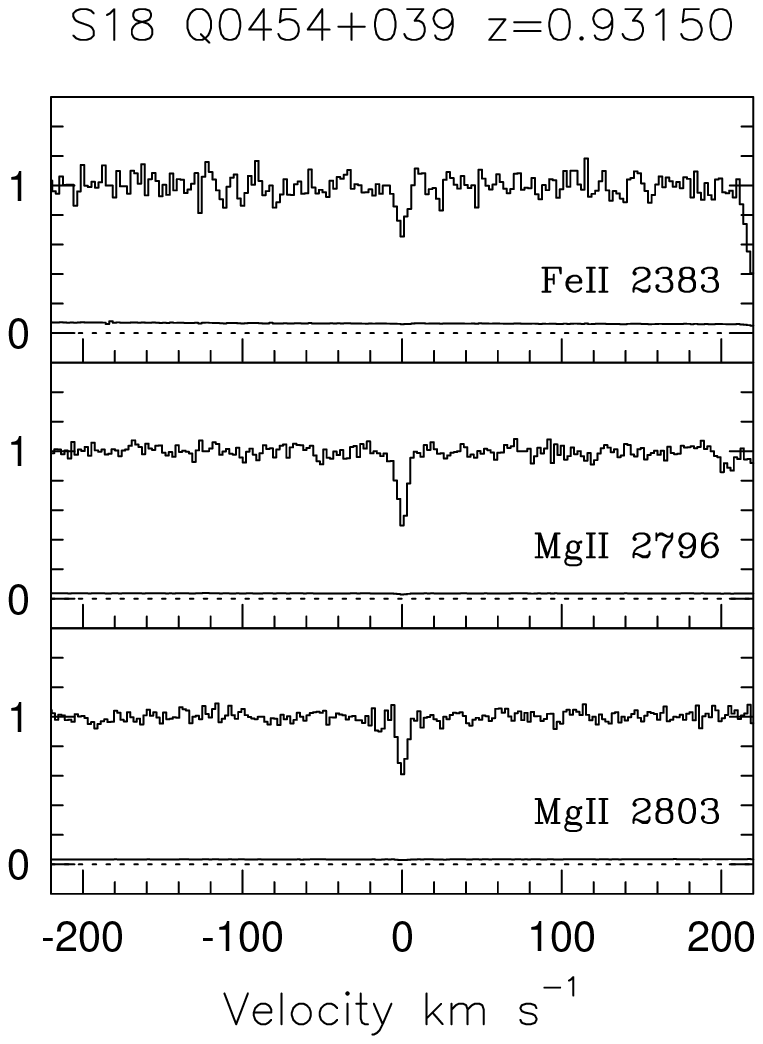}
\protect\caption
{Same as for figure \ref{fig:s1}}
\label{fig:s18}
\end{figure}

\begin{figure}[ht]
\myplottwo{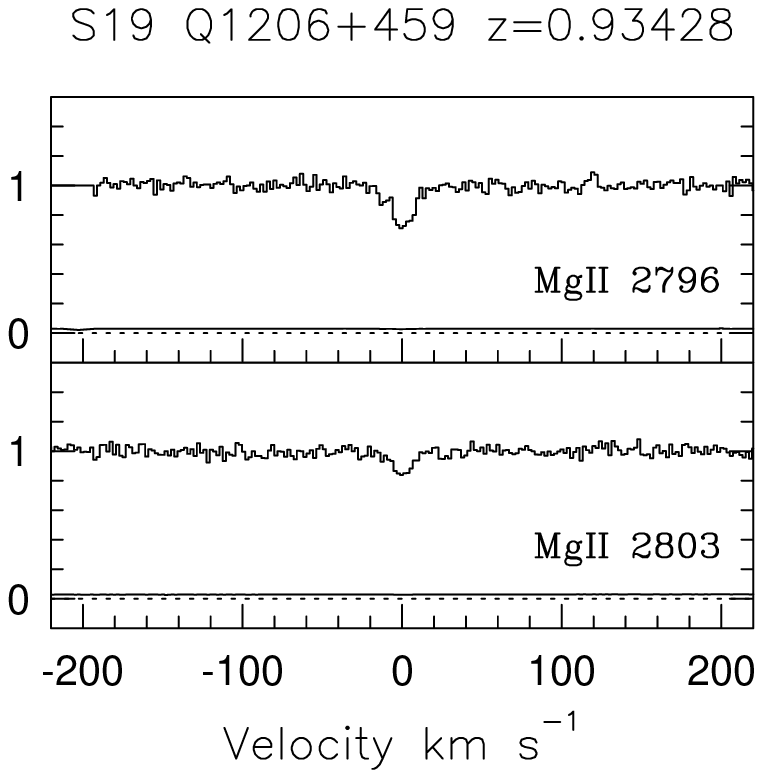}
\protect\caption
{Same as for figure \ref{fig:s1}}
\label{fig:s19}
\end{figure}

\begin{figure}[ht]
\myplottwo{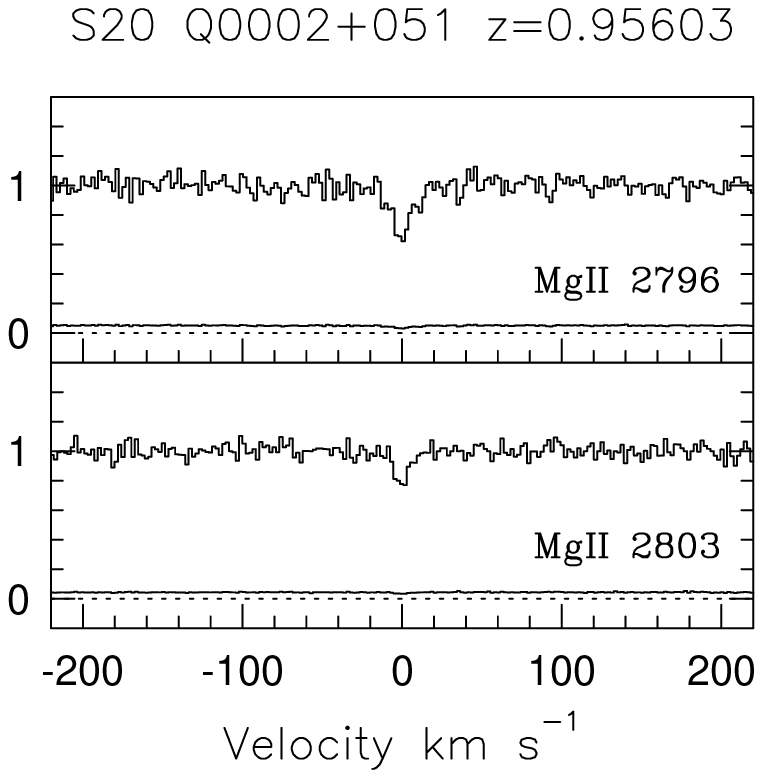}
\protect\caption
{Same as for figure \ref{fig:s1}}
\label{fig:s20}
\end{figure}
 
\begin{figure}[ht]
\myplottwo{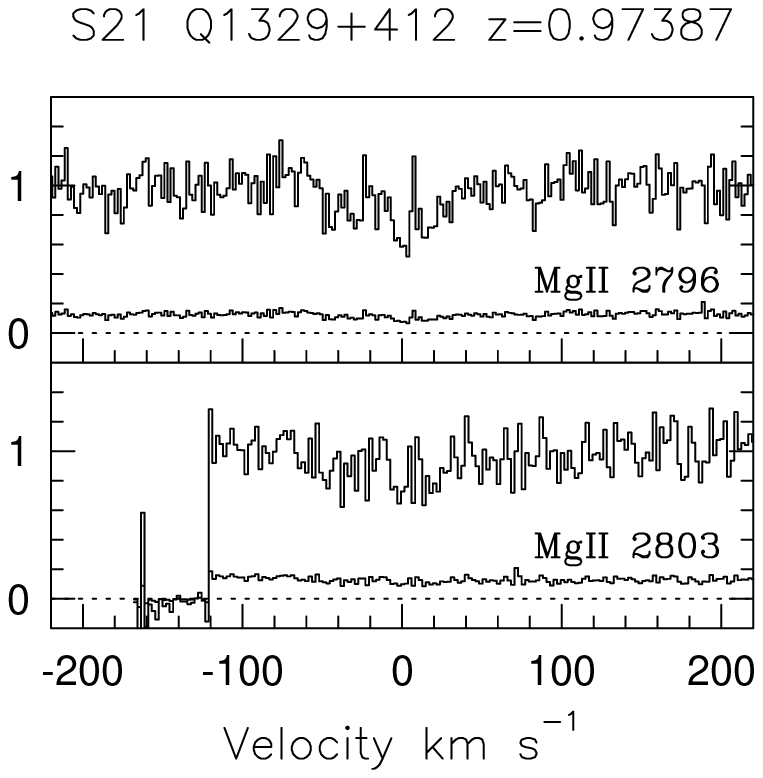}
\protect\caption
{Same as for figure \ref{fig:s1}}
\label{fig:s21}
\end{figure}

\begin{figure}[ht]
\myplotfive{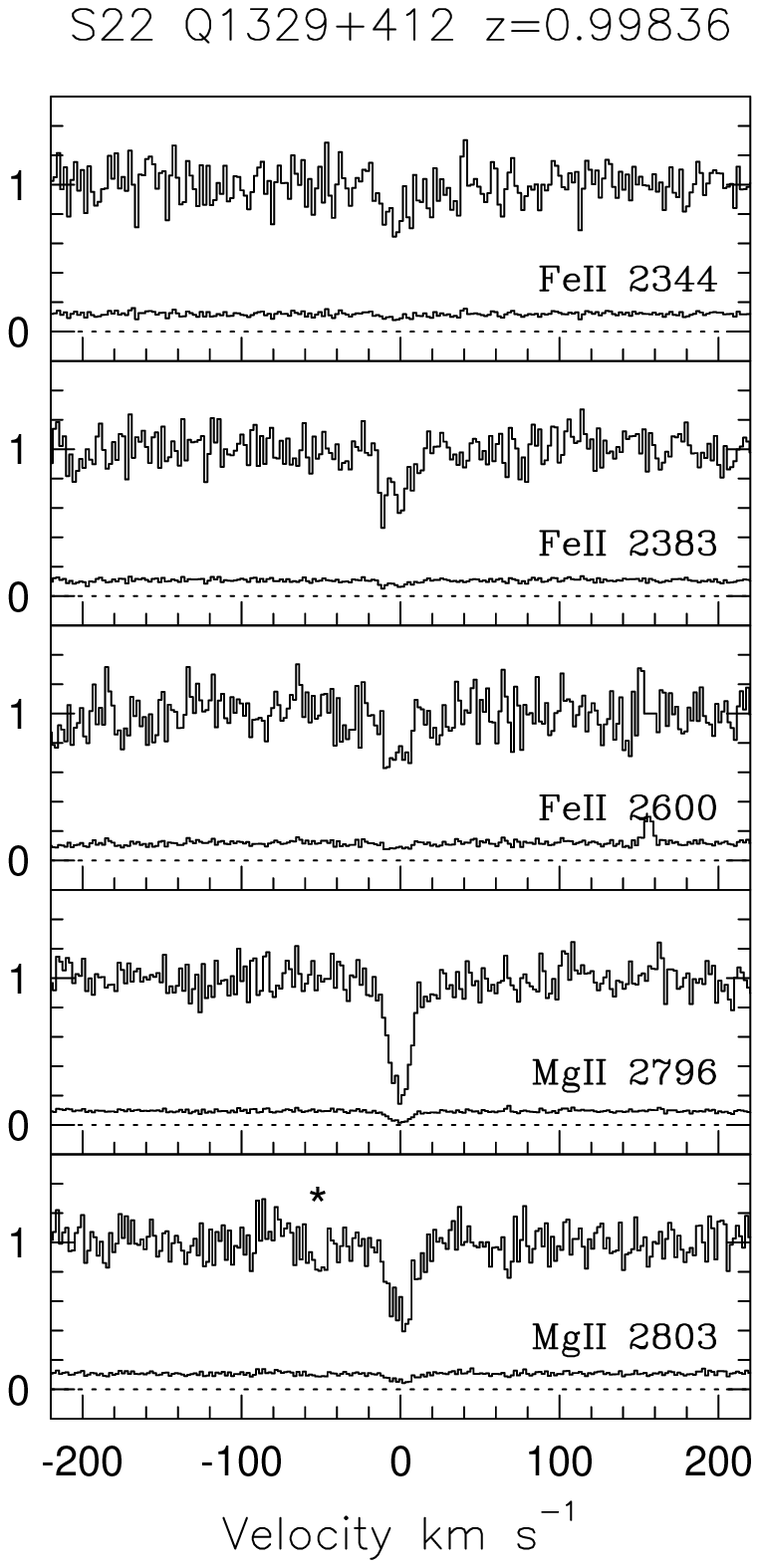}
\protect\caption
{Same as for figure \ref{fig:s1}}
\label{fig:s22}
\end{figure}

\begin{figure}[ht]
\myplottwo{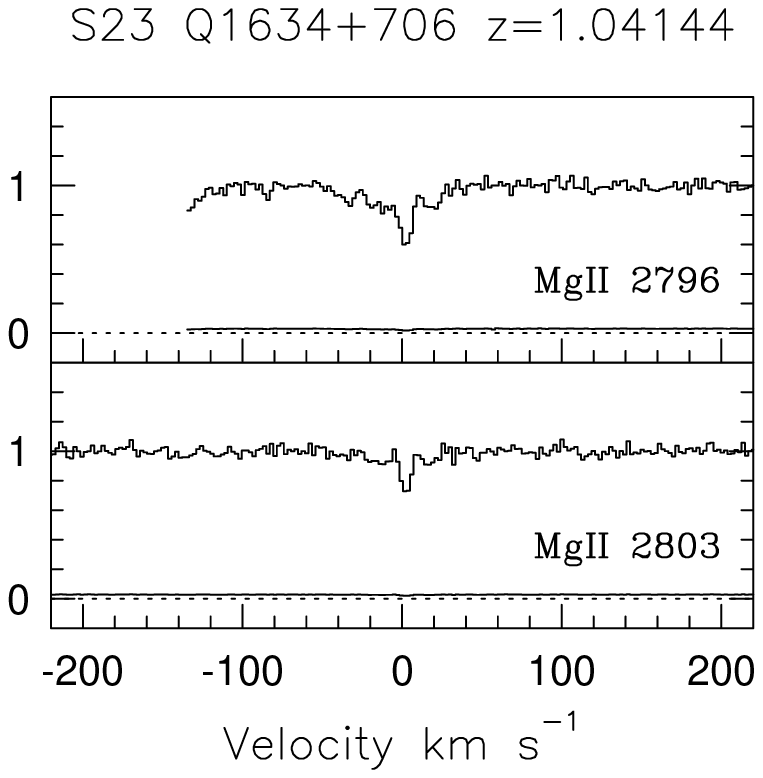}
\protect\caption
{Same as for figure \ref{fig:s1}}
\label{fig:s23}
\end{figure}
 
\clearpage

\begin{figure}[hb]
\myplottwo{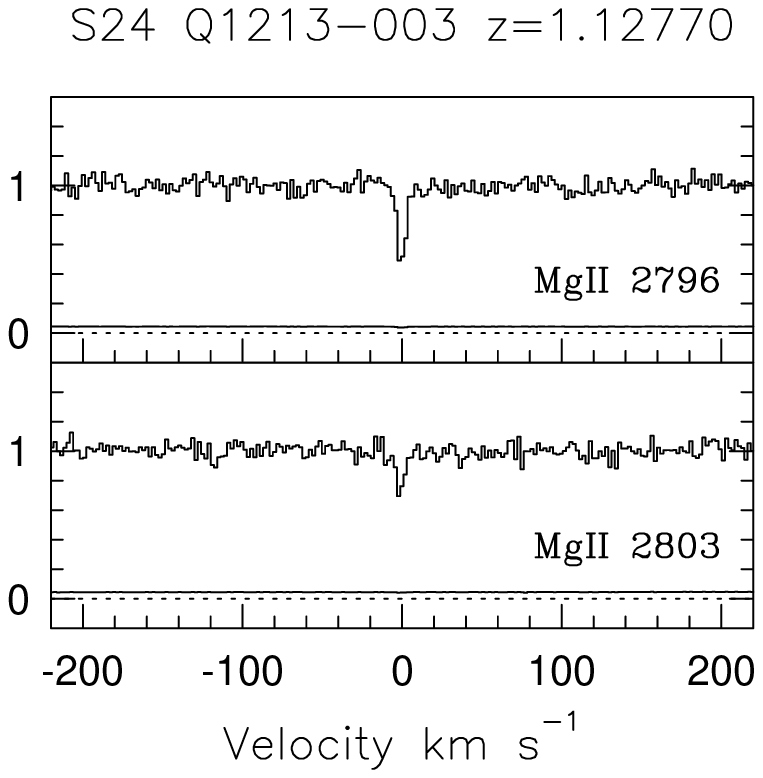}
\protect\caption
{Same as for figure \ref{fig:s1}}
\label{fig:s24}
\end{figure}
 
\begin{figure}[ht]
\myplottwo{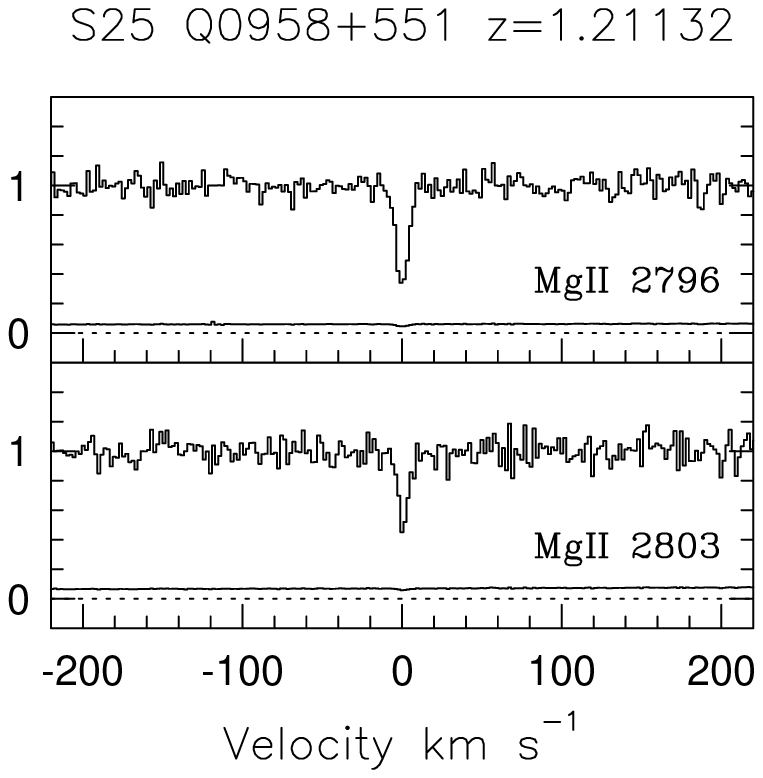}
\protect\caption
{Same as for figure \ref{fig:s1}}
\label{fig:s25}
\end{figure}
 
\begin{figure}[ht]
\myplotseven{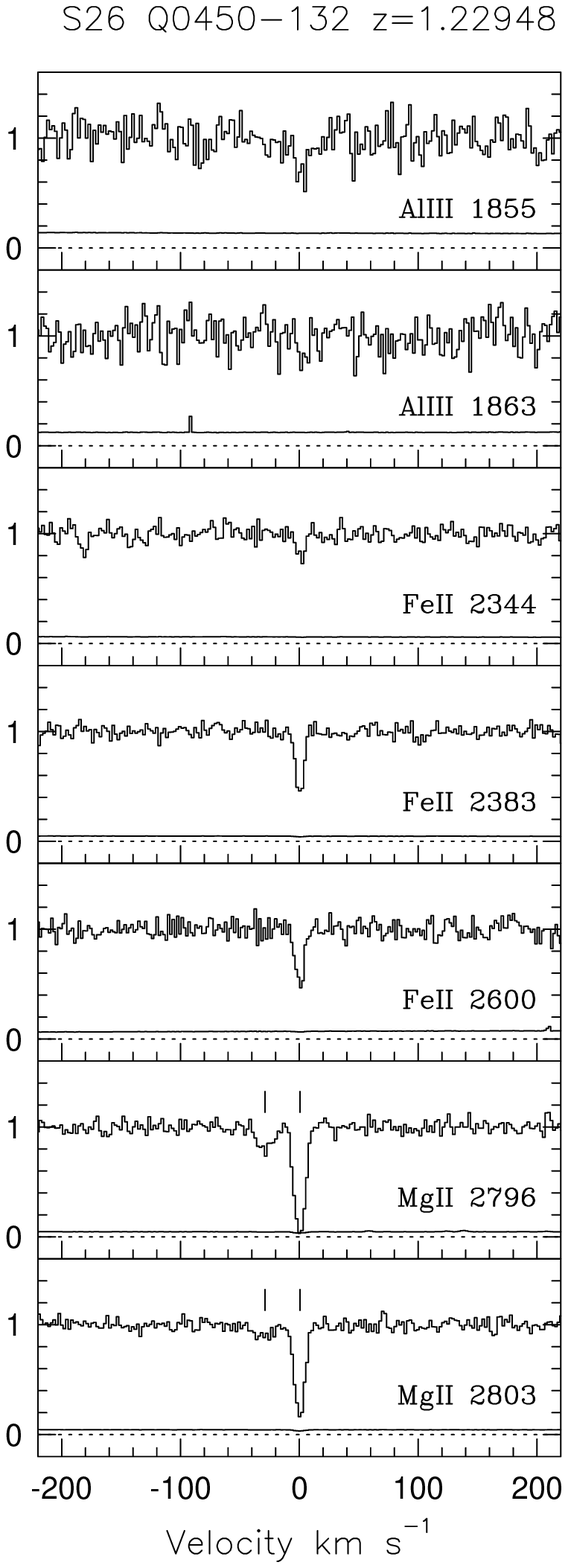}
\protect\caption
{Same as for figure \ref{fig:s1}}
\label{fig:s26}
\end{figure}
 
\alphfigtwo
 
\begin{figure}[ht]
\myplotfive{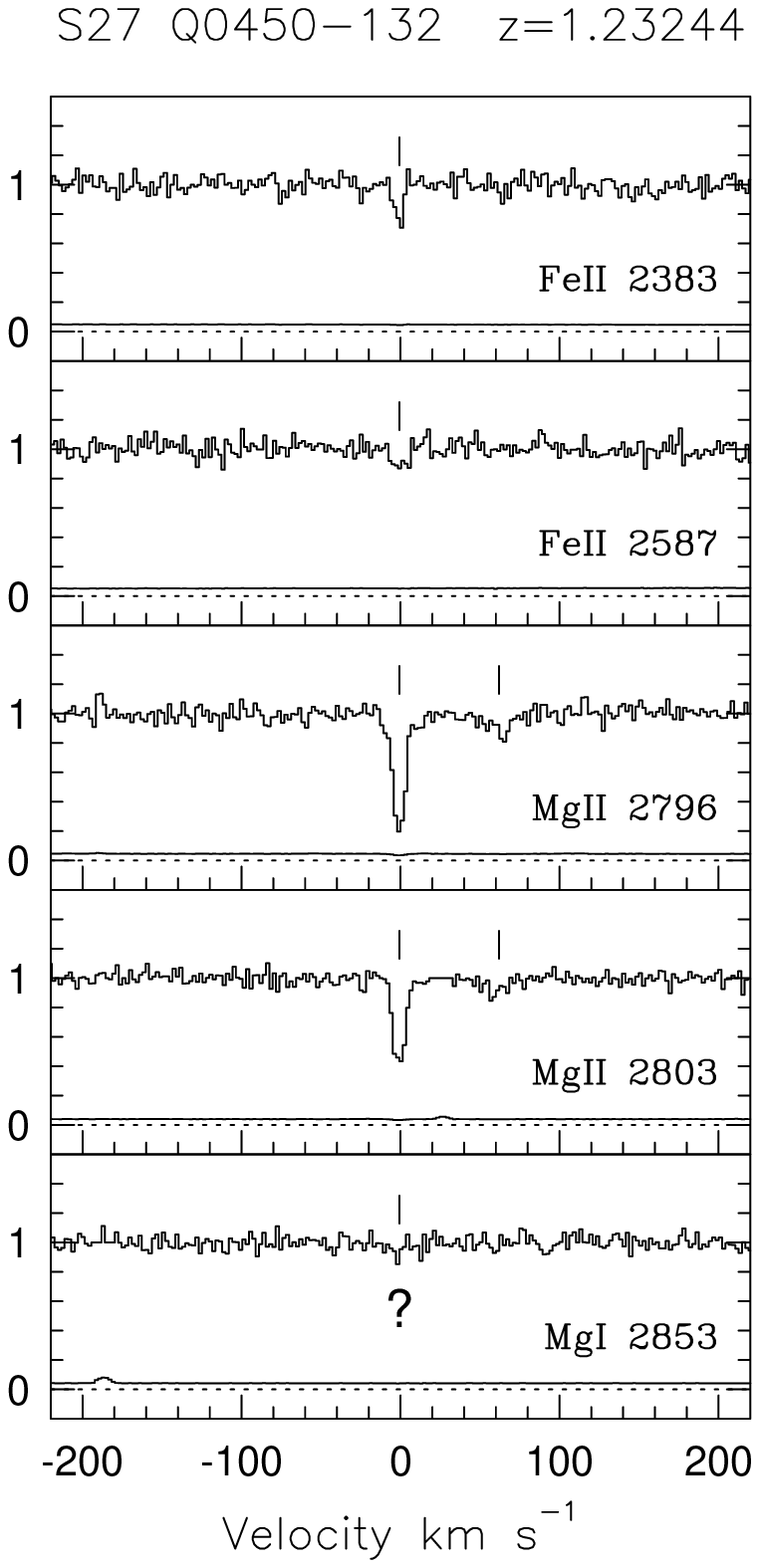}
\protect\caption
{Same as for figure \ref{fig:s1}}
\label{fig:s27}
\end{figure}
 
\begin{figure}[ht]
\myplotfive{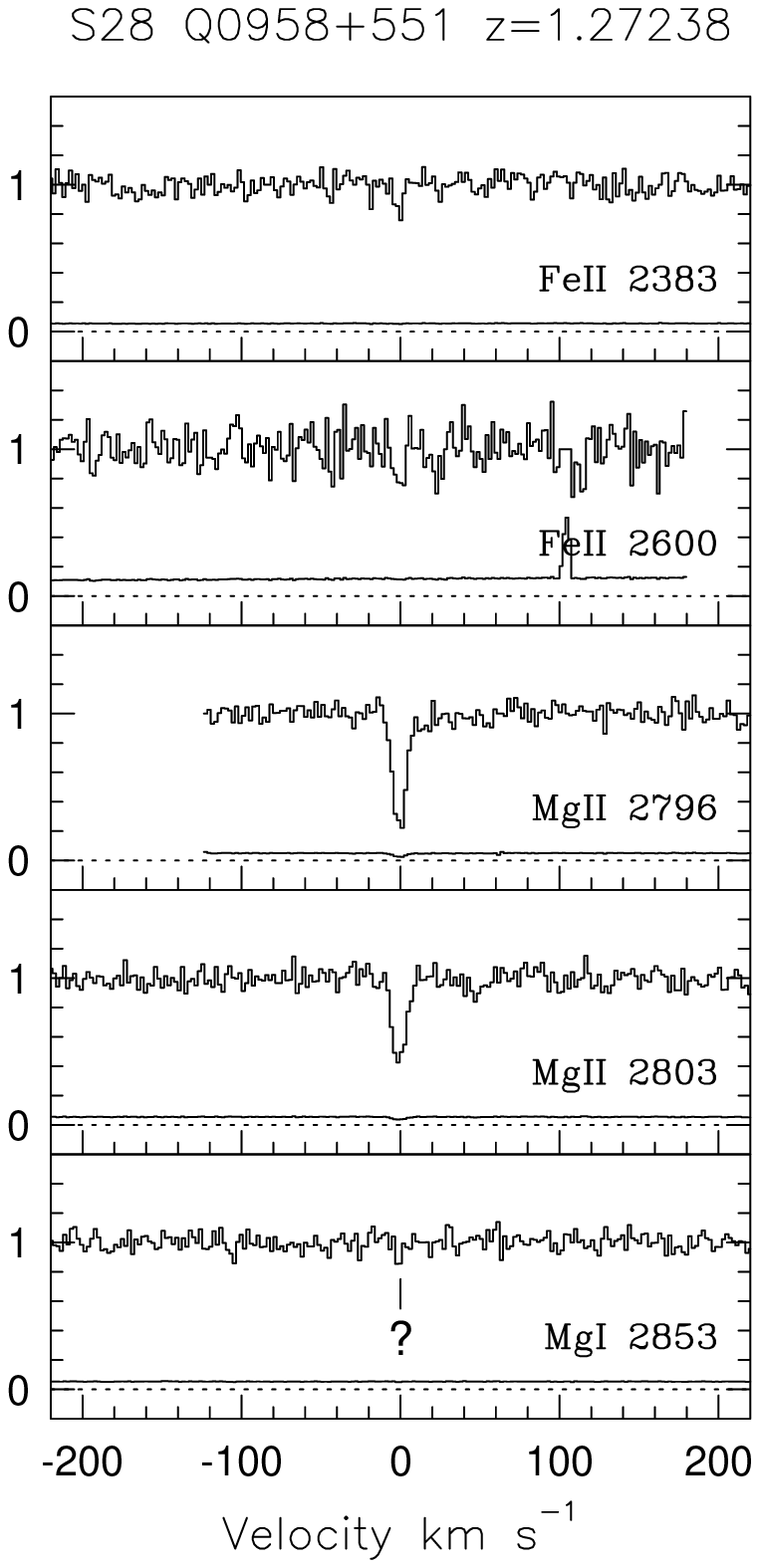}
\protect\caption
{Same as for figure \ref{fig:s1}}
\label{fig:s28}
\end{figure}
 
\begin{figure}[ht]
\myploteight{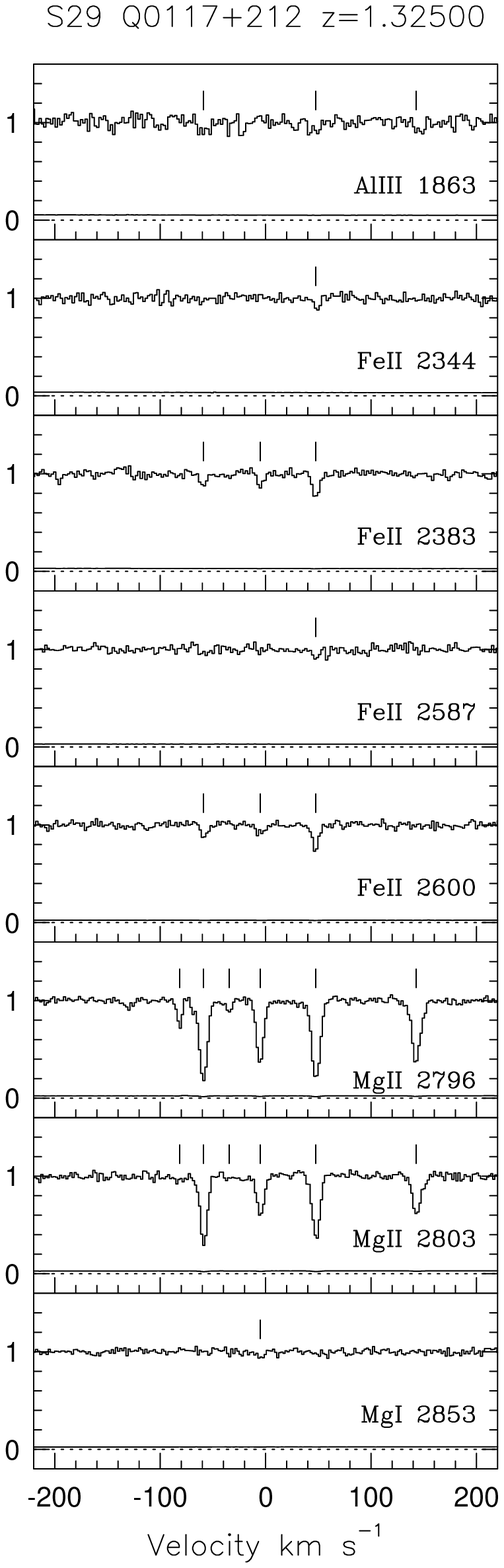}
\protect\caption
{Same as for figure \ref{fig:s1}}
\label{fig:s29}
\end{figure}
 
\begin{figure}[ht]
\myplotsix{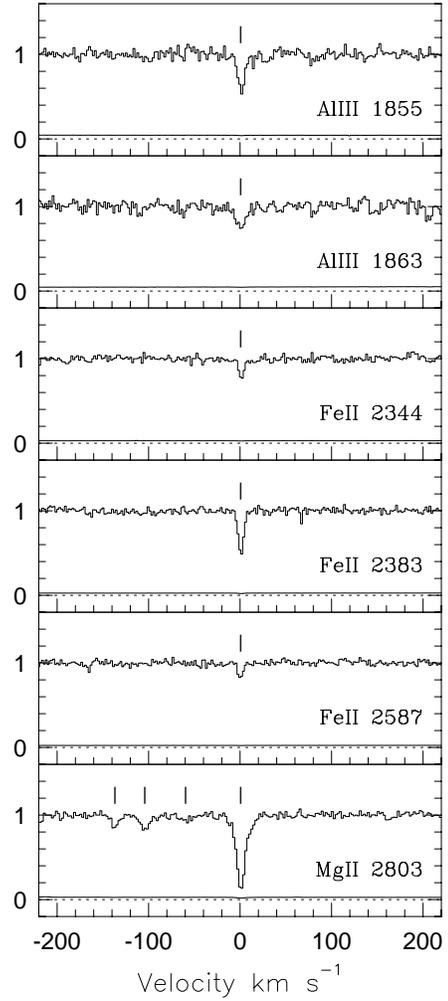} 
\protect\caption
{Same as for figure \ref{fig:s1}}
\label{fig:s30}
\end{figure}
 
\clearpage

\subsubsection{\rm S4  (Q$1222+228$; $z_{\rm abs}=0.550202$)}

There is no previous report of S4.
Though {\FeII} and {\MgI} were captured by the CCD, only the {\MgII}
doublet was detected.
However, {\FeII} $\lambda 2600$ may be present at the $2.7\sigma$
level.
A {\CIV} detection was ambiguous in the FOS/{\it HST\/} spectrum (also
see \cite{impey96}); we conservatively quote a non--restrictive limit.
There is no galaxy candidate (C. Steidel, private communication).

\subsubsection{\rm S5  (Q$1241+174$; $z_{\rm abs}=0.55844$)}

There is no previous report of S5.
Though {\FeII} and {\MgI} were captured by the CCD, only the {\MgII}
doublet was detected.
{\CIV} was tentatively detected in the FOS/{\it HST\/} spectrum;
however, the $\lambda 1548$ line may be {\Lya} (\cite{kp13}).
There is no galaxy candidate (C. Steidel, private
communication).
 
\subsubsection{\rm S6  (Q$0002+051$; $z_{\rm abs}=0.59149$)}

S6 is known to be associated with a red $\simeq 1.3L^{\ast}_{K}$
galaxy with impact parameter $\simeq 24 h^{-1}$~kpc (\cite{csv96}).
Though multiple {\FeII} transitions were captured by the CCD, none were
detected.
If {\MgI} is present, it was detected only at the $2\sigma$
significance level.
The equivalent widths may be biased by a zero--point uncertainty,
which has not been included in the error measurement.
{\CIV} was not detected in the FOS/{\it HST\/} spectrum (also see
\cite{kp13}).
 
\subsubsection{\rm S7  (Q$0454+036$; $z_{\rm abs}=0.64283$)}

S7 was studied by Churchill \& Le~Brun (1998\nocite{cl98}) and has 
no detectable {\CIV} in the FOS/{\it HST\/} spectrum.
There is no galaxy at this redshift in this well studied field.
Here, we also report a $3.3\sigma$ detection of {\FeII} $\lambda
2587$.  
{\MgI} was not detected.

\subsubsection{\rm S8  (Q$0823-223$; $z_{\rm abs}=0.705472$)}

There is no previous report of S8.
Though {\FeII} and {\MgI} were captured by the CCD, only the {\MgII}
doublet was detected.
{\CIV} was not detected in the FOS/{\it HST\/} spectrum.

\subsubsection{\rm S9  (Q$0058+019$; $z_{\rm abs}=0.72518$)}

There is no previous report of S9.
Relatively strong {\MgI} was detected.
There is a $3\sigma$ detection of {\FeII} $\lambda 2600$, which we
present in Table~\ref{tab:ews}, but this detection is insecure.
A FOS/{\it HST\/} spectrum of this QSO was not available.
There is no galaxy candidate, though there is a galaxy
at $z\simeq 0.68$ (C. Steidel, private communication).

\subsubsection{\rm S10  (Q$0117+212$; $z_{\rm abs}=0.72907$)}

S10, a multi--component system, is associated with a red and massive
$\simeq 3.7 L_{K}^{\ast}$ galaxy at an impact parameter of $\simeq 36
h^{-1}$ kpc (Churchill \etal 1996\nocite{csv96}).
The absorption at $v \simeq -120$~{\kms} near the $\lambda 2803$
transition is a {\TiII} transition from a damped {\Lya} absorber at
$z_{\rm abs} = 0.5764$.
Also, {\FeII} $\lambda 2600$ is nearly blended with {\MgI}
from this damped {\Lya} absorber. 
{\FeII} has been detected in four of the five {\MgII}
components and {\MgI} was detected in the strongest one.
{\CIV} was not detected in the FOS/{\it HST\/} spectrum (also see
\cite{kp13}).

\subsubsection{\rm S11  (Q$1548+093$; $z_{\rm abs}=0.77065$)}

SSB reported a weak, but unambiguous, {\MgII} absorbing system at
$z_{\rm abs} = 0.7708$.
This system is associated with a reddish low mass galaxy with $\simeq
0.1L^{\ast}_{K}$ at an impact parameter of $\simeq 23h^{-1}$~kpc
(C. Steidel, private communication). 
Though the HIRES spectrum is noisy, both {\FeII} and {\MgI} were
detected.
A FOS/{\it HST\/} spectrum of this QSO was not available.

\subsubsection{\rm S12  (Q$1634+706$; $z_{\rm abs}=0.81816$)}

There is no previous report of S12.
The {\MgII} is very weak, with $W_{\rm r}(2796) = 0.03$~{\AA}.
There is no detection of {\MgI}, where the signal--to--noise ratio is
quite high, nor of {\FeII}.
{\CIV} was not detected in the FOS/{\it HST\/} spectrum (also see
\cite{kp7}).

\subsubsection{\rm S13  (Q$1421+331$; $z_{\rm abs}=0.84325$)}

There is no previous report of S13.
However, Uomoto (1984\nocite{uomoto84}) reported two unidentified 
weak lines in his low resolution spectrum;  these two lines correspond
to {\FeII} $\lambda 2600$ and $\lambda 2374$ at this redshift.
This system is very rich in {\FeII} transitions.
Also, {\MgI} was detected.
A FOS/{\it HST\/} spectrum of this QSO was not available.

\resetfig

\begin{figure*}[bth]
\plotfiddle{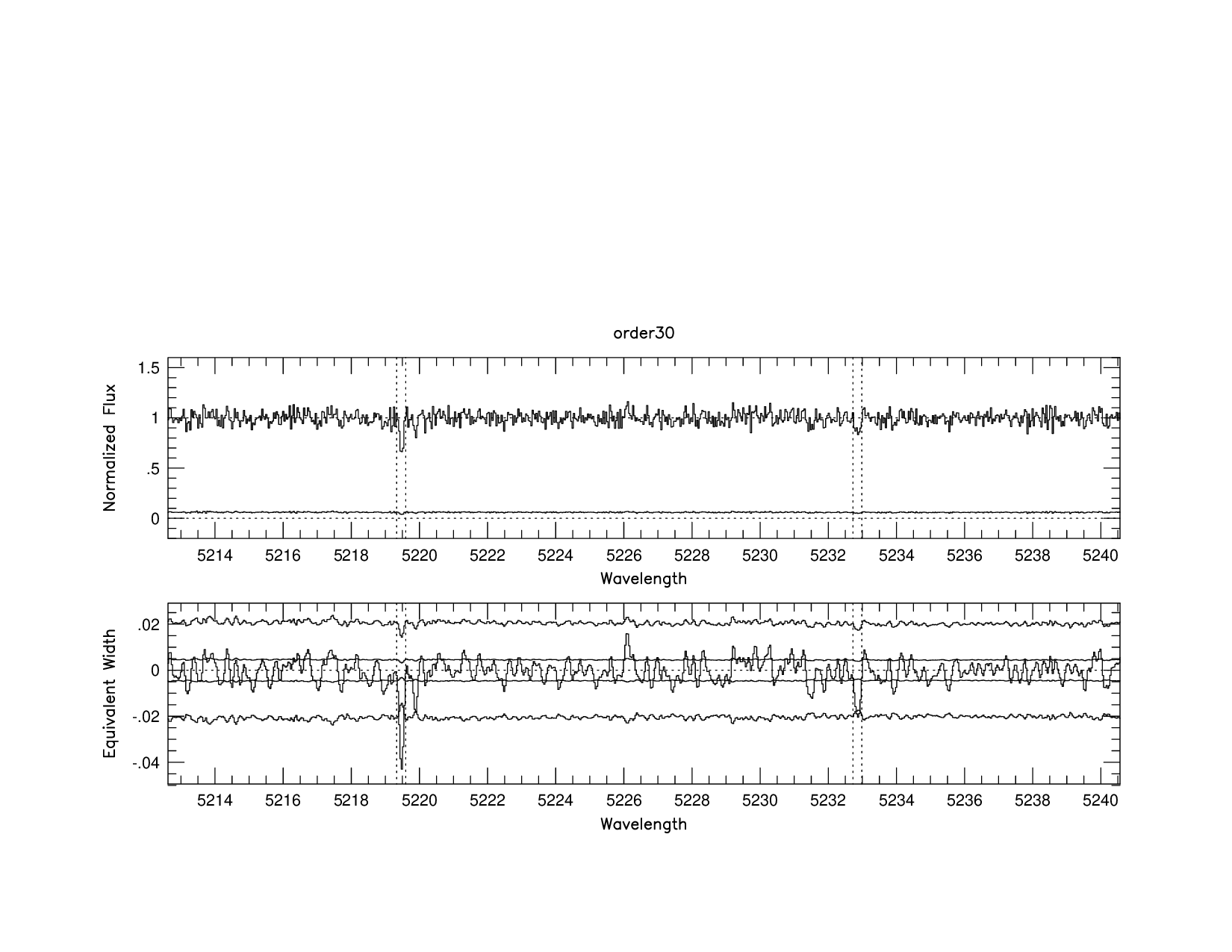}{3.3in}{0}{72.}{72.}{-286}{-40}
\protect\caption
{The S15 {\MgII} doublet detection ($5\sigma$) illustrated from our
detection software.  S15 is the weakest system in the sample and the
(unresolved) profile shapes of the $\lambda 2796$ and $\lambda 2803$
transitions are quite different, with the  $\lambda 2803$ profile
being broader.  This system provides an example of the detection
sensitivity of our doublet searching algorithm. The limiting observed
equivalent width on this order is roughly 0.02~{\AA}, which
corresponds to 0.011~{\AA} in the rest frame of the $\lambda 2796$
transition at these wavelengths. The top panel shows the spectrum and
the uncertainty spectrum. The lower panel shows the equivalent width
spectrum (average of zero). Pixels with positive equivalent widths are
emission features and those with negative equivalent widths are
absorption features.
The uncertainty in the equivalent width spectrum is shown at both
$1\sigma$ (inner) and $5\sigma$ (outer) levels.  
A feature is objectively identified when a pixel has an equivalent
width that is larger than the $5\sigma$ uncertainty.
The vertical dashed lines are to illustrate the locations of the
identified features. \label{fig:S15detection}}
\end{figure*}

\subsubsection{\rm S14  (Q$1248+401$; $z_{\rm abs}=0.85455$)}

There is no previous report of S14, which exhibits multiple components.
{\FeII} was detected only in the strongest {\MgII} component.
{\CIV} was reported by Jannuzi \etal (1998\nocite{kp13}) in the
FOS/{\it HST\/} spectrum and is confirmed in our search.
There is no galaxy candidate (C. Steidel, private communication).

\subsubsection{\rm S15  (Q$0002+051$; $z_{\rm abs}=0.86653$)}

There is no previous report of S15.
This is the weakest system in our survey, with $W_{\rm r}(2796)
= 0.018$~{\AA}.  
In Figure~\ref{fig:S15detection}, we show the detection of this
system.
{\FeII} and {\MgI} were captured by the CCD, but neither was detected.
The equivalent widths may be biased by a zero--point uncertainty,
which has not been included in the error measurement.
{\CIV} was not detected in the FOS/{\it HST\/} spectrum (also see
\cite{kp13}).
The Q$0002+051$ field has been studied in detail and there is no
galaxy (to roughly $0.2L_{K}^{\ast}$) observed at this redshift within
20{\arcsec} of the QSO (C. Steidel, private communication). 

\subsubsection{\rm S16  (Q$1241+174$; $z_{\rm abs}=0.89549$)}

There is no previous report of S16.
Neither {\FeII} nor {\MgI} were detected.
{\CIV} was not detected in the FOS/{\it HST\/} spectrum (also see
\cite{kp13}).
There is no galaxy candidate (C. Steidel, private communication).
 
\subsubsection{\rm S17  (Q$1634+706$; $z_{\rm abs}=0.90555$)}

There is no previous report of detected {\MgII} in S17, though 
{\CIV} was reported by Bergeron (1994\nocite{bergeron94}) and Bahcall
\etal (1996\nocite{kp7}).
{\MgI} was not detected, where the signal--to--noise ratio is high,
nor was {\FeII}.
 
\subsubsection{\rm S18  (Q$0454+036$; $z_{\rm abs}=0.93150$)}

S18 was studied by Churchill \& Le~Brun (1998\nocite{cl98}), who found
no {\CIV} in the FOS/{\it HST\/} spectrum and no galaxy at this
redshift. 
{\FeII} $\lambda 2383$ was detected, and {\FeII} $\lambda 2600$
would likely have been detected, but the region of the spectrum was
compromised by the pen mark on the HIRES CCD.
There is a tentative detection ($2.7\sigma$) of {\FeII} $\lambda
2587$.
{\MgI} was not detected.

\subsubsection{\rm S19  (Q$1206+456$; $z_{\rm abs}=0.93428$)}

S19 was studied by Churchill \& Charlton (1998\nocite{q1206}).
No {\FeII} transitions were detected, nor was {\MgI}.
S19 may be a member of a small group of galaxies.
Kirhakos \etal (1992\nocite{kirhakos92}) identified ten galaxies
within a 100{\arcsec} of the QSO, three of which are within
10{\arcsec}.
Thimm (1995\nocite{thimm95}) found strong {\OII} $\lambda 3727$
emission at $z=0.93$ from one of these galaxies.
In the FOS/{\it HST\/} spectrum, {\CIV} and {\OVI} are clearly present
(see Jannuzi \etal 1998\nocite{kp13}; Churchill \& Charlton
1998\nocite{q1206}).

\subsubsection{\rm S20  (Q$0002+051$; $z_{\rm abs}=0.95603$)}

There is no previous report of S20.
Neither {\FeII} nor {\MgI} was detected.
The equivalent widths may be biased by a zero--point uncertainty,
which has not been included in the error measurement.
{\CIV} was detected in the FOS/{\it HST\/} spectrum (also see
\cite{kp13}).
The Q$0002+051$ field has been studied in detail and there is no
galaxy (to roughly $0.2L_{K}^{\ast}$) observed at this redshift within
20{\arcsec} of the QSO (C. Steidel, private communication). 

\subsubsection{\rm S21  (Q$1329+412$; $z_{\rm abs}=0.97387$)}

There is no previous report of S21.
The $\lambda 2803$ transition is near the CCD edge.
{\FeII} was not detected.
{\MgI} may have been detected, but only at the $2.5\sigma$ level.
A non--restrictive limit was placed in {\CIV}, which falls in the
{\Lya} forest in the FOS/{\it HST\/} spectrum.
There is no galaxy candidate (C. Steidel, private communication).

\subsubsection{\rm S22  (Q$1329+412$; $z_{\rm abs}=0.99836$)}

There is no previous report of S22.
The system has strong {\FeII} absorption.
{\MgI} was not captured by the CCD.
{\CIV} was not detected in the FOS/{\it HST\/} spectrum.
There is no galaxy candidate (C. Steidel, private communication).

\subsubsection{\rm S23  (Q$1634+706$; $z_{\rm abs}=1.04144$)}

There is no previous report of detected {\MgII} in S23, though {\CIV}
was reported by Bergeron \etal (1994\nocite{bergeron94}) and Bahcall
\etal (1996\nocite{kp7}).
Even at very high signal--to--noise ratio, neither {\FeII}  nor {\MgI}
was detected.

\subsubsection{\rm S24  (Q$1213-003$; $z_{\rm abs}=1.12770$)}

There is no previous report of S24.
This system had a ``false alarm'' probability of $P_{\rm fa} = 0.009$,
the largest in the sample.
{\FeII} $\lambda 2383$ may have been detected at the $3\sigma$ level,
but there are two nearby $3\sigma$ features. 
We conservatively quote a limit.
A ground--based spectrum covering {\CIV} was not found in the
literature.

\subsubsection{\rm S25  (Q$0958+551$; $z_{\rm abs}=1.21132$)}

There is no previous report of S25.
Neither {\FeII} nor {\MgI} was detected.
A ground--based spectrum covering {\CIV} was not found in the
literature.

\subsubsection{\rm S26  (Q$0450-132$; $z_{\rm abs}=1.22948$)}

S26, a multi--component system, was reported by SS92\nocite{ss92}. 
The system is strong in {\FeII} absorption in the strongest {\MgII}
component.
The {\AlIII} $\lambda \lambda 1855, 1863$ doublet was detected in the
strongest component (also see Petitjean, Rauch, \& Carswell
1994\nocite{prc94}). 
This is the lowest redshift system in which {\AlIII} was covered.
{\MgI} was not detected.
A ground--based spectrum covering {\CIV} was not found in the
literature.

\subsubsection{\rm S27  (Q$0450-132$; $z_{\rm abs}=1.23244$)}

There is no previous report of S27.
The {\MgII} profiles is in two components.
{\FeII} and {\MgI}  were detected in the narrow component.
However, the {\MgI} detection is ambiguous.
A ground--based spectrum covering {\CIV} was not found in the
literature.

\subsubsection{\rm S28  (Q$0958+551$; $z_{\rm abs}=1.27238$)}

There is no previous report of S28.
{\FeII} was detected; however, the $\lambda 2600$ equivalent width is
unphysically large with respect to the more robust $\lambda 2383$
equivalent width.
{\MgI} was detected, but is deemed uncertain.
Sargent, Boksenberg, \& Steidel (1988\nocite{sbs88}) reported {\CIV}
at this redshift.

\subsubsection{\rm S29  (Q$0117+212$; $z_{\rm abs}=1.32500$)}

S29 was reported by SS92\nocite{ss92}.
The system is comprised of five distinct components.
{\FeII} was detected in three of the five components.
{\AlIII} $\lambda 1863$ was also detected in three components; 
{\AlIII} $\lambda 1855$ was not captured by the CCD.
{\MgI} was detected in the weakest {\MgII} component, and in this
component there is no {\AlIII}.
{\CIV} absorption was reported by SS92\nocite{ss92}, though the
doublet was not resolved.
There are candidate galaxies for this system (C. Steidel, private
communication).

\subsubsection{\rm S30  (Q$0117+212$; $z_{\rm abs}=1.34297$)}

S30 was also reported by SS92\nocite{ss92}.
Only the $\lambda 2803$ transition of the {\MgII} doublet was
captured. 
As such, this system would not have been detected in our unbiased
doublet search.
It is not a member of our adopted sample nor was it included in 
any of the system statistics.
The {\MgII} profile is comprised of four components.
{\MgI} was not detected.
{\FeII} was detected in the dominant, but narrow, {\MgII} component.
Also, {\AlIII} was found in this component.
{\CIV} was reported by SS92\nocite{ss92}, though the doublet was not
resolved.
There are candidate galaxies for this system (C. Steidel, private
communication).

\subsection{Survey Completeness}
\label{sec:completeness}

To evaluate the completeness of the survey as a function of redshift
we have adopted the formalism used by SS92\nocite{ss92} and
LTW\nocite{ltw87},
namely the ``redshift path density'', $g(W,z)$. 
This function gives the number of sight lines along which a {\MgII}
$\lambda 2796$ transition at redshift $z$ and with rest--frame
equivalent width greater than or equal to $W$ could have been
discovered.
Because of the high resolution of the spectra, we have slightly
modified the computation of the redshift path density to include 
sensitivity to the {\MgII} doublet ratio.
Thus, we have $g(W,z,{\rm DR})$, the number of sight lines along which
a {\MgII} {\it doublet\/} at redshift $z$ with  $\lambda 2796$
rest--frame equivalent width greater than or equal to $W$, and with
doublet ratio less than DR could have been detected.

\begin{figure}[th]
\plotfiddle{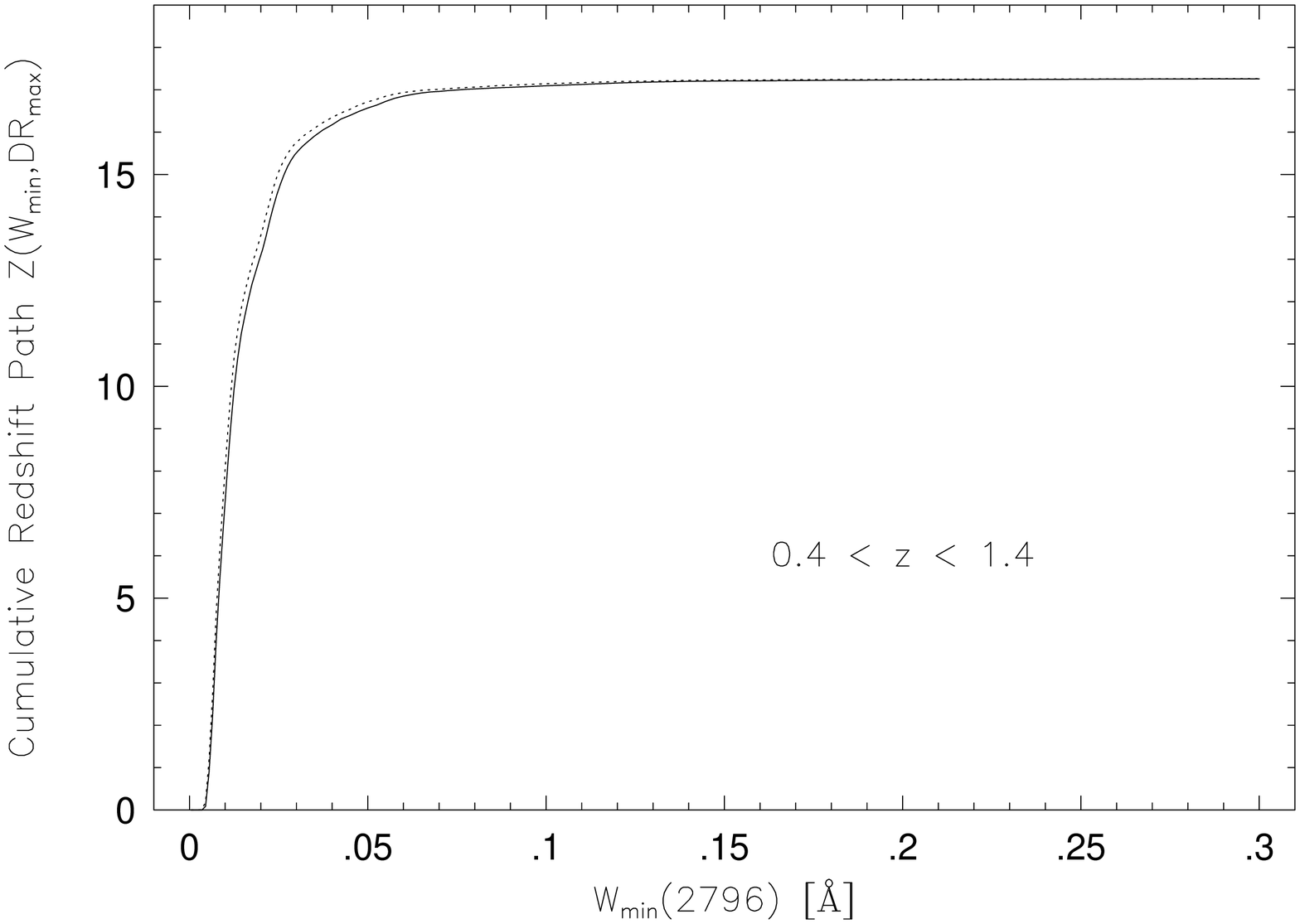}{2.2in}{0}{32.}{32.}{-122}{-8}
\protect\caption
{The redshift path of the survey over the redshift interval $0.4 \leq
z \leq 1.4$ verses the rest--frame {\MgII} $\lambda 2796$ equivalent
width threshold for a $5\sigma$ detection level.  The dotted curve is
for unit doublet ratio and the solid curve is for ${\rm DR} =
2$. \label{fig:ZofWDR}}
\end{figure}

The cumulative redshift path length covered by the survey over a
given redshift interval is then
\begin{equation}
Z(W_{\rm r},{\rm DR}) = \int _{z_{1}} ^{z_{2}} g(W_{\rm r},z,{\rm DR})
   dz,
\label{eq:redpath}
\end{equation}
where we have chosen not to integrate from $0 \leq z \leq \infty$
because our $g(W_{\rm r},z,{\rm DR})$ drops dramatically for $z_{1} <
0.4$ and for $z_{2} > 1.4$.
In Figure~\ref{fig:ZofWDR}, we have plotted the cumulative redshift
path of the survey as a function of $W^{\rm min}_{\rm r}$ for ${\rm
DR}^{\rm max} = 1$ (dotted curve) and for ${\rm DR}^{\rm max} = 2$
(solid curve).
It is apparent that the redshift path, and thus the survey
completeness, is not sensitive to the doublet ratio.  
We are 91\% 
complete at $W_{\rm r}(2796) = 0.03$~{\AA} ($5\sigma$), and 80\%
complete at $W_{\rm r}(2796) = 0.02$~{\AA}.
For comparison, SS92\nocite{ss92} were 83\% 
complete at $W_{\rm r}(2796) = 0.3$~{\AA}.


\section{The Statistical Properties of Weak {\MgII} Absorbers}
\label{sec:statistics}

\subsection{Redshift Number Density}
\label{sec:dndz}

\begin{figure}[bht]
\plotfiddle{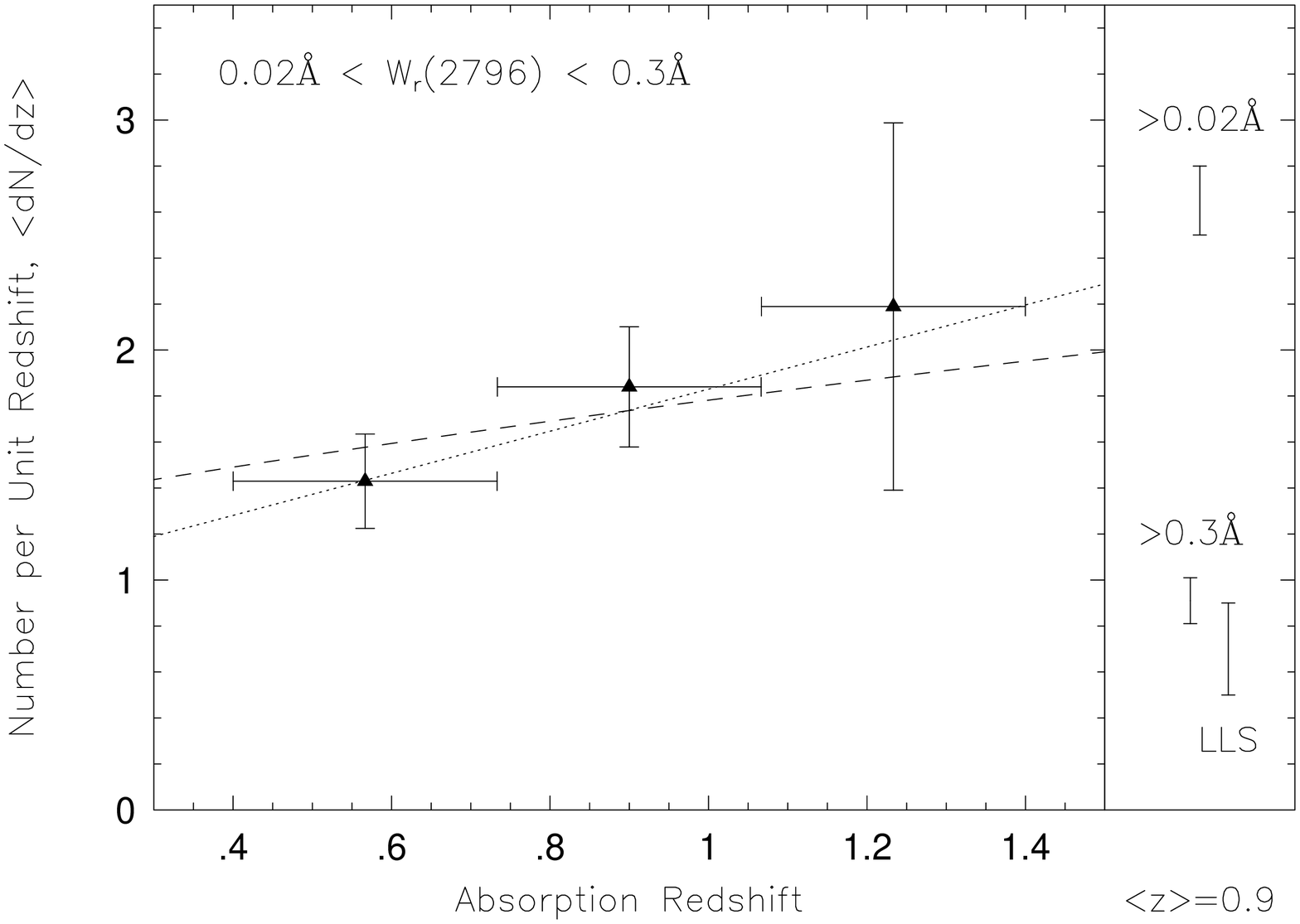}{2.2in}{0}{32.}{32.}{-122}{-8}
\protect\caption
{(left panel) The number of {\MgII} systems per unit redshift,
$dN/dz$, with $0.02 \leq W_{\rm r}(2796) < 0.3$~{\AA} for three
redshift bins over the interval $0.4 \leq z \leq 1.4$.  The vertical
error bars are the Poisson uncertainties in the $dN/dz$ and the
horizontal give the redshift bins.  The dotted curve is the
no--evolution expectations for $q_{0} = 0$ and the dashed curve is for
$q_{0} = 0.5$, where the curves have been normalized at $z=0.9$ (see
text). --- (right panel) A comparison of the {\MgII} $dN/dz$ for
different $\lambda 2796$ equivalent width cut offs with the LLS
$dN/dz$.  The $W_{\rm r}(2796) \geq 0.3$~{\AA} number is taken from
SS92, and the Lyman limit data are taken from Stengler--Larrea \etal
(1995). \label{fig:dNdz}}
\end{figure}

Since we are unbiased only for $W_{\rm r}(2796) < 0.3$~{\AA}, we
calculated the number of absorbers per unit redshift, $dN/dz$, for
this limited range using the formalism of LTW\nocite{ltw87}.
From Eq.~\ref{eq:redpath}, we computed the redshift path length,
$Z(W_{i},DR_{i})$, over which the $i$th system could have been
detected in our survey.
The values are presented in Table~\ref{tab:systems}. 
The number per unit redshift path is simply the sum of the reciprocal
of the cumulative redshift path lengths,
\begin{equation}
\frac{dN}{dz}  =
\sum _{i}^{N_{\rm sys}} \left[ Z(W_{i},DR_{i}) \right] ^{-1} .
\end{equation}
The variance in $dN/dz$ is given by
\begin{equation}
\sigma ^{2}_{dN/dz} = \sum _{i}^{N_{\rm sys}}
    \left[ Z(W_{i},DR_{i}) \right] ^{-2} .
\end{equation}
Over the redshift range $0.4 \leq z \leq 1.4$, we obtained
\begin{equation}
\frac{dN}{dz} = 1.74 \pm 0.11 \qquad \hbox{for}~~
0.02 \leq W_{\rm r}(2796) < 0.3~\hbox{\AA}, 
\end{equation}
where $\left< z \right> = 0.9$.
In the left hand panel of Figure~\ref{fig:dNdz}, we have plotted 
$dN/dz$ verses redshift for $0.02 \leq W_{\rm r}(2796)
< 0.3$~{\AA}, for three redshift bins, $[0.40,0.74]$, $[0.74,1.07]$,
and $[1.07,1.40]$.
The curves represent the no--evolution expectations for $q_{0} = 0.5$
(dashed curve) and for $q_{0} = 0$ (dotted curve) normalized to 
$dN/dz = 1.74$ at $z = 0.9$.
We have assumed the standard parameterization, $dN/dz = N_0 (1+z)
^{\gamma}$, where $\gamma = 1$ for $q_{0} = 0$ and $\gamma = 0.5$ for 
$q_{0} = 0.5$.
A formal fit yielded $\gamma = 1.3\pm0.9$ and $N_0 = 0.8\pm 0.4$.
The data are not inconsistent with the no--evolution expectations for
either $q_0$.
In the right hand panel of Figure~\ref{fig:dNdz}, we have plotted 
the mean $dN/dz$ and its uncertainty at $\left< z \right> = 0.9$ for:
(1) the $W_{\rm r} ^{\rm min}(2796) = 0.3$~{\AA} MG1 sample
of SS92\nocite{ss92}, which has $dN/dz = 0.91\pm0.10$; 
(2) the combined $W_{\rm r} ^{\rm min}(2796) = 0.02$~{\AA} sample,
which has $dN/dz = 2.65\pm0.15$;
and (3), the {\it HST\/} Key Project results for a sample of Lyman
limit systems (LLS), which has $dN/dz = 0.7\pm0.2$ over the interval
$0.4\leq z \leq 1.4$ (\cite{stengler-larrea}).

{\it Taken at face value, these numbers imply that weak {\MgII}
absorbers comprise $\sim 65$\%  of the total {\MgII} absorber 
population and that the vast majority of them must arise in sub--LLS
environments}. 
We return to this point in \S\ref{sec:llsdiscuss}.
The $dN/dz$ of weak {\MgII} absorbers is roughly $5-7$\% of that of
the {\Lya} forest with $W_{\rm r}({\Lya}) \geq 0.1$~{\AA}
(\cite{kp13}).
We tentatively suggest that $\sim 5$\% of $z \leq 1$ ``{\Lya} forest
clouds'' with  $0.1 \leq W_{\rm r}({\Lya}) \leq 1.6$~{\AA} will
exhibit {\MgII} absorption to a 5$\sigma$ $W_{\rm r}(2796)$
detection limit of 0.02~{\AA}.
The two {\MgII} systems found by Churchill \& Le~Brun
(1998\nocite{cl98}) in a search through 28 forest clouds in the
spectrum of PKS~$0454+039$ are consistent with these expectations.

\subsection{Equivalent Width Distribution}
\label{sec:ewdist}

The distribution function, $n(W)$, is defined as the number of {\MgII}
absorption systems with equivalent width $W$ per unit equivalent width
per unit redshift path.
It has been customary to parameterize the distribution by either an
exponential,
\begin{equation}
n(W) dW = \left( \frac{N_{\ast}}{W^{\ast}} \right) 
\exp \left( - \frac{W}{W^{\ast}}\right) dW ,
\label{eq:nWexp}
\end{equation}
or a power law,
\begin{equation}
n(W) dW = CW^{-\delta} dW ,
\label{eq:nWpwrlaw}
\end{equation}
where $N^{\ast}$ and $W^{\ast}$ (for the exponential)
and  $C$ and $\delta$ (for the power law) are parameters obtained by 
fitting the data.
TBSYK\nocite{tytler87} fit the distribution with a power law of
$\delta \sim 2$ for a sample with $W_{\rm r}^{\rm min}(2796) =
0.25$~{\AA} and mean redshift $\left< z \right> \simeq 0.5$.
For $W_{\rm r}^{\rm min}(2796) = 0.3$~{\AA} systems at $\left< z
\right> \simeq 1.6$, LTW\nocite{ltw87} fit the distribution to both an
exponential and a power--law distribution (the latter in agreement with
TBSYK\nocite{tytler87}).
LTW\nocite{ltw87} concluded that both adequately represented the data,
with the exponential distribution slightly favored.
SS92\nocite{ss92} found that $n(W)$ could be parameterized tolerably
well by either an exponential, with $N^{\ast} \sim 1.5$ and $W^{\ast}
\sim 0.66$, or by a power law, with $C \sim 0.4$ and $\delta \simeq
1.65$.
SS92\nocite{ss92} noted that the exponential underpredicted the number
of $W_{\rm r}(2796) \leq 0.5$~{\AA} systems whereas the power law
underpredicted the number with ``intermediate'' equivalent widths,
those with $0.7 \leq  W_{\rm r}(2796) \leq 1.3$~{\AA}.

\begin{figure*}[bth]
\vglue 0.5in
\plotfiddle{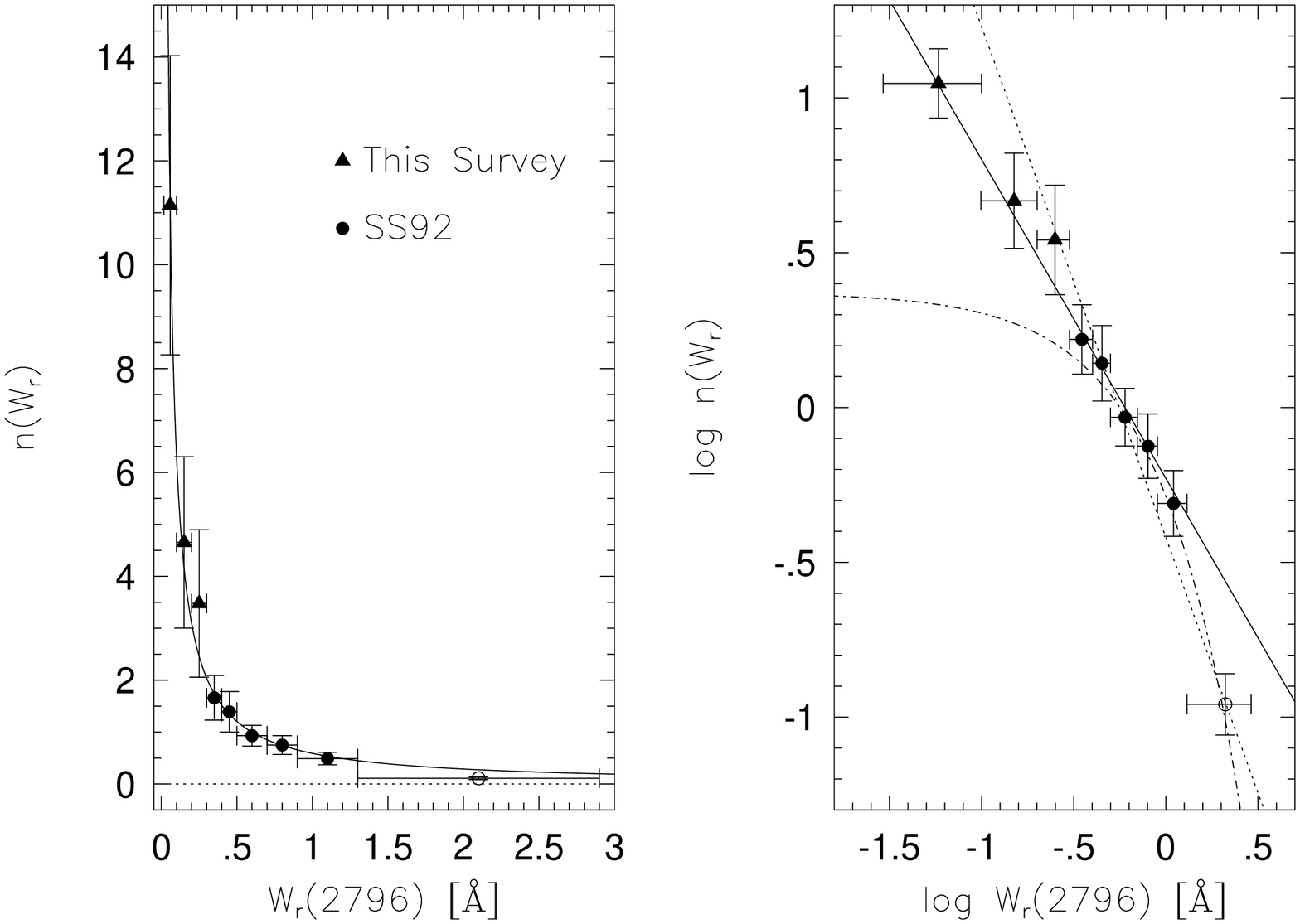}{3.5in}{0.}{60.}{60.}{-250}{0}
\vglue -0.35in
\protect\caption
{The equivalent width distribution verses the rest--frame equivalent
width of the {\MgII} $\lambda 2796$ transition. The left hand panel
illustrates the rapid increase in the number of small $W_{\rm
r}(2796)$ systems.  Solid triangles are from this study and solid
circles are taken from SS92.  The right hand panel shows a comparison
of our power--law fit (solid line) to the combined data set from this
study and SS92.   Also shown are the fits presented by SS92 for a
power law (dot--dot) and an exponential (dash--dot).  The largest
equivalent widths [$W_{\rm r}(2796) \geq 1.3$~{\AA}; shown with an
open dot] from the SS92 data are predominantly from redshifts higher
than $z=1.4$, our upper cut off.  Due to the evolution of the strongest
systems, this binned data point would be lower if the SS92 data were
limited to our redshift coverage (see the text).
\label{fig:ewdist}}
\end{figure*}

In Figure~\ref{fig:ewdist}a and b, we present $n(W_{\rm r})$ 
for $W_{\rm r}(2796) \geq 0.0165$~{\AA}.
We are 70\% complete to this equivalent width.
Solid triangles are the results from our survey.   
The data have been binned accounting for the redshift path length over
which the equivalent widths could have been detected.
The three equivalent width bins are $[0.0165, 0.1]$, $[0.1, 0.2]$, and 
$[0.2, 0.3]$~{\AA}.
Shown as solid circles are the binned equivalent width data from
SS92\nocite{ss92} (see their Figure~6), which span the range $0.3 \leq
W_{\rm r}(2796) \leq 2.8$~{\AA}.
Figure~\ref{fig:ewdist}a illustrates the dramatic increase in
$n(W_{\rm r})$ with decreasing equivalent width.  
{\it There is no turnover or break in the equivalent width
distribution for $0.02 \leq W_{\rm r}(2796) < 0.3$~\hbox{\rm {\AA}}
at $\left< z \right> = 0.9$}.

The lack of a turnover is further illustrated in
Figure~\ref{fig:ewdist}b, which shows $\log n(W_{\rm r})$ verses $\log
W_{\rm r}(2796)$.
The solid curve is a power law with $\delta = 1.04$, which was
obtained by minimizing the absolute deviation between the curve
(Eq.~\ref{eq:nWpwrlaw}) and the binned data.
The absolute deviation for the presented fit is 0.07 (in log--log).
This fit is also presented in Figure~\ref{fig:ewdist}a.
It would be proper to fit the combined data of this survey and
SS92\nocite{ss92} using the maximum likelihood technique employed by
LTW\nocite{ltw87}, TBSYK\nocite{tytler87}, and SS92\nocite{ss92}.
This would require that we invoke the $g(W,z)$ function of
SS92\nocite{ss92} and reanalyze the SS92\nocite{ss92} data over the
redshift interval $0.4 \leq z \leq 1.4$.
However, with this work, it is our intent to clearly demonstrate the
absence of a turnover in the equivalent width distribution at small
equivalent widths and to discern between the exponential and power--law
parameterizations of the distribution.
A reanalysis of the SS92\nocite{ss92} data was not required to
demonstrate these points.

Neither Eq.~\ref{eq:nWexp} nor \ref{eq:nWpwrlaw} accounts for redshift
evolution in the distribution.
However, such evolution has been observed.
TBSYK\nocite{tytler87} found evidence for more large equivalent width
absorbers at high redshift than at low redshift. 
PB90\nocite{pb90} noted that the ratio of ``weak'' to ``strong''
absorbers, demarcated by $W_{\rm r}(2796) = 0.6$~{\AA}, increased with
decreasing redshift.
SS92\nocite{ss92} measured how the number density evolution of {\MgII}
systems changed as a function of $W_{\rm r}^{\rm min}(2796)$.
As $W_{\rm r}^{\rm min}(2796)$ is increased, evolution becomes
pronounced; large equivalent width systems evolve away with time.
When $W_{\rm r}^{\rm min}(2796) = 0.3$~{\AA} is applied, the
population of {\MgII} absorbers is consistent with no--evolution
expectations.
It appears that the evolution of the strongest systems, those with
$W_{\rm r}(2796) \geq 0.6$~{\AA}, may have strongly biased the
SS92\nocite{ss92} fits to the equivalent width distribution (a point
they address in their paper).
Slightly more than 60\% of the equivalent widths\footnote{The estimate
of 60\% is based upon the observed numbers.  
Accounting for the relative number of sight lines observed by
SS92\nocite{ss92} above and below $z = 1.4$ would raise this estimate
and accentuate our point.}  contributing to the largest bin ($[1.3,
2.8]$~{\AA}) taken from SS92\nocite{ss92} arise at $z\geq 1.4$. 
We have plotted this bin as an open circle in
Figures~\ref{fig:ewdist}a and \ref{fig:ewdist}b.
If we had corrected this bin for our limited redshift coverage, it
would be roughly 40\% of its plotted value ($\log n \sim -1.46$, which
is off the bottom of Figure~\ref{fig:ewdist}b). 
There is no compelling evidence of such strong evolution in the
equivalent width distribution for $0.3 \leq W_{\rm r}(2796) \leq
1.3$~{\AA}.
Thus, we have omitted the [1.3,2.8]~{\AA} bin from our fit, and
include only the bins for which evolution is thought to be
negligible.
To emphasize our point, we note that SS92\nocite{ss92} obtained a
significantly steeper power law ($\delta = 1.65$) than the one we
quote here ($\delta = 1.04$).
This is likely due to the fact that the distribution of
SS92\nocite{ss92} was fit over the full redshift range $0.2 \leq z
\leq 2.2$, resulting in a bias from the relative paucity of large
equivalent width systems at $z\leq 1.4$.
This is illustrated in Figure~\ref{fig:ewdist}b, in which we have
plotted both the power--law fit (dot--dot) and the exponential fit
(dash--dot) from SS92\nocite{ss92}.
Note that the $W_{\rm r}(2796) \leq 1.3$~{\AA} data of
SS92\nocite{ss92} also appear to be best described by the $\delta =
1.04$ power law distribution.

In summary, there is no turnover in the equivalent width distribution
for $W_{\rm r}(2796) < 0.3$~{\AA}, but there is a strong break above
$W_{\rm r}(2796) \simeq 1.3$~{\AA} for $z \leq 1.4$.
The upper limit on the slope for $W_{\rm r}(2796) \geq 1.3$~{\AA} is
$\delta = 2.3$.

\subsection{Clustering and the Issue of a Biased Sample}
\label{sec:clustering}

The spectra used for this study are biased toward strong {\MgII} absorbers.
If the weak {\MgII} systems tend to cluster around the stronger
systems, then it is difficult to argue that the sample of weak systems
is unbiased.
On the other hand, if they do not cluster about the strong systems,
we can conclude that the QSO sight lines are not biased toward an
overabundance of weak systems.

\begin{figure}[b]
\plotfiddle{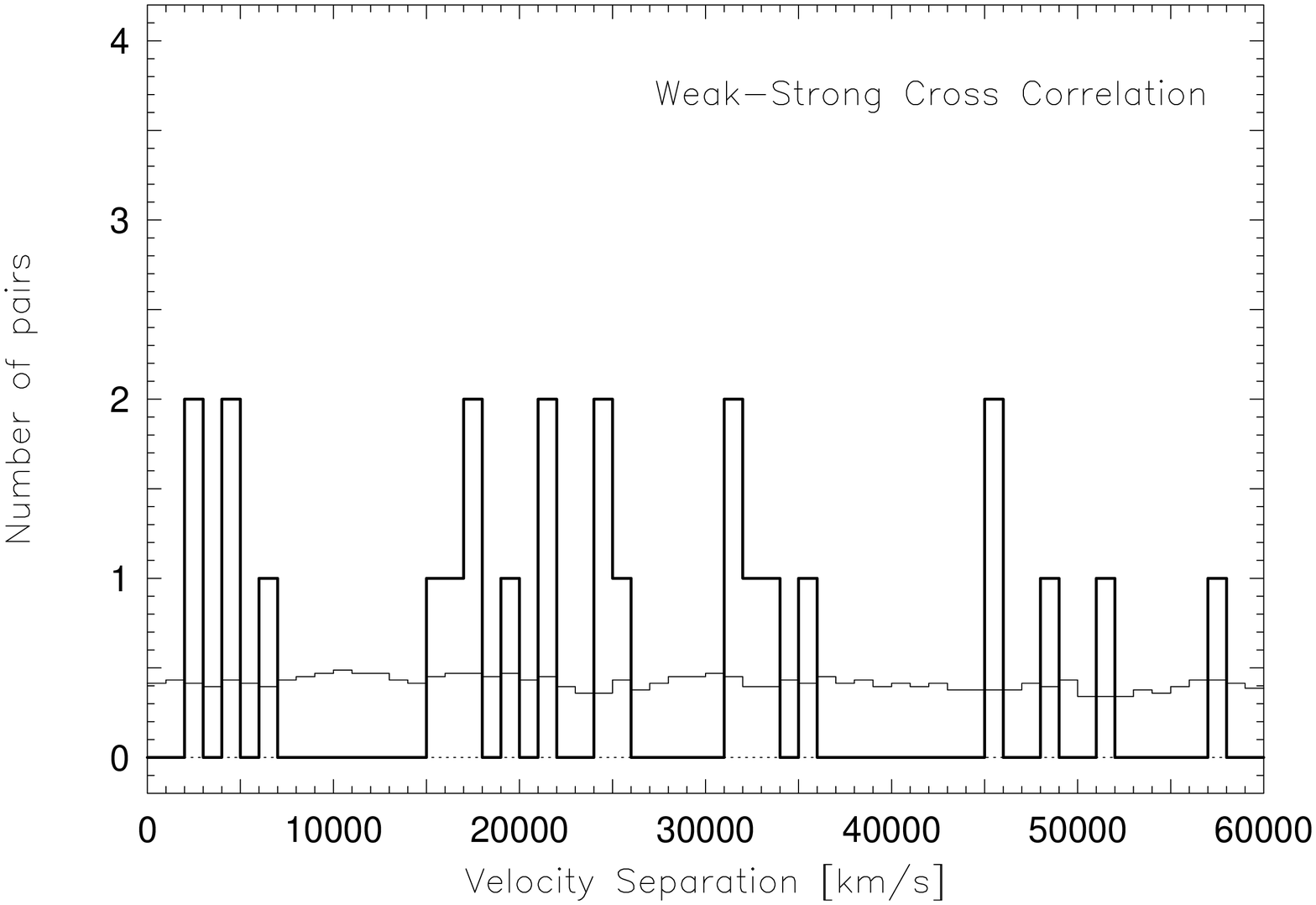}{2.2in}{0}{32.}{32.}{-122}{-8}
\protect\caption
{The two--point velocity cross--correlation function of weak and
strong systems.  Each bin gives the number of weak systems with
velocity separations $\Delta v$ [{\kms}] from strong systems along the
same QSO sight line.  The expected number in each bin for a random
distribution is given by the thin histogram.  There is no evidence
that the weak systems cluster in velocity about the strong systems.
This implies that the QSO sight lines, though biased for the strong
systems, are in fact unbiased for the presence of weak
systems. \label{fig:vcorr}}
\end{figure}

We have computed the two--point velocity correlation
function, which is presented in Figure~\ref{fig:vcorr}, where the
thick histogram distribution is the cross--correlation function of
weak systems with respect to the strong systems.
There are no weak systems with velocity separations less than
1000~{\kms} from the strong ones, and there is no apparent signal in
the velocity separations.
To test if the weak systems are distributed like a random population
with respect to the strong systems, we have computed the relative 
probability, $P(\Delta v)$, of detecting a $\Delta v$ separation from
each strong absorption system.
Following SS92\nocite{ss92}, we have limited our co--moving velocity
difference to $\Delta v = 60,000$~{\kms}, and have normalized the
probability integral to the observed number of weak systems.
Formally, the observed $\Delta v$ distribution is not inconsistent
with a random distribution; a $\chi^{2}$ test on the binned data
(1000~{\kms} bins) yielded a probability of 0.18 that the two
distribution were drawn from the same parent population.

We examined the redshift clustering of the weak systems with respect
to one another and found that it is not inconsistent with a random
distribution.  
The $\chi ^{2}$ probability was 0.25.
These results suggest that the weak systems are statistically
consistent with a random cosmological distribution.
We conclude that the QSO sight lines surveyed are unbiased
for the presence of weak {\MgII} absorbers.

\subsection{Absorption Properties}
\label{sec:properties}

A more detailed examination of the cloud to cloud chemical and
ionization conditions will be presented in a companion paper
(Churchill \etal 1998, \cite{paper2}).
In the HIRES spectra, only {\MgI} $\lambda 2853$, {\FeII} $\lambda
2344$, 2374, 2383, 2587, and 2600, and the {\AlIII} $\lambda \lambda
1855, 1863$ doublet were detected.
In the FOS/{\it HST\/} spectra we have limited our search to {\CIV}
and supplemented this with measurements from ground--based
observations taken from the literature.
The statistics are as follows: 13 of 29 have detected {\FeII}
(either $\lambda 2383$ or $\lambda 2600$), seven of 29 have detected
{\MgI} $\lambda 2853$ and each of these also has {\FeII}, three of
four have detected {\AlIII}, and nine of 22 have {\CIV}.
The $3\sigma$ average equivalent width threshold for {\FeII} is
0.01~{\AA} for {\FeII} $\lambda 2600$ and 0.008~{\AA} for {\FeII}
$\lambda 2383$.
The average threshold for {\MgI} absorption is 0.006~{\AA}, for
{\AlIII} is 0.01~{\AA}, and for {\CIV} is 0.16~{\AA}.

\begin{figure}[thb]
\plotfiddle{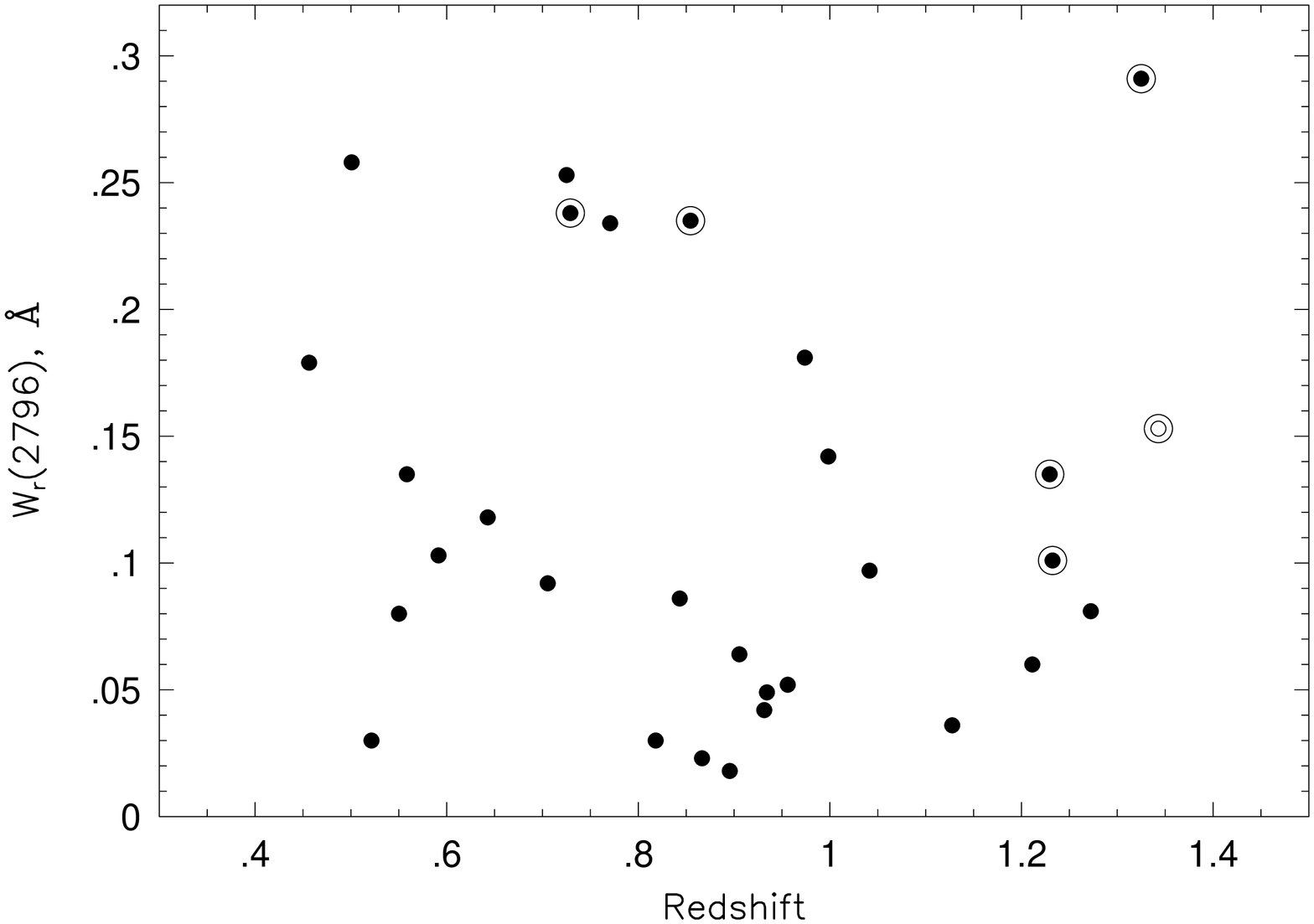}{2.2in}{0}{32.}{32.}{-122}{-8}
\protect\caption
{The full system rest--frame equivalent width of the {\MgII} $\lambda
2796$ transition verses the absorption redshift. There is no evidence
of any trend in $W_{\rm r}(2796)$ with redshift. There is an
indication that most all individual {\MgII} absorbing clouds likely
have $W_{\rm r}(2796) \leq 0.15$~{\AA}.  Those systems that are
comprised of multiple absorption features are marked with a concentric
circle; the individual ``cloud'' $W_{\rm r}(2796)$ are $\sim
0.15$~{\AA} or less. The three remaining systems with $W_{\rm r}(2796)
\sim 0.25$~{\AA} have relatively poor signal--to--noise ratios; they
are likely to be unresolved multiple clouds.  See text for
discussion. \label{fig:ew_vs_zabs}}
\end{figure}

\begin{figure}[thb]
\plotfiddle{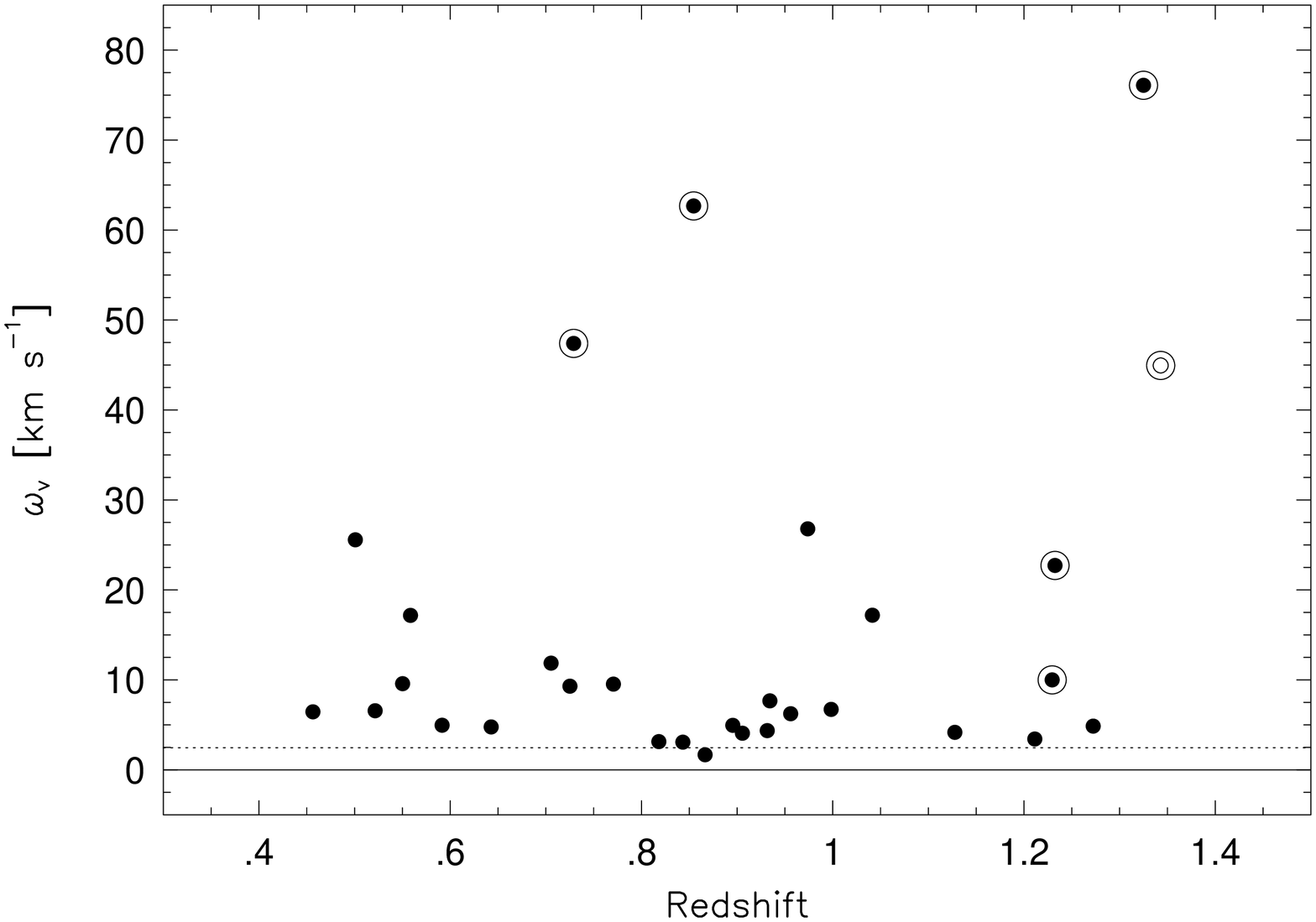}{2.2in}{0}{32.}{32.}{-122}{-8}
\protect\caption
{The velocity width, $\omega _{v}$ (equivalent Gaussian width), of the
{\MgII} $\lambda 2796$ transition as a function of absorption
redshift. The dotted line gives the limit for unresolved absorption
features. The data points with ``rings'' are those systems that have
multiple absorption features separated by continuum.  The open data
point (with ring) is measured from the $\lambda 2803$ transition
because the  $\lambda 2796$ transition was unavailable.  There is no
evidence for any trend in $\omega _{v}$ with
redshift. \label{fig:b_vs_zabs}}
\end{figure}
 
Plotted in Figure~\ref{fig:ew_vs_zabs} are the rest--frame {\MgII}
$\lambda 2796$ equivalent widths as a function of redshift.
There appears to be no trend in the distribution of weak absorbers
with redshift.
We have run Spearman--Kendall (SK) non--parametric rank correlation
tests to explore if any correlations are present among the detected
absorption properties (limits were not included).
We tested redshifts, velocity widths, equivalent widths, and doublet
ratios against one another and found no correlations.  
The most suggestive ranking was an anti--correlation between 
$W_{\rm r}(2796)$ and the {\MgII} doublet ratio at the $1.4\sigma$
level.
As seen in Figure~\ref{fig:ew_vs_zabs}, most, if not all, of
the {\it individual clouds\/} in these weak absorbers have $W_{\rm
r}(2796) \leq 0.15$~{\AA}.
The three $W_{\rm r}(2796) \geq 0.2$~{\AA} systems, S2, S9, and S11,
are likely comprised of two or more blended ``clouds''.
In the case of S2, the profile clearly has multiple clouds.

Several of the systems have multiple components.
These systems are S10, S14, S26, S27, S29, and S30.
The detection of these multiple features is {\it not\/} correlated
with the signal--to--noise ratio in the spectra, since they could be
detected to $W_{\rm r}(2796) = 0.02$~{\AA} in 80\% of the systems.
Plotted in Figure~\ref{fig:b_vs_zabs} are the velocity widths, $\omega
_{v}$, of the full {\MgII} $\lambda 2796$ profiles as a function
redshift.
The dotted line at $\omega _{v} = 2.46$~{\kms}, which is the Gaussian
width of the instrumental profile, shows the threshold for fully
unresolved features.
Systems that have been resolved into multiple individual ``clouds'' are
marked with a concentric circle.
Note that the many single ``cloud'' systems are unresolved or only
marginally resolved.
The same holds true for the individual clouds in the multiple cloud
systems.
The average velocity width, $\omega _{v}$, of the individual
``clouds'' for the full sample is $\sim 4$~{\kms}, which implies an
average temperature of $\sim 25,000$~K for thermal broadening.
Of the seven systems in which {\MgI} is detected, three have $W_{\rm
r}(2796) \leq 0.11$~{\AA}.
Two of the {\MgI} ``clouds'' are in multiple component systems, S10
and S29, and these clouds have $W_{\rm r}(2796) =  0.11$~{\AA} and
$0.05$~{\AA}, respectively.
There does not appear to be a clear threshold for the presence
of {\MgI} with the equivalent width of {\MgII} down to $W_{\rm
r}(2796) \sim 0.1$~{\AA}.  
It appears that {\MgI} can survive in sub--LLS environments.

\begin{figure}[htb]
\plotfiddle{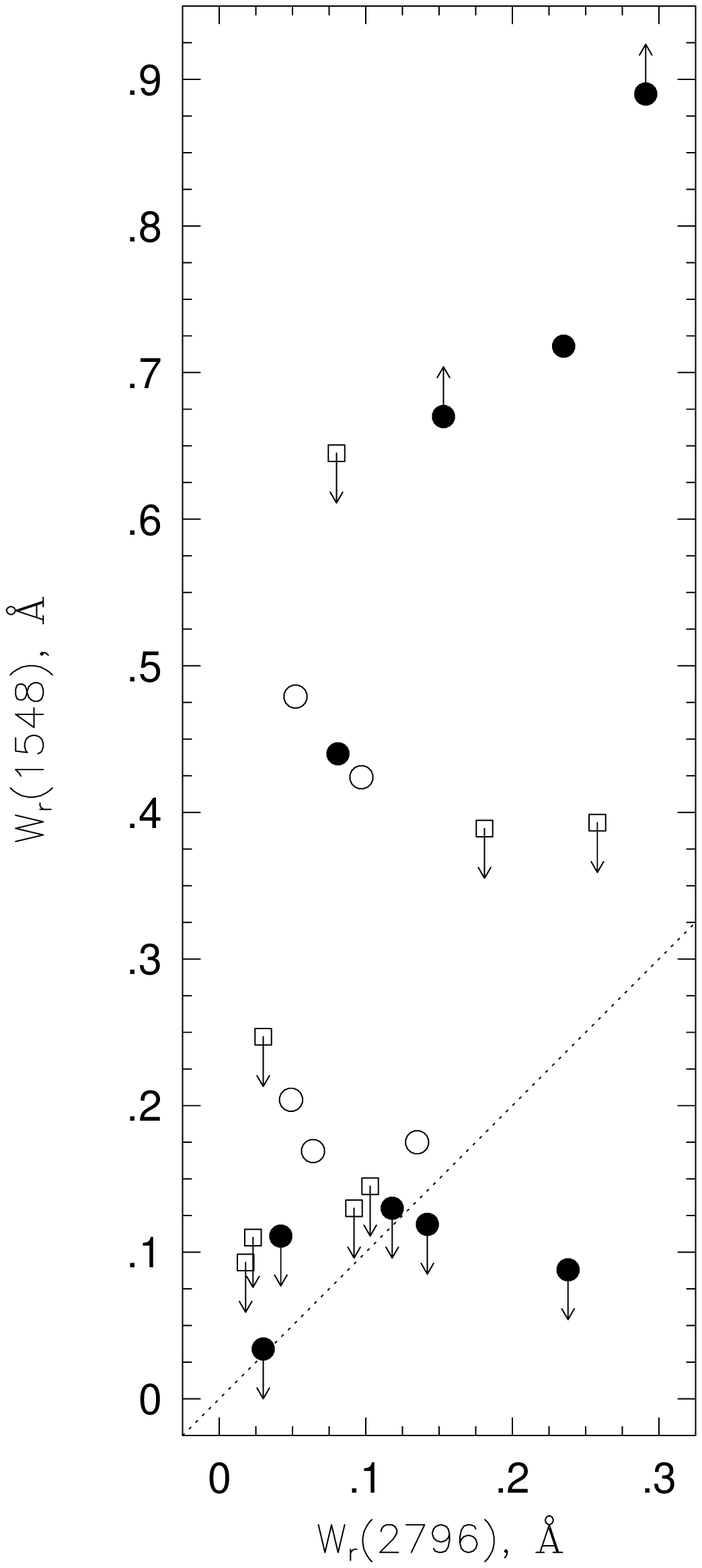}{6.3in}{0}{70.}{70.}{-211}{-37}
\protect\caption
{$W_{\rm r}(1548)$ verses $W_{\rm r}(2796)$.  Based upon
photoionization models, the data are designated as low or multi--phase
ionization conditions (solid circles), high ionization (open circles),
or undetermined (open boxes).  See text for definitions.  Downward
arrows give upper limits on $W_{\rm r}(1548)$.  The two upward arrows
are lower limits for an assumed optically thick doublet ratio in
unresolved optical data taken from the literature. The dotted line is 
$W_{\rm r}(1548) = W_{\rm r}(2796)$. \label{fig:civfig}}
\end{figure}

In Figure~\ref{fig:civfig}, we present $W_{\rm r}(1548)$ verses
$W_{\rm r}(2796)$ from the data presented in Table~\ref{tab:civcomp}.
Included in this table are the number of absorption components (not VP
components) seen in {\MgII} absorption, and $W_{\rm r}(2796)$, $W_{\rm
r}(2600)$, and  $W_{\rm r}(1548)$.
Upper limits are denoted with downward arrows.
A tentatively suggested ionization condition, either L (low),
H (high), or M (multi--phase), is given.
Filled circles are those designated as L or M, open circles as H, and
boxes as undetermined ionization conditions.
We elaborate on these inferred conditions in \S\ref{sec:civdiscuss},
where the terms L, H, and M are defined and photoionization models are
presented.

\clearpage

In summary, we find that a typical individual {\MgII} absorbing cloud
is characterized by $W_{\rm r}(2796) \leq 0.15$~{\AA}, and a
temperature of $\sim 25,000$~K.
The presence or non--presence of {\MgI} does not appear have a
threshold dependence upon $W_{\rm r}(2796)$ down to $0.1$~{\AA}, and
the ionization conditions appear to cover a broad range as inferred
from the equivalent width ratios of {\CIV}, {\FeII} and {\MgII}.


\section{On the Nature of Weak {\MgII} Absorbers}
\label{sec:discuss}

\subsection{Comparison with Lyman Limit Systems}
\label{sec:llsdiscuss}

A corresponding Lyman limit break is almost always found
in UV spectra at the redshift of a strong {\MgII} absorber
(\cite{lanzetta88}).
Thus, the majority of strong {\MgII} absorbers are believed to arise
in LLS environments.
We are led to conclude that {\it virtually all weak  
{\MgII} absorbers arise in sub--LLS environments}, as can be inferred
directly from Figure~\ref{fig:dNdz} (right hand panel).
The $dN/dz$ of $W_{\rm r}^{\rm min} (2796) = 0.3$~{\AA} {\MgII}
absorbers and of LLS absorbers are consistent within uncertainties,
with the number density of the strong {\MgII} absorbers being slightly
higher (SS92\nocite{ss92}; \cite{stengler-larrea}).
However, for the $W_{\rm r}^{\rm min} (2796) = 0.02$~{\AA} {\MgII}
absorbers (combined sample of this work and SS92\nocite{ss92}), $dN/dz
= 2.65$, which is a factor of $3.8\pm1.1$ greater than that of the
strong {\MgII}--LLS absorbers. 

That weak {\MgII} systems arise in sub--LLS environments is also
consistent with the expectations for photoionized clouds.
From preliminary Voigt profile fits, we have found $10^{11.8} \leq
N({\MgII}) \leq 10^{13.2}$~{\cm2} for the individual clouds.
We built a grid of photoionization models using CLOUDY
(\cite{ferland}), where we have assumed a Haardt \& Madau
(1996\nocite{hm96}) extra--galactic UV background spectrum normalized at
$z=1$ and a solar abundance pattern with $[Z/Z_{\odot}] = -1$ (which
can be interpreted as the ``gas--phase metallicity'', accounting for
possible dust depletion of {\MgII}).
The model clouds are constant density plane--parallel slabs, each
defined by its neutral hydrogen column density, $N({\HI})$, and
ionization parameter, $U = n_{\gamma}/n_{\rm H}$, where $n_{\gamma}$
and $n_{\rm H}$ are the number density of photons capable of ionizing
hydrogen and the total hydrogen number density, respectively.

\begin{figure}[th]
\plotfiddle{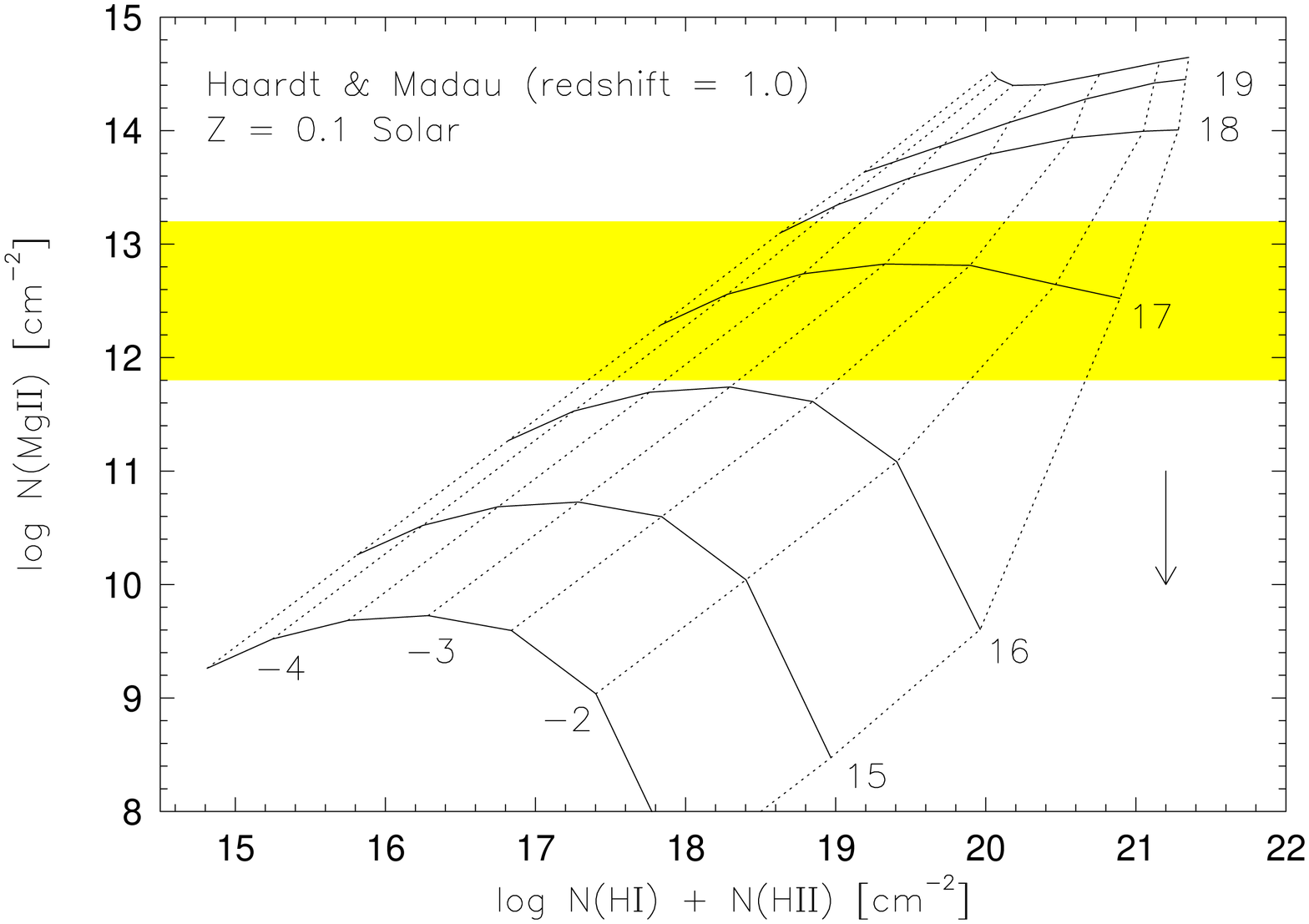}{2.2in}{0}{32.}{32.}{-122}{-8}
\protect\caption
{{\MgII} column density verses total hydrogen column density for a
grid of CLOUDY photoionization models, assuming a Haardt \& Madau
(1996) UV flux at $z=1$.  A $Z= 0.1Z_{\odot}$ abundance pattern has
been assumed.  The solid curves are contours of constant $N({\HI})$
and the dot--dot curves are contours of constant ionization parameter,
$\log U$.  The shaded region gives the locus of measured $N({\MgII})$
for the sample, based upon Voigt profile fits to the data
(Paper II).  For this abundance pattern, clouds which have
$N({\FeII})/N({\MgII})\sim 1$ must have low ionization, $\log U\sim
-3.5$.  The arrow shows how the grid would move if the metallicity of
the models were decreased by 1 dex. \label{fig:cloudy}} 
\end{figure}

In Figure~\ref{fig:cloudy}, the grid of clouds is plotted for
$N({\MgII})$ verses the total hydrogen column density, $N({\HI}) +
N({\HII})$.  
Solid curves are lines of constant $N({\HI})$, and dotted curves
are lines of constant $\log U$ presented at intervals of 0.5 dex.
The range of observed weak {\MgII} absorber column densities is
crudely represented by the shaded region.
Note that the majority of the absorbers are predicted to arise in
sub--LLS clouds, those with $N({\HI}) \leq 10^{17.3}$~{\cm2}.
A main point to be gleaned from Figure~\ref{fig:cloudy}, however,
arises due to the metallicity dependence of the photoionization grid.
If the metallicity of the model clouds were reduced by 1 dex, the
entire grid would move approximately 1 dex downward along the 
$N({\MgII})$ axis.
Thus, if the metallicity of the clouds were much lower than
$[Z/Z_{\odot}] = -1$, the majority of the inferred neutral hydrogen
column densities would lie above the Lyman limit value of
$10^{17.3}$~{\cm2}.
This is not allowed because most all weak {\MgII} absorbers must arise
in sub--LLS environments as constrained by the $dN/dz$ of LLS
absorbers.
Our inferred upper limit on the metallicity would scale roughly in
direct inverse proportion to any $\alpha$--group abundance
enhancement.

In conclusion, if the weak {\MgII} absorbers are photoionized by a
Haardt \& Madau--like spectrum, then the cloud metallicities are 
(1) $[Z/Z_{\odot}] \geq -1$ with $\hbox{[$\alpha$/Fe]} \sim
\hbox{[$\alpha$/Fe]}_{\odot}$, or 
(2) $[Z/Z_{\odot}]$ slightly below $-1$ with $\hbox{[$\alpha$/Fe]} >
\hbox{[$\alpha$/Fe]}_{\odot}$ by a factor of a few.
Such abundance patterns are consistent with those observed in the
Galaxy (\cite{jtl}; \cite{savagearaa}). 
The metallicity cannot typically be significantly lower than
$[Z/Z_{\odot}] = -1$, since that would require an implausibly large
enhancement of $\hbox{[$\alpha$/Fe]}$ with respect to the solar ratio.

\subsection{Comparison with {\CIV} Systems: Ionization Conditions}
\label{sec:civdiscuss}

Sargent, Boksenberg, \& Steidel (1988\nocite{sbs88}) found $dN/dz =
1.76\pm0.33$ for {\CIV} at $\left< z \right> = 1.5$ for a sample
complete to $W_{\rm r}^{\rm min}(1548) = 0.3$~{\AA}.
We note that this value is consistent with the number per unit
redshift of the weak {\MgII} systems at $\left< z \right> = 0.9$.
Does this imply that the population of systems selected by weak
{\MgII} absorption are, in essence, the same population as selected by
the presence of {\CIV} with $W_{\rm r}^{\rm min}(1548) = 0.3$~{\AA}?
It is not expected that all weak {\MgII} systems at $\left< z \right >
= 0.9$ would have associated {\CIV} with $W_{\rm r}(1548) \geq
0.3$~{\AA}, since the {\CIV} number density is seen to decrease with
decreasing redshift for this equivalent width threshold. 
Bergeron \etal (1994\nocite{bergeron94}) found $dN/dz = 0.87\pm0.43$
for {\CIV} with $W_{\rm r}^{\rm min}(1548) = 0.3$~{\AA} at $\left< z
\right> = 0.3$.
We have confirmed this expectation with our search for {\CIV} in
FOS/{\it HST\/} spectra.
Of the 13 systems for which {\CIV} was not detected in FOS/{\it HST\/}
spectra, all but three have a $3\sigma$ $W_{\rm r}(1548)$ threshold 
below $0.3$~{\AA}.
Of course, weaker {\CIV} could always be present.

\begin{figure*}[ht]
\vglue 0.5in
\plotfiddle{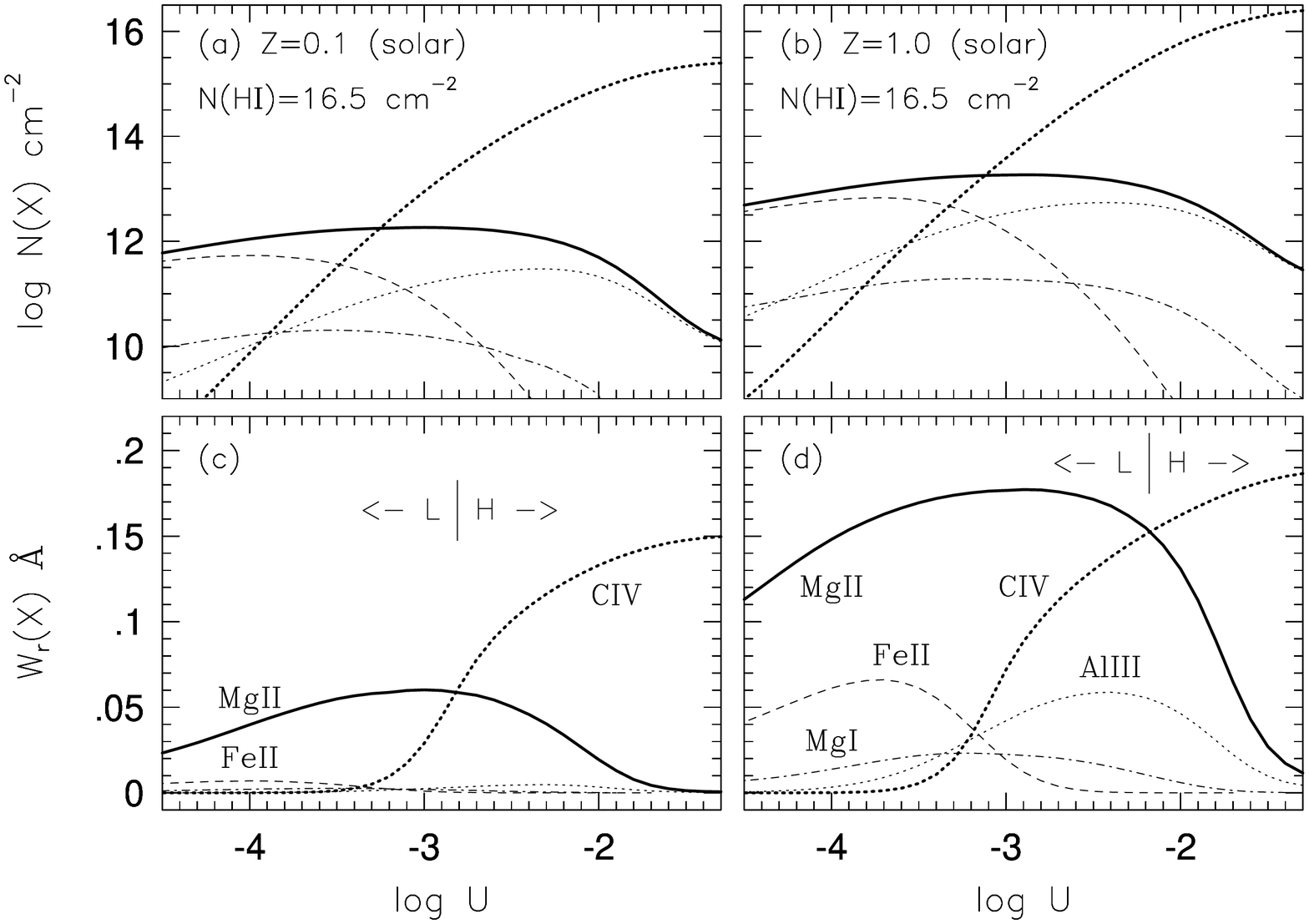}{3.5in}{0.}{60.}{60.}{-250}{0}
\vglue -0.35in
\protect\caption
{Photoionization models with $N({\HI}) = 10^{16.5}$~{\cm2} for solar
abundance patterns with $[Z/Z_{\odot}] = -1$ (left panels) and
$[Z/Z_{\odot}] = 0$ (right panels).  The upper panels show the
observed column densities as a function of ionization parameter and
the lower panels show the rest--frame equivalent widths for
$b=6$~{\kms} for {\CIV} $\lambda 1548$, {\MgII} $\lambda 2796$,
{\FeII} $\lambda 2600$, {\MgI} $\lambda 2853$, and {\AlIII} $\lambda
1855$. Low and high  ionization (``L'' and ``H'', respectively) are
defined by the ratio of $W({\CIV})/W({\MgII})$. \label{fig:ewcloudy}}
\end{figure*}

In and of itself, {\CIV} absorption is an important indicator of the
ionization level in absorbers selected by weak {\MgII} absorption.
A more powerful indicator, however, is the relative absorption
strengths of {\CIV} and {\FeII}.
In the upper panels (a and b) of Figure~\ref{fig:ewcloudy}, we present
the column densities of {\MgII}, {\FeII}, {\MgI}, {\AlIII}, and {\CIV}
as a function of ionization parameter for CLOUDY (\cite{ferland}) models.
We have assumed $N({\HI}) = 10^{16.5}$~{\cm2} for two metallicities,
$[Z/Z_{\odot}] = -1$ and $0$, respectively, with a Haardt \& Madau
(1996\nocite{hm96}) UV background flux normalized at $z=1$.
The bottom panels (c and d) show the rest--frame equivalent widths for
a Doppler parameter of $b=6$~{\kms}, the median velocity width of the
{\it individual\/} {\MgII} absorption components\footnote{Inspection
of the {\MgII} profiles with $\omega _{v} \geq 10$~{\kms} (for
example, S2, S4, S5, S21, S23) reveals that they are blends of narrower
components.  Preliminary Voigt Profile decomposition of the profiles
is consistent with $b=6$~{\kms} clouds.}.
The models presented have {\MgII} $\lambda 2796$ equivalent widths
consistent with the observed range for individual components (see
Figure~\ref{fig:ew_vs_zabs}).
For these sub--LLS clouds, the models show no temperature or
ionization structure with cloud depth; the kinetic temperature is the
same for all ionization species.
For the modeled column densities, {\MgI} $\lambda 2853$, {\FeII}
$\lambda 2600$, and {\AlIII} $\lambda 1855$ are on the linear part of
the curve of growth ($b$ independent).
For {\MgII} $\lambda 2796$ and {\CIV} $\lambda 1548$, the value of the
$b$ parameter becomes important for column densities greater than
$\sim 10^{11}$~{\cm2} and $\sim 10^{13}$~{\cm2}, respectively. 
We have neglected thermal scaling of the $b$ parameters, which would
yield slightly elevated {\CIV} equivalent widths.

The demarcation between low and high ionization absorbers (L and H,
respectively, as marked on panels c and d of
Figure~\ref{fig:ewcloudy}) is given by the ratio $W_{\rm
r}(1548)/W_{\rm r}(2796)$, where greater than unity designates high
ionization (Bergeron \etal 1994\nocite{bergeron94}).
For these single--phase photoionization models, high ionization clouds 
have $N({\FeII}) \ll N({\MgII})$ and $N({\FeII}) \ll N({\CIV})$.
For low ionization gas there is a range of $N({\FeII})$ as compared to
$N({\CIV})$ and $N({\MgII})$.
For very low ionization conditions, $N({\FeII})$ can be comparable to 
$N({\MgII})$, with both much larger than $N({\CIV})$.
We would not expect {\CIV} to be reported in the literature nor
found in FOS/{\it HST\/} spectra for clouds with $W_{\rm r}(2600) /
W_{\rm r}(2796) > 0.1 $ (roughly).
In fact, for no case in which we have detected {\FeII} would we expect
a high--ionization single--phase cloud as defined by $W_{\rm r} (1548) /
W_{\rm r} (2796) \geq 1$.
The point is that the presence of {\FeII} absorption at a
non--negligible level is a strong indicator of low ionization
conditions for single--phase clouds.
The above arguments remain valid for $N({\HI}) = 10^{17}$~{\cm2}
cloud models in which some ionization structure is present, resulting in 
roughly constant $N({\MgII})$ as a function of $\log U$, but identical
behavior in $N({\FeII})$ and $N({\CIV})$.

As seen in Table~\ref{tab:civcomp}, of the 22 systems for which both
{\CIV} and {\FeII} was available, four have {\FeII} but no {\CIV}
(S7, S10, S18, and S22).  
These are low ionization systems.  
Five of 22 systems have {\CIV} but no {\FeII} (S5, S17, S19, S20,
and S23) to $\left< W_{\rm r}^{\rm min}(2600) \right>
\sim 0.01$~{\AA}; these may be high ionization systems.
Four of 22 systems have both {\CIV} and {\FeII} (S14, S28, S29,
and S30), two of which (S28 and S29) also have {\MgI}.
{\it The simultaneous presence of\/ {\FeII} and\/ {\CIV} implies that a
single cloud or a single--phase absorption model is not adequate}.
From this, we speculate that S11, S13, and S26 are either low or
multi--phase systems due to their having substantial {\FeII}, even
though we have no information on {\CIV}.
Overall, it appears that the population of metal--line absorbers
selected by weak {\MgII} absorption have a range of ionization
conditions, possibly including multi--phase conditions.
In Figure~\ref{fig:civfig}, we have illustrated these suggested
ionization conditions.

\subsection{Evidence for Multi--Phase Ionization}

From Figure~\ref{fig:ewcloudy}, we see that it is difficult
to obtain $W_{\rm r}(1548) > 0.2$~{\AA} for
single--phase photoionization models with profile widths (i.e.\ $b$
parameters) consistent with those observed for individual {\MgII}
components.
This is because of the very flat dependence of $W_{\rm r}(1548)$ on
$N({\CIV})$ for the small $b$ parameters implied by the individual
{\MgII} components.
This upper limit holds for higher $N({\HI})$ cloud models.
The upshot is that if $W_{\rm r}(1548)$ is significantly greater than
$\sim 0.2$~{\AA} in a single component absorber with $W_{\rm r}(2796)
\leq 0.2$~{\AA}, one of two possibilites can be inferred:
(1) Either the absorber has multiple ionization phases, such that the
{\MgII} arises in a cooler low--ionization region embedded in a higher
ionization medium, or 
(2) additional multiple components (spatially distinct) are observed
in {\CIV} but not in {\MgII}, in which case a high resolution spectrum
of the {\CIV} profile would be required to determine the ionization
level of the components seen in {\MgII}.

Three of the four systems exhibiting both {\CIV} and {\FeII}
absorption, S14, S29, and S30, have multiple unblended components
spread over $\sim 100$~{\kms}.  
It is quite plausible that they are multi--phase, that low ionization
{\MgII} components are embedded in a high ionization ``halo''.
Consider S29, which has six components.
We estimated $W_{\rm r}(1548) \simeq 0.9$~{\AA} (lower
limit) from the unresolved {\CIV} equivalent width reported by
SS92\nocite{ss92}.
For $N({\HI}) \leq 10^{17}$~{\cm2}, the {\CIV} equivalent width
could be produced as the component sum of high ionization clouds
(near--solar metallicity required), each component contributing $\sim
0.2$~{\AA} to the total {\CIV} equivalent width.
However, this is unlikely because the {\FeII} absorption enforces
low ionization conditions in the phase giving rise to {\MgII}
absorption.
The mean equivalent width of the individual {\MgII} $\lambda 2796$
components is 0.05~{\AA} and that of {\FeII} $\lambda 2600$ is
0.015~{\AA}.
Refering to Figure~\ref{fig:ewcloudy}, the mean ratio of $W_{\rm r}(2600)/
W_{\rm r}(2796) \simeq 0.3$ for $W_{\rm r}(2796) \simeq 0.05$~{\AA}
implies ionization parameters in the range $-3.5 \leq \log U \leq -4$;
these are low ionization clouds.
Thus, the overall absorption properties of S29 are more consistent
with low ionization {\MgII} components embedded in a high ionization
medium or juxtaposed with additional spatially distinct high
ionization components.

\subsection{The Physical Sizes of the Absorbers}

The measured $dN/dz$ can be used to constrain their physical sizes,
assuming that all absorbers are associated with luminous
galaxies (\cite{steidel95}; Churchill \etal 1996\nocite{csv96}).
For a Schechter function (1976\nocite{schechter76}), 
$ 
\Phi (L)dL = \Phi ^{\ast} \left( {L}/{L^{\ast}} \right) ^{\alpha}
  \exp \left( - {L}/{L^{\ast}} \right) dL ,
\label{eq:schechter}
$
there are two important parameters: the faint--end slope, $\alpha$,
and the density of $L^{\ast}$ galaxies, $\Phi ^{\ast}$.
Second, a Holmberg (1975\nocite{holmberg75}) dependence,
$
R(L) = R^{\ast} \left( {L}/{L^{\ast}} \right) ^{\beta} ,
\label{eq:holmberg}
$
of the spatial extent of the absorbing gas on the luminosity of the
galaxy is assumed, where $R^{\ast}$ is the fiducial size of an
$L^{\ast}$ galaxy.
The product $\sigma n$, is given by
$
\sigma n = \pi \int ^{\infty}_{L_{\rm min}} \Phi (L) R^{2}(L)dL ,
\label{eq:sigman}
$ which simplifies to 
\begin{equation}
\sigma n = \pi \Phi ^{\ast} {R^{\ast}}^{2} \, \, \Gamma ( \alpha + 2
\beta + 1 ,  L_{\rm min} / L^{\ast} ) ,
\label{eq:fullglory}
\end{equation}
where $\Gamma$ is the incomplete gamma function, which accounts for
the additional assumption of a minimum luminosity, $L_{\min}$, at
which a galaxy will no longer exhibit {\MgII} absorption to a
well--defined detection level at any impact parameter.
Assuming no evolution in the redshift path density of absorbing
galaxies, the relation between $\sigma n$ and the measured $dN/dz$ 
is 
\begin{equation}
\frac{dN}{dz} = \frac{c \sigma n}{H_{0}}
\left( 1+z \right) \left( 1 + 2q_{0}z \right) ^{-1/2} .
\label{eq:pathdensity}
\end{equation}

For $W_{\rm r}^{\rm min}(2796) = 0.3$~{\AA}, Steidel
(1995\nocite{steidel95}) directly measured $\beta = 0.15$ and
$R^{\ast} = 38h^{-1}$~kpc for the Holmberg relation and
$L_{\rm min}/L^{\ast} \sim 0.06$ for $K$--band luminosities.
From this, Steidel \etal (1994\nocite{sdp94}) deduced a faint--end
slope of $\alpha = -1$ and a number density of $\Phi^{\ast} =
0.03h^{3}$~Mpc$^{-3}$ for galaxies selected by the presence of {\MgII}
absorption.
These values are in good agreement with the values $\alpha = -0.9$ and
$\Phi ^{\ast} = 0.033h^{3}$~Mpc$^{-3}$ measured for the Canada--France
Redshift Survey (CFRS) sample of galaxies over the similar redshift
range of $0.0 \leq z \leq 1.3$ (\cite{lilly95}).
However, it is not entirely clear that the radius--luminosity
relationship measured by Steidel (1995\nocite{steidel95}) is
applicable when the equivalent width detection threshold is lowered to
$W_{\rm r}^{\rm min}(2796) = 0.02$~{\AA}.
Nor is there {\it a priori\/} reason to assume that the $L_{\rm
min}/L_{K}^{\ast} = 0.06$ cut off should apply for a population of
galaxies selected by $W_{\rm r}(2796) < 0.3$~{\AA}
absorption.
It is possible that low surface brightness (LSB) galaxies 
contribute to the measured $dN/dz$ of weak {\MgII}
absorbers\footnote{We comment that is interesting that LSB galaxies
have not yet been identified with strong {\MgII} absorption, except
possibly in the cases of a few damped {\Lya} systems.} 
(\cite{impey89}; \cite{cl98}).
At redshift $z\sim 0$, Dalcanton \etal (1997\nocite{dalcanton})
has measured $\Phi ^{\ast} = 0.08h^{3}$~Mpc$^{-3}$ for LSB galaxies
with central surface brightnesses of $23 \leq \mu _{0} (V) \leq
25$~mag arcsec$^{-2}$.
This is a factor of $\sim 2.5$ higher than the number density of
CFRS and {\MgII}--selected galaxies, which exhibit the Freeman
(1970\nocite{freeman}) central surface brightness of $\mu _{0}(B) =
21.7$ mag arcsec$^{-2}$. 
Thus, we do not simply assume that the luminosity function of galaxies
selected by weak {\MgII} absorption would have the same number density
as that found by Steidel \etal (1994\nocite{sdp94}) for strong
systems.

For $W_{\rm r}^{\rm min}(2796) = 0.02$~{\AA}, we found $dN/dz = 2.71$
at  $\left< z \right> = 0.9$.
Writing Eq.~\ref{eq:fullglory} to obtain $R^{\ast}$ as a
function of $\Phi ^{\ast}$, $L_{\rm min}/L^{\ast}$, and $\beta$,
gives, 
\begin{equation}
R^{\ast} = 79.5 h^{-1} \,
  \left[ \left( \Phi^{\ast} / \Phi^{\ast}_{\sc (cfrs)}  \right)
 \, \Gamma ( 2 \beta , L_{\rm min} / L^{\ast} ) 
  \right] ^{-1/2} \quad \hbox{kpc},
\end{equation} 
where it is assumed that the absorbing gas is spherically distributed
and has unity covering factor\footnote{If one assumes that only some
fraction, $f_{\rm abs}$, of the galaxies contribute to absorption, and
that the gas covering factor, $f_{\rm c}$, is less than unity, then
the quoted values of $R^{\ast}$ should be scaled by $(f_{\rm
abs}f_{\rm c})^{-1/2}$.  If one assumes the geometry is somewhat
flattened, then an additional factor of $\kappa ^{-1/2}$ can be
introduced, where $\kappa= 0.5$ for infinitely thin ``disks''.}.
For $L_{\rm min}/L^{\ast}= 0.06$, $\beta = 0.15$, and $\Phi
^{\ast}/\Phi^{\ast}_{\sc cfrs} \sim 1$, the inferred $R^{\ast}$ is
$63h^{-1}$~kpc.
If LSB galaxies have the same absorbing properties as do those
observed by Steidel (1995\nocite{steidel95}), then the combined galaxy
population would have $\Phi ^{\ast}/\Phi^{\ast}_{\sc cfrs} \sim 3$
(\cite{dalcanton}), which yields $R^{\ast} \sim 35h^{-1}$~kpc.
This is surprisingly consistent with the $R^{\ast}$ measured by
Steidel (1995\nocite{steidel95}) for $W_{\rm r}(2796) > 0.3$~{\AA}
absorption selected galaxies.
For $L_{\rm min}/L^{\ast}= 0$, $\beta= 0.15$, and $\Phi
^{\ast}/\Phi^{\ast}_{\sc cfrs} \sim 1$, the inferred size is
$R^{\ast} = 46h^{-1}$~kpc.
If $\beta \sim 0$ with $L_{\rm min}/L^{\ast}= 0.06$, then $R^{\ast} =
53h^{-1}$~kpc.
If $\beta = 0.4$ (the historically invoked ``Holmberg'' value), then
$R^{\ast} = 77h^{-1}$~kpc.

\subsection{Evolution of $n(W_{\rm r})$}

If the absorbing gas is photoionized by the extra--galactic UV 
background, then ionization conditions are expected to be higher
at higher redshifts.
In fact, Bergeron \etal (1994\nocite{bergeron94}) have found that the
ratio of ``high'' to ``low'' ionization metal--line absorbers
increases with increasing redshift at a rate that is not inconsistent
with a factor of $\sim 5$ increase in the UV background
(\cite{kulkarni-fall}).
Our observations place no constraint on the evolution of the weak
{\MgII} absorbers over this redshift range.

In principle, it might be that at $z\sim2$ the {\MgII} equivalent
width distribution exhibits a cut off at the smallest equivalent
widths.
It is also expected that the power--law slope of the distribution is
flatter for larger equivalent widths at $z\sim 2$ [there are more
large $W_{\rm r}(2796)$ absorbers at high redshift] than that measured
in this work at $z \sim 1$.
This implies a very curious effect in the inferred evolution of
{\MgII} absorption strengths as the minimum equivalent width of the
sample covering $0.4\leq z \leq 2.2$ is continually decreased.
It is well established that for $W_{\rm r}^{\rm min}(2796) = 0.3$~{\AA},
the population is consistent with no--evolution expectations, and that
the lack of evolution is dominated by the lower end of the equivalent
width distribution to this $W_{\rm r}^{\rm min}(2796)$.
When $W_{\rm r}^{\rm min}(2796)$ is increased, the evolution becomes
pronounced in that the ratio of ``large'' to ``small'' $W_{\rm
r}(2796)$ absorbers increases with increasing redshift
(SS92\nocite{ss92}).
Since the UV background is more intense at higher redshifts, one
would expect that the presence of {\MgII} is increasingly dependent
upon photon shielding by neutral hydrogen; at $z \sim 2$
(and above), it would be expected that {\MgII} would not survive in
sub--LLS environments (also see discussion in \S7 of
SS92\nocite{ss92}).
Therefore, as $W_{\rm r}^{\rm min}(2796)$ is decreased from 0.3~{\AA}
to $\sim 0.02$~{\AA} for a $0.4\leq z \leq 2.2$ sample, {\it evolution
should again become apparent, this time due to a paucity of weak
systems at the higher redshifts}.
This trend is tentatively suggested by the fact that at $\left< z
\right> = 0.9$ there is a 30\% 
difference between the $dN/dz$ of $W_{\rm r}^{\rm min}(2796) =
0.3$~{\AA} absorbers and LLS absorbers, whereas there is only an 8\%
difference at $\left< z \right> = 2$ (SS92\nocite{ss92};
\cite{stengler-larrea}).


\section{Summary}
\label{sec:conclude}

We searched for weak {\MgIIdblt} doublets, those with rest--frame
equivalent widths $W_{\rm r}(2796) < 0.3$~{\AA}, in HIRES/Keck
spectra of 26 QSOs.
The QSO sight lines are unbiased for these weak systems.
The cumulative redshift path was $Z \sim 17$ over the range $0.4 \leq z
\leq 1.4$ and $Z \sim 0.7$ for $1.4 \leq z \leq 1.7$.  
The survey was complete to $W_{\rm r}(2796) = 0.06$~{\AA}, and 80\%
complete to $W_{\rm r}(2796) = 0.02$~{\AA}, where we have enforced a
$5\sigma$ detection limit.
A total of 30 systems were detected, of which 23 were discovered in
these spectra.
The {\MgI} $\lambda 2853$ transition was detected in seven of the
systems and {\FeII}, especially the $\lambda 2600$ and/or the $\lambda
2383$ transition, was detected in half of the systems.
When {\AlIII} $\lambda \lambda 1855, 1863$ was covered, we found it in
three of four systems.
From a literature search and a search we conducted in archival
FOS/{\it HST\/} spectra, we detected {\CIV} in nine of 22 covered
systems to $3\sigma$ equivalent width threshold of $0.03$ to
$0.3$~{\AA} (rest--frame).
No systems were found at $1.4 \leq z \leq 1.7$, though this is
consistent with expectations when we extrapolate from lower $z$, given
that the cumulative redshift path was only $Z \sim 0.7$ over this
redshift range.
We have combined our sample with the $W_{\rm r}(2796) \geq 0.3$~{\AA}
sample MG1 of SS92\nocite{ss92}, taken over the redshift interval $0.4
\leq z \leq 1.4$, and measured the redshift number density and
equivalent width distribution of {\MgII} absorbers with $W_{\rm
r}(2796) \geq 0.02$~{\AA}.

Main results from this work include:

\begin{enumerate}

\item
The redshift path density of weak {\MgII} absorbers was measured to be
$dN/dz = 1.74 \pm 0.11$.  
There is no evidence for evolution in the redshift path density, but
the measured value of $\gamma = 1.3 \pm 0.9$ is not constraining. 
Incorporating the MG1 sample of SS92\nocite{ss92} [$W_{\rm r}^{\rm
min}(2796) = 0.3$~{\AA}], we find that the {\it total\/} number
density per unit redshift for systems with $W_{\rm r} ^{\rm min}(2796)
= 0.02$~{\AA} at $\left< z \right> = 0.9$ is $dN/dz = 2.65\pm0.15$.
The $dN/dz$ of weak {\MgII} absorbers is roughly $5-7$\% of that of
the {\Lya} forest with $W_{\rm r}({\Lya}) \geq 0.1$~{\AA}
(\cite{kp13}).
Thus, it is plausible that $\sim 5$\% of $z \sim 0.9$ ``{\Lya}
clouds'' will have detectable {\MgII} absorption to $W_{\rm r}(2796) =
0.02$~{\AA}.

\item
For $W_{\rm r}^{\rm min} (2796) =  0.02$~{\AA}, {\MgII} absorbers at
$\left< z \right> = 0.9$ outnumber LLS absorbers by a factor of
$3.8\pm1.1$.
That the populations of strong {\MgII} absorbers and LLS are
indistinguishable (\cite{lanzetta88}) strongly suggests that virtually
all of the weak {\MgII} systems arise in sub--LLS environments.
It is possible that the weak {\MgII} systems have high metallicities,
whether their ionization conditions are low or high (see
Figures~\ref{fig:cloudy} and \ref{fig:ewcloudy}).
Photoionization models, using Ferland's CLOUDY, are consistent with
this conclusion; many weak {\MgII} absorbers probably have
$[Z/Z_{\odot}] \geq -1$.
Lower metallicities require $N({\HI})$ above the Lyman limit value,
which is not allowed because the ratio of $dN/dz$ of weak {\MgII}
absorbers to that of LLS absorbers requires that most all weak {\MgII}
absorbers have $N({\HI})$ near or below the Lyman limit value.

\item
The equivalent width distribution was found to follow a power law with
slope $\delta \sim 1.0$ down to $W_{\rm r}(2796) = 0.017$~{\AA}.
{\it There is no turnover or break in the equivalent width
distribution for $W_{\rm r}(2706) < 0.3$~\hbox{\rm {\AA}} at
$\left< z \right> = 0.9$}.
A single power--law slope does not describe the equivalent width
distribution over the full redshift range $0.3 \leq z \leq 2.2$. 
For $z<1.4$, there is a break (a significantly steeper slope) in the
distribution for $W_{\rm r}(2796) > 1.3$~{\AA}.
The upper limit on this slope is $\delta = 2.3$.

\item
Most, if not all, of the {\it individual clouds\/} in these weak
absorbers have $W_{\rm r}(2796) \leq 0.15$~{\AA}.
Their line widths are narrow, with an average of $\sim 4$~{\kms},
implying temperatures of $\sim 25,000$~K.
The presence of {\MgI} is not correlated with $W_{\rm r}(2796)$ or with
the {\MgII} doublet ratio.
In fact, {\MgI} absorption is present in at least one component having
$W_{\rm r}(2796)$ as small as $0.08$~{\AA}.
Statistically, there appears to be no $W_{\rm r}(2796)$ threshold for
the presence of {\MgI} down to $W_{\rm r}(2796)\sim 0.1$~{\AA}.

\item
The weak {\MgII} absorbers may comprise a diverse population,
including low, high, and multi--phase ionization systems.
The low ionization systems would be immediately recognizable by the
presence of $W_{\rm r}(2600)/W_{\rm r}(2796) \geq 0.1$.
In a single {\MgII} component with $b \leq 6$~{\kms} and $W_{\rm
r}(2796) < 0.2$~{\AA}, the presence of {\CIV} with $W_{\rm r}(1548)
\geq 0.2$~{\AA}, and especially the simultaneous presence of {\FeII}
and {\CIV}, may be due to multi--phase absorption or additional
spatially distinct high ionization components in which {\MgII} has not
been detected.
In an absorber with $N_{\rm c}$ components in which the individual
components have  $b \leq 6$~{\kms} and $W_{\rm r}(2796) < 0.2$~{\AA},
multi--phase or additional high ionization components are likely if
the total system {\CIV} equivalent width is larger than $\sim N_{\rm c}
\times 0.2$~{\AA}.
Again, this is especially true if the individual {\MgII} components
have $W_{\rm r}(2600)/W_{\rm r}(2796) \geq 0.1$.

\item 
The velocity clustering of weak {\MgII} absorbers around strong ones
is consistent with a random distribution; a $\chi^{2}$ test yielded a
probability of 0.18 that clustering was drawn from a random
distribution.
Furthermore, we examined the redshift clustering of the weak systems
with respect to one another and found that the systems are also not
inconsistent with a random distribution.  
The $\chi ^{2}$ probability was 0.25.
We conclude that the QSO sight lines surveyed for this work are
unbiased for the presence of weak {\MgII} absorbers.

\item
Assuming the radius--luminosity relationship between gas and galaxies
and the cut off, $L_{\rm min}$, in the {\MgII} absorbing galaxy
luminosity function (Steidel 1995\nocite{steidel95}), we inferred an
$R^{\ast}$ for $W_{\rm r}^{\rm min}(2796) = 0.02$~{\AA} {\MgII}
absorbers of $65h^{-1}$~kpc.
If there is no radius--luminosity dependence, then $R^{\ast} \simeq
55h^{-1}$~kpc.
If there is no luminosity cut off, then $R^{\ast} \simeq 45h^{-1}$~kpc.
If LSB galaxies have the same absorbing properties as do the {\MgII}
selected galaxies observed by Steidel (1995\nocite{steidel95}), then
the combined galaxy population would have $R^{\ast} \sim
35h^{-1}$~kpc.
It is not inconsistent with the data to suggest that a non--negligible
fraction of the weak {\MgII} absorbers are arising in LSB galaxies.

\end{enumerate}

\subsection{Objects Selected by Weak {\MgII} Absorption}

The question remains; what luminous objects could be associated with
the large numbers of weak {\MgII} systems at $0.4 \leq z \leq 1.4$?
Do the galaxies selected by {\MgII} absorption with $W_{\rm r}(2796)
\geq 0.3$~{\AA} (\cite{steidel95}; Steidel \etal 1994\nocite{sdp94})
tell the whole story or do they represent the ``tip of the {\MgII}
iceberg''.
Are weak {\MgII} absorbers analogous to a bulk of material that lies
hidden below the ``galactic water line''?
That is, are a substantial fraction not directly associated with
bright galaxies, but rather with intergalactic material or ``failed''
galaxies?
Given that only four of 20 have obvious galaxy counterparts, it is not
clear that the surface brightnesses of the star forming environments
associated with weak {\MgII} absorbers are above the Freeman
(1970\nocite{freeman}) value.
It could be that the successful broad--band imaging technique for
identifying the luminous counterparts of strong {\MgII} absorbers in
QSO fields will be less effective for locating many of the
counterparts associated with weak {\MgII} absorbers.

If weak {\MgII} absorption arises in the extended regions of {\MgII}
absorption--selected galaxies, it likely is most often occurring at
large impact parameters (a region of $D \geq 40h^{-1}$~kpc;
beyond the well probed regions within 10{\arcsec} of the QSO).
It is possible that some fraction of the {\Lya} lines seen to
arise beyond $\sim 40$~kpc of large extended halos or disks of
galaxies (\cite{lanzetta95}; \cite{lebrunlya}) give rise to weak
{\MgII} absorption.
An alternative possibility is that some fraction of the weak
{\MgII} population is selecting sub--LLS environments around LSB
galaxies or low luminosity galaxies, those with $L_{K} <
0.06K_{K}^{\ast}$.
That is, some fraction of weak systems could arise in ``normal'' high
surface brightness galaxies (perhaps at greater impact parameters) and
some fraction could arise in dwarf or large LSB galaxies.

Conceivably, LSB galaxies could significantly contribute to the weak
{\MgII} absorption cross section.
At the low redshifts studied in this work, most sub--LLS absorbers are
thought to be associated with LSB galaxies (\cite{salpeter};
\cite{linder}).
Impey \& Bothun (1989\nocite{impey89}), upon reexamining the selection
effects and assumptions that go into the calculations of galaxy cross
sections from QSO absorbers, found that LSB galaxies are expected to
dominate the absorption cross section.
If so, why would it be that strong {\MgII} absorbers are not
associated with LSB galaxies more often than or as often as are normal
bright galaxies?
And, when strong {\MgII} absorption is associated with an LSB galaxy,
why is it that the absorbers are damped {\Lya} systems
(\cite{steidel94}; \cite{lebrundla})?
LSB galaxies typically have {\HI} surface densities a factor of two
lower than normal high surface brightness galaxies (\cite{deblok})
and lower metallicities (\cite{mcgaugh}).
Further, the inner $15h^{-1}$~kpc of LSB galaxies are found to have
$N({\HI})$ of a $\hbox{few} \times 10^{20}$~{\cm2} (de~Blok \etal
1996\nocite{deblok}), i.e.\ they are damped {\Lya} absorbers.
It could be that some weak {\MgII} absorbers arise in LSB galaxies at
galactocentric distances greater than $\sim 15h^{-1}$~kpc, where
the {\HI} environment is sub--Lyman limit and clouds cannot have large
{\MgII} column densities.
Thus, it is possible that the strong {\MgII} (and {\FeII}) associated
with damped {\Lya} absorbers in LSB galaxies arise in their inner
regions, whereas only weak {\MgII} absorbing clouds can survive in
their outer regions.

Other possibilities include isolated star forming or post star forming
dwarf galaxies and/or pre--galaxy fragments (\cite{yanny-york}), 
or the remnant material left over from the formation of galaxies
and/or small galaxy groups (Bowen \etal 1995\nocite{bbp95};
\cite{vangorkom96}; Le~Brun \etal 1996\nocite{lebrunlya}).
Narrow band imaging of emission lines at the absorber redshifts might
provide a test for the former scenario, while charting the
distribution of galaxies at the absorber redshifts in wide fields
($\sim 100${\arcsec}) centered on the QSOs might provide tests for the
latter.
A more detailed determination of the chemical and ionization
conditions and identification of the luminous objects associated
selected by weak {\MgII} absorption is the next logical step toward
their further exploration.

\acknowledgements
This work has been supported in part by the National Science
Foundation Grants AST--9617185 and AST--9529242, by NASA LTSA grant
NAG5--6399, and by a Sigma--Xi Grant in Aid of Research.
CWC acknowledges support as a Eberly School of Science Distinguished
Postdoctoral Fellowship.
We especially thank Sofia Kirhakos, Buell Jannuzi, and Don Schneider
for permission to publish results from our {\CIV} search using the
archival FOS/{\it HST\/} from our collaboration.
Thanks are extended to Adam Dobrzycki for kindly assisting us with a
preliminary search for {\CIV} through his FOS/{\it HST\/} archival
spectra.
Thanks to G.~Ferland for making CLOUDY a public tool.
We thank an anonymous referee and the scientific editor for comments
that led to an improved manuscript.
This paper is dedicated to the life of Julius L. Nelson.


 
\resetfig
 


 
\begingroup
\small
\begin{deluxetable}{lccccccc}
\tablewidth{0pc}
\tablecaption{HIRES/Keck and FOS/{\it HST\/} Data}
\tablehead
{
 & &
\multicolumn{3}{c}{HIRES Spectra} & &
\multicolumn{2}{c}{FOS Spectra} \\
\cline{3-5} \cline{7-8}
\colhead{Object} &
\colhead{$z_{\rm em}$} &  
\colhead{Date [UT]} & 
\colhead{Exp [s]\tablenotemark{a}} & 
\colhead{$\lambda $ Range [\AA]\tablenotemark{b}} & &
\colhead{Source} & 
\colhead{Grating} 
} 
\startdata
$0002+051$ & 1.899 & 1994 Jul 05 &  2700 & $3655.7-6079.0$ & & KP & 270H \nl 
$0058+019$ & 1.959 & 1996 Jul 18 &  3000 & $3766.2-5791.3$ & & \nodata & \nodata \nl 
$0117+212$ & 1.491 & 1995 Jan 23 &  5400 & $4317.7-6775.1$ & & KP & 270H \nl
$0420-014$ & 0.915 & 1995 Jan 23 &  3600 & $3810.5-6304.9$ & & \nodata & \nodata \nl
$0450-132$ & 2.253 & 1995 Jan 24 &  5400 & $3986.5-6424.5$ & & \nodata & \nodata \nl  
$0454+036$ & 1.343 & 1995 Jan 22 &  4500 & $3765.8-6198.9$ & & AR & 270H \nl  
$0454-220$ & 0.534 & 1995 Jan 22 &  5400 & $3765.8-6198.9$ & & AR & 270H \nl  
$0823-223$ & $\cdots$ & 1995 Jan 24 &  3600 & $3977.8-6411.8$ & & AR & 270H \nl
$0958+551$ & 1.755 & 1995 Jan 23 &  3600 & $5400.0-7830.0$ & & KP & 270H \nl
$1148+384$ & 1.299 & 1995 Jan 24 &  5400 & $3986.5-6424.5$ & & \nodata & \nodata \nl 
$1206+456$ & 1.155 & 1995 Jan 23 &  3600 & $3810.5-6304.9$ & & KP & 270H \nl  
$1213-003$ & 2.691 & 1995 Jan 24 &  5200 & $5008.1-7356.7$ & & \nodata & \nodata \nl  
$1222+228$ & 2.040 & 1995 Jan 23 &  3600 & $3810.5-6304.9$ & & AR & 270H \nl 
$1225+317$ & 2.219 & 1995 Jan 24 &  2400 & $5737.5-8194.7$ & & \nodata & \nodata \nl 
$1241+174$ & 1.282 & 1995 Jan 22 &  2400 & $3765.8-6189.9$ & & KP & 270H \nl 
$1248+401$ & 1.032 & 1995 Jan 22 &  4200 & $3765.8-6189.9$ & & KP & 270H \nl  
$1254+044$ & 1.018 & 1995 Jan 22 &  2400 & $3765.8-6189.9$ & & KP & 270H \nl   
$1317+274$ & 1.014 & 1995 Jan 23 &  3600 & $3810.5-6304.9$ & & AR & 270H \nl   
$1329+412$ & 1.937 & 1996 Jul 18 &  6300 & $3766.2-5791.3$ & & AR & 270H \nl
$1354+193$ & 0.719 & 1995 Jan 22 &  3600 & $3765.8-6189.9$ & & KP & 270H \nl 
$1421+331$ & 1.906 & 1995 Jan 23 &  3600 & $3818.6-6316.9$ & & \nodata & \nodata \nl    
$1548+092$ & 2.749 & 1996 Jul 19 &  3600 & $3766.2-5791.3$ & & \nodata & \nodata \nl
$1622+235$ & 0.927 & 1994 Jul 04 & 16200 & $3726.9-6191.0$ & & AR & 270H \nl
           &       & 1994 Jul 05 &  3040 &                  & &  &  \nl
$1634+704$ & 1.335 & 1994 Jul 04 &  2700 & $3723.3-6185.7$ & & AR & 270H \nl
           &       & 1994 Jul 05 &  5400 &                  & &  & \nl  
$2128-123$ & 0.501 & 1996 Jul 19 &  3900 & $3766.2-5791.3$ & & KP & 270H \nl
$2145+064$ & 0.999 & 1996 Jul 18 &  4500 & $3766.2-5791.3$ & & AR & 270H \nl
\enddata
\tablenotetext{a}{Total exposure time is sum of combined frames.}
\tablenotetext{b}{Above 5100~{\AA} there are small gaps in the
wavelength coverage.}
\tablecomments{``AR'' denotes FOS spectrum taken from the {\it HST\/}
archive and ``KP'' denotes FOS spectrum obtained from the Key Project.}
\label{tab:obsjournal}
\end{deluxetable}

\begin{deluxetable}{lclcccc}
\tablewidth{0pc}
\tablecaption{Sample of Absorbers}
\tablehead
{
\colhead{ID} &
\colhead{$z_{\rm abs}$} &
\colhead{QSO} &
\colhead{$\omega_{v}$} &
\colhead{W$_{\rm r}(\lambda 2796)$} &
\colhead{DR} &
\colhead{$Z(W_{\rm r},{\rm DR})$}   \\
\colhead{ } & 
\colhead{ } & 
\colhead{ } & 
\colhead{[km~s$^{-1}$]} &
\colhead{[{\AA}]} &
\colhead{ } & 
\colhead{ }  
}
\startdata 
S1                  & 0.456415 & $1421+331$ & $6.45\pm0.98$ & $0.179\pm0.019$ & $1.15\pm0.09$ & 17.24\nl
S2                  & 0.500786 & $1329+412$ & $25.57\pm1.39$ & $0.258\pm0.035$ & $1.33\pm0.41$ & 17.26 \nl
S3\tablenotemark{a} & 0.521498 & $1354+193$ & $6.57\pm1.14$ & $0.030\pm0.007$ & $1.31\pm0.68$ & 15.74 \nl
S4                  & 0.550202 & $1222+228$ & $9.58\pm1.25$ & $0.080\pm0.014$ & $1.30\pm0.35$ & 17.05 \nl
S5                  & 0.558443 & $1241+174$ & $17.17\pm1.10$ & $0.135\pm0.014$ & $2.03\pm0.62$ & 17.19 \nl
S6                  & 0.591485 & $0002+051$ & $4.96\pm0.73$ & $0.103\pm0.008$ & $1.62\pm0.31$ & 17.13 \nl
S7                  & 0.642827 & $0454+039$ & $4.77\pm0.57$ & $0.118\pm0.008$ & $1.46\pm0.19$ & 17.18 \nl
S8\tablenotemark{a} & 0.705472 & $0823-223$ & $11.86\pm0.68$ & $0.092\pm0.007$ & $2.09\pm0.70$ & 17.06 \nl
S9                  & 0.725175 & $0058+019$ & $9.30\pm0.45$ & $0.253\pm0.012$ & $1.26\pm0.12$ & 17.26 \nl
S10                 & 0.729071 & $0117+212$ & $47.4\pm1.39$ & $0.238\pm0.009$ & $1.71\pm0.18$ & 17.25 \nl
S11                 & 0.770646 & $1548+092$ & $9.53\pm0.85$ & $0.234\pm0.024$ & $1.19\pm0.19$ & 17.25 \nl
S12                 & 0.818157 & $1634+706$ & $3.14\pm1.56$ & $0.030\pm0.018$ & $1.68\pm0.61$ & 17.67 \nl
S13                 & 0.843247 & $1421+331$ & $3.08\pm0.75$ & $0.086\pm0.008$ & $1.01\pm0.13$ & 15.09 \nl
S14                 & 0.854554 & $1248+401$ & $62.68\pm9.42$ & $0.235\pm0.014$ & $1.76\pm0.25$ & 17.25 \nl
S15\tablenotemark{a}& 0.866529 & $0002+051$ & $1.68\pm0.54$ & $0.023\pm0.008$ & $1.62\pm0.35$ & 14.44 \nl
S16\tablenotemark{a}& 0.895493 & $1241+174$ & $4.95\pm1.21$ & $0.018\pm0.005$ & $1.87\pm0.41$ & 12.71 \nl
S17                 & 0.905554 & $1634+706$ & $4.07\pm0.64$ & $0.064\pm0.004$ & $1.42\pm0.17$ & 16.96 \nl
S18\tablenotemark{a}& 0.931502 & $0454+039$ & $4.36\pm0.89$ & $0.042\pm0.005$ & $1.84\pm0.28$ & 16.34 \nl
S19                 & 0.934283 & $1206+459$ & $7.67\pm0.83$ & $0.049\pm0.005$ & $2.02\pm0.50$ & 16.53 \nl
S20                 & 0.956029 & $0002+051$ & $6.24\pm0.98$ & $0.052\pm0.007$ & $2.07\pm0.61$ & 16.60 \nl
S21                 & 0.973867 & $1329+412$ & $26.79\pm2.72$ & $0.181\pm0.035$ & $1.41\pm0.62$ & 17.24 \nl
S22                 & 0.998358 & $1329+412$ & $6.72\pm0.65$ & $0.142\pm0.010$ & $1.51\pm0.26$ & 17.22 \nl
S23                 & 1.041443 & $1634+706$ & $17.19\pm1.07$ & $0.097\pm0.008$ & $2.10\pm0.44$ & 17.08 \nl
S24                 & 1.127698 & $1213-003$ & $4.17\pm1.07$ & $0.036\pm0.006$ & $1.86\pm0.69$ & 16.03 \nl
S25                 & 1.211316 & $0958+551$ & $3.43\pm1.08$ & $0.060\pm0.007$ & $1.36\pm0.34$ & 16.93 \nl
S26                 & 1.229475 & $0450-132$ & $10.00\pm0.93$ & $0.135\pm0.010$ & $1.29\pm0.16$ & 17.21 \nl
S27                 & 1.232440 & $0450-132$ & $22.72\pm2.45$ & $0.101\pm0.009$ & $1.44\pm0.21$ & 17.13 \nl
S28                 & 1.272377 & $0958+051$ & $4.86\pm0.62$ & $0.081\pm0.007$ & $1.45\pm0.24$ & 17.04 \nl
S29                 & 1.325004 & $0117+212$ & $76.09\pm1.46$ & $0.291\pm0.011$ & $1.60\pm0.11$ & 17.26 \nl
S30\tablenotemark{b}& 1.342969 & $0117+212$ & $44.95\pm3.67$ & $0.153\pm0.008$ & \nodata & \nodata \nl
\enddata 
\tablenotetext{a}{Quantites obtained using Gaussian fitting to data.}
\tablenotetext{b}{The $\lambda 2796$ transition was not covered.  All
quantities based upon $\lambda 2803$ transition.}
\label{tab:systems}
\end{deluxetable}

\begin{deluxetable}{rcc}
\tablewidth{275pt}
\tablecolumns{3}
\tablecaption{System Properties}
\tablehead
{
\colhead{Ion/Tran} &
\colhead{$\lambda _{\rm abs}$\tablenotemark{a}} &
\colhead{EW$_{\rm r}$} \\
\colhead{} &
\colhead{[{\AA}]} &
\colhead{[{\AA}]}
}
\startdata
\cutinhead{S1 \hfil Q$1421+331$ \hfil $z_{\rm abs}=0.45642$}
 {\MgII}~2796    & 4072.649  &  $ 0.179\pm0.019$  \nl
 {\MgII}~2803    & 4083.105  &  $ 0.155\pm0.020$  \nl
 {\MgI}~2853     & 4155.100  &  $< 0.014$ \nl
\cutinhead{S2 \hfil Q$1329+412$ \hfil $z_{\rm abs}=0.50079$}
 {\FeII}~2587    & 3882.008  &  $< 0.092$ \nl
 {\FeII}~2600    & 3902.303  &  $< 0.100$ \nl
 {\MgII}~2796    & 4196.726  &  $ 0.258\pm0.035$  \nl
 {\MgII}~2803    & 4207.500  &  $ 0.194\pm0.054$  \nl
 {\MgI}~2853     & 4281.688  &  $< 0.038$ \nl
\cutinhead{S3 \hfil Q$1354+193$ \hfil $z_{\rm abs}=0.52149$}
 {\FeII}~2587    & 3935.583  &  $< 0.016$ \nl
 {\FeII}~2600    & 3956.158  &  $< 0.012$ \nl
 {\MgII}~2796\tablenotemark{b}  & 4254.644  &  $ 0.030\pm0.007$  \nl
 {\MgII}~2803 & 4265.567  &  $ 0.023\pm0.010$  \nl
 {\MgI}~2853     & 4340.779  &  $< 0.007$ \nl
\cutinhead{S4 \hfil Q$1222+228$ \hfil $z_{\rm abs}=0.55020$}
 {\FeII}~2587    & 4009.830  &  $< 0.011$ \nl
 {\FeII}~2600    & 4030.793  &  $< 0.009$ \nl
 {\MgII}~2796    & 4334.910  &  $ 0.080\pm0.014$  \nl
 {\MgII}~2803    & 4346.039  &  $ 0.061\pm0.013$  \nl
 {\MgI}~2853     & 4422.670  &  $< 0.008$ \nl
\cutinhead{S5 \hfil Q$1241+174$ \hfil $z_{\rm abs}=0.55844$}
 {\FeII}~2587\tablenotemark{c}  & 4031.147  &  $\cdots$  \nl
 {\FeII}~2600    & 4052.221  &  $< 0.012$ \nl
 {\MgII}~2796    & 4357.955  &  $ 0.135\pm0.014$  \nl
 {\MgII}~2803    & 4369.143  &  $ 0.066\pm0.019$  \nl
 {\MgI}~2853     & 4446.182  &  $< 0.008$ \nl
\tablebreak
\cutinhead{S6 \hfil Q$0002+051$ \hfil $z_{\rm abs}=0.59149$}
 {\FeII}~2344    & 3730.781  &  $< 0.026$ \nl
 {\FeII}~2374    & 3778.919  &  $< 0.032$ \nl
 {\FeII}~2383    & 3792.135  &  $< 0.020$ \nl
 {\FeII}~2587    & 4116.615  &  $< 0.014$ \nl
 {\FeII}~2600    & 4138.136  &  $< 0.012$ \nl
 {\MgII}~2796    & 4450.352  &  $ 0.103\pm0.008$  \nl
 {\MgII}~2803    & 4461.778  &  $ 0.064\pm0.011$  \nl
 {\MgI}~2853     & 4540.449  &  $<0.010$  \nl
\cutinhead{S7 \hfil Q$0454+036$ \hfil $z_{\rm abs}=0.64283$}
 {\FeII}~2344    & 3851.138  &  $< 0.031$ \nl
 {\FeII}~2374    & 3900.829  &  $< 0.019$ \nl
 {\FeII}~2383    & 3914.471  &  $ 0.029\pm0.005$  \nl
 {\FeII}~2587\tablenotemark{d}  & 4249.418  &  $ 0.014\pm0.004$  \nl
 {\FeII}~2600    & 4271.634  &  $ 0.037\pm0.014$  \nl
 {\MgII}~2796    & 4593.923  &  $ 0.118\pm0.008$  \nl
 {\MgII}~2803    & 4605.716  &  $ 0.081\pm0.009$  \nl
 {\MgI}~2853     & 4686.926  &  $< 0.005$ \nl
\cutinhead{S8 \hfil Q$0823-223$ \hfil $z_{\rm abs}=0.70547$}
 {\FeII}~2344    & 3997.991  &  $< 0.019$ \nl
 {\FeII}~2374    & 4049.577  &  $< 0.016$ \nl
 {\FeII}~2383    & 4063.739  &  $< 0.017$ \nl
 {\FeII}~2587    & 4411.459  &  $< 0.008$ \nl
 {\FeII}~2600    & 4434.522  &  $< 0.008$ \nl
 {\MgII}~2796    & 4769.100  &  $ 0.092\pm0.007$  \nl
 {\MgII}~2803    & 4781.344  &  $ 0.044\pm0.011$  \nl
 {\MgI}~2853     & 4865.650  &  $< 0.005$ \nl
\cutinhead{S9 \hfil Q$0058+019$ \hfil $z_{\rm abs}=0.72518$}
 {\FeII}~2344    & 4044.179  &  $< 0.026$ \nl
 {\FeII}~2374    & 4096.361  &  $< 0.020$ \nl
 {\FeII}~2383    & 4110.687  &  $< 0.018$ \nl
 {\FeII}~2587    & 4462.424  &  $< 0.013$ \nl
 {\FeII}~2600\tablenotemark{d}  & 4485.753  &  $0.017\pm0.005$  \nl
 {\MgII}~2796    & 4824.197  &  $ 0.253\pm0.012$  \nl
 {\MgII}~2803    & 4836.582  &  $ 0.201\pm0.016$  \nl
 {\MgI}~2853     & 4921.862  &  $ 0.041\pm0.026$  \nl
\tablebreak
\cutinhead{S10 \hfil Q$0117+212$ \hfil $z_{\rm abs}=0.72907$}
 {\FeII}~2587\tablenotemark{d}  & 4472.502  &  $ 0.012\pm0.002$  \nl
 {\FeII}~2600    & 4495.884  &  $ 0.075\pm0.016$  \nl
 {\MgII}~2796    & 4835.091  &  $ 0.238\pm0.009$  \nl
 {\MgII}~2803    & 4847.504  &  $ 0.132\pm0.013$  \nl
 {\MgI}~2853     & 4932.977  &  $ 0.013\pm0.009$  \nl
\cutinhead{S11 \hfil Q$1548+093$ \hfil $z_{\rm abs}=0.77065$}
 {\FeII}~2374    & 4204.330  &  $< 0.190$ \nl
 {\FeII}~2383    & 4219.033  &  $< 0.131$ \nl
 {\FeII}~2587\tablenotemark{d}  & 4580.041  &  $ 0.143\pm0.006$  \nl
 {\FeII}~2600    & 4603.986  &  $ 0.101\pm0.027$  \nl
 {\MgII}~2796    & 4951.349  &  $ 0.234\pm0.024$  \nl
 {\MgII}~2803    & 4964.061  &  $ 0.197\pm0.025$  \nl
 {\MgI}~2853     & 5051.589  &  $ 0.051\pm0.059$  \nl
\cutinhead{S12 \hfil Q$1634+706$ \hfil $z_{\rm abs}=0.81816$}
 {\FeII}~2344    & 4262.149  &  $< 0.019$ \nl
 {\FeII}~2374    & 4317.143  &  $< 0.017$ \nl
 {\FeII}~2383    & 4332.241  &  $< 0.019$ \nl
 {\FeII}~2587    & 4702.936  &  $< 0.007$ \nl
 {\FeII}~2600    & 4727.523  &  $< 0.008$ \nl
 {\MgII}~2796    & 5084.207  &  $ 0.030\pm0.005$  \nl
 {\MgII}~2803    & 5097.260  &  $ 0.018\pm0.006$  \nl
 {\MgI}~2853     & 5187.136  &  $< 0.004$ \nl
\cutinhead{S13 \hfil Q$1421+331$ \hfil $z_{\rm abs}=0.84325$}
 {\FeII}~2344    & 4320.965  &  $ 0.041\pm0.015$  \nl
 {\FeII}~2374    & 4376.718  &  $ 0.028\pm0.018$  \nl
 {\FeII}~2383    & 4392.024  &  $ 0.068\pm0.011$  \nl
 {\FeII}~2587    & 4767.835  &  $ 0.048\pm0.009$  \nl
 {\FeII}~2600    & 4792.761  &  $ 0.071\pm0.014$  \nl
 {\MgII}~2796    & 5154.367  &  $ 0.086\pm0.008$  \nl
 {\MgII}~2803    & 5167.600  &  $ 0.085\pm0.007$  \nl
 {\MgI}~2853     & 5258.717  &  $ 0.022\pm0.009$  \nl
\tablebreak
\cutinhead{S14 \hfil Q$1248+401$ \hfil $z_{\rm abs}=0.85455$}
 {\FeII}~2344    & 4347.471  &  $ 0.024\pm0.010$  \nl
 {\FeII}~2374    & 4403.567  &  $< 0.023$ \nl
 {\FeII}~2383    & 4418.966  &  $ 0.035\pm0.010$  \nl
 {\FeII}~2587    & 4797.082  &  $ 0.010\pm0.009$  \nl
 {\FeII}~2600    & 4822.161  &  $ 0.031\pm0.007$  \nl
 {\MgII}~2796    & 5185.986  &  $ 0.253\pm0.014$  \nl
 {\MgII}~2803    & 5199.300  &  $ 0.136\pm0.017$  \nl
 {\MgI}~2853     & 5290.976  &  $< 0.019$ \nl
\cutinhead{S15 \hfil Q$0002+051$ \hfil $z_{\rm abs}=0.86653$}
 {\FeII}~2344    & 4375.543  &  $< 0.004$ \nl
 {\FeII}~2374    & 4432.001  &  $< 0.009$ \nl
 {\FeII}~2383    & 4447.500  &  $< 0.006$ \nl
 {\FeII}~2587    & 4828.057  &  $< 0.007$ \nl
 {\FeII}~2600    & 4853.298  &  $< 0.010$ \nl
 {\MgII}~2796    & 5219.472  &  $ 0.023\pm0.008$  \nl
 {\MgII}~2803    & 5232.872  &  $ 0.014\pm0.007$  \nl
 {\MgI}~2853     & 5325.140  &  $< 0.007$ \nl
\cutinhead{S16 \hfil Q$1241+174$ \hfil $z_{\rm abs}=0.89549$}
 {\FeII}~2344    & 4443.441  &  $< 0.006$ \nl
 {\FeII}~2374    & 4500.775  &  $< 0.006$ \nl
 {\FeII}~2383    & 4516.514  &  $< 0.005$ \nl
 {\FeII}~2587    & 4902.977  &  $< 0.005$ \nl
 {\FeII}~2600    & 4928.610  &  $< 0.005$ \nl
 {\MgII}~2796    & 5300.466  &  $ 0.018\pm0.005$  \nl
 {\MgII}~2803    & 5314.073  &  $ 0.010\pm0.005$  \nl
 {\MgI}~2853     & 5407.773  &  $< 0.004$ \nl
\cutinhead{S17 \hfil Q$1634+706$ \hfil $z_{\rm abs}=0.90555$}
 {\FeII}~2344    & 4467.026  &  $< 0.008$ \nl
 {\FeII}~2374    & 4524.664  &  $< 0.008$ \nl
 {\FeII}~2383    & 4540.487  &  $< 0.010$ \nl
 {\FeII}~2587    & 4929.001  &  $< 0.005$ \nl
 {\FeII}~2600    & 4954.770  &  $< 0.005$ \nl
 {\MgII}~2796    & 5328.600  &  $ 0.064\pm0.004$  \nl
 {\MgII}~2803    & 5342.280  &  $ 0.045\pm0.005$  \nl
 {\MgI}~2853     & 5436.477  &  $< 0.004$ \nl
\tablebreak
\cutinhead{S18 \hfil Q$0454+036$ \hfil $z_{\rm abs}=0.93150$}
 {\FeII}~2344    & 4527.854  &  $< 0.005$ \nl
 {\FeII}~2374    & 4586.277  &  $< 0.004$ \nl
 {\FeII}~2383    & 4602.315  &  $ 0.030\pm0.008$  \nl
 {\FeII}~2587    & 4996.120  &  $< 0.004$ \nl
 {\FeII}~2600\tablenotemark{e}  & 5022.239  &  $\cdots$  \nl
 {\MgII}~2796    & 5401.159  &  $ 0.043\pm0.005$  \nl
 {\MgII}~2803    & 5415.026  &  $ 0.023\pm0.005$  \nl
 {\MgI}~2853     & 5510.506  &  $< 0.005$ \nl
\cutinhead{S19 \hfil Q$1206+456$ \hfil $z_{\rm abs}=0.93428$}
 {\FeII}~2344    & 4534.373  &  $< 0.005$ \nl
 {\FeII}~2374\tablenotemark{c}  & 4592.880  &  $\cdots$  \nl
 {\FeII}~2383    & 4608.942  &  $< 0.005$ \nl
 {\FeII}~2587    & 5003.313  &  $< 0.004$ \nl
 {\FeII}~2600    & 5029.470  &  $< 0.004$ \nl
 {\MgII}~2796    & 5408.936  &  $ 0.049\pm0.005$  \nl
 {\MgII}~2803    & 5422.822  &  $ 0.024\pm0.005$  \nl
 {\MgI}~2853     & 5518.440  &  $< 0.004$ \nl
\cutinhead{S20 \hfil Q$0002+051$ \hfil $z_{\rm abs}=0.95603$}
 {\FeII}~2344    & 4585.351  &  $< 0.005$ \nl
 {\FeII}~2374    & 4644.515  &  $< 0.007$ \nl
 {\FeII}~2383    & 4660.757  &  $< 0.009$ \nl
 {\FeII}~2587    & 5059.562  &  $< 0.008$ \nl
 {\FeII}~2600    & 5086.014  &  $< 0.005$ \nl
 {\MgII}~2796    & 5469.746  &  $ 0.052\pm0.007$  \nl
 {\MgII}~2803    & 5483.788  &  $ 0.025\pm0.007$  \nl
 {\MgI}~2853     & 5580.480  &  $< 0.005$ \nl
\cutinhead{S21 \hfil Q$1329+412$ \hfil $z_{\rm abs}=0.97387$}
 {\FeII}~2344    & 4627.167  &  $< 0.025$ \nl
 {\FeII}~2374    & 4686.871  &  $< 0.032$ \nl
 {\FeII}~2383    & 4703.261  &  $< 0.027$ \nl
 {\FeII}~2587    & 5105.703  &  $< 0.026$ \nl
 {\FeII}~2600    & 5132.395  &  $< 0.028$ \nl
 {\MgII}~2796    & 5519.627  &  $ 0.181\pm0.035$  \nl
 {\MgII}~2803    & 5533.797  &  $ 0.128\pm0.050$  \nl
 {\MgI}~2853     & 5631.371  &  $< 0.024$ \nl
\tablebreak
\cutinhead{S22 \hfil Q$1329+412$ \hfil $z_{\rm abs}=0.99836$}
 {\FeII}~2344    & 4684.579  &  $ 0.044\pm0.016$  \nl
 {\FeII}~2374    & 4745.024  &  $< 0.011$ \nl
 {\FeII}~2383    & 4761.617  &  $ 0.061\pm0.013$  \nl
 {\FeII}~2587    & 5169.053  &  $< 0.011$ \nl
 {\FeII}~2600    & 5196.076  &  $ 0.058\pm0.017$  \nl
 {\MgII}~2796    & 5588.112  &  $ 0.142\pm0.010$  \nl
 {\MgII}~2803    & 5602.459  &  $ 0.094\pm0.015$  \nl
\cutinhead{S23 \hfil Q$1634+706$ \hfil $z_{\rm abs}=1.04144$}
 {\FeII}~2344    & 4785.579  &  $< 0.008$ \nl
 {\FeII}~2374    & 4847.327  &  $< 0.006$ \nl
 {\FeII}~2383    & 4864.279  &  $< 0.005$ \nl
 {\FeII}~2600    & 5308.105  &  $< 0.038$ \nl
 {\MgII}~2796    & 5708.593  &  $ 0.097\pm0.008$  \nl
 {\MgII}~2803    & 5723.249  &  $ 0.046\pm0.009$  \nl
 {\MgI}~2853     & 5824.163  &  $< 0.003$ \nl
\cutinhead{S24 \hfil Q$1213-003$ \hfil $z_{\rm abs}=1.12770$}
 {\FeII}~2374    & 5052.136  &  $< 0.007$ \nl
 {\FeII}~2383    & 5069.804  &  $ <0.014$ \nl
 {\FeII}~2587    & 5503.610  &  $< 0.008$ \nl
 {\FeII}~2600    & 5532.383  &  $< 0.010$ \nl
 {\MgII}~2796    & 5949.793  &  $ 0.036\pm0.006$  \nl
 {\MgII}~2803    & 5965.067  &  $ 0.019\pm0.006$  \nl
 {\MgI}~2853     & 6070.246  &  $< 0.005$ \nl
\cutinhead{S25 \hfil Q$0958+551$ \hfil $z_{\rm abs}=1.21132$}
 {\FeII}~2587    & 5719.901  &  $< 0.011$ \nl
 {\FeII}~2600    & 5749.804  &  $< 0.006$ \nl
 {\MgII}~2796    & 6183.618  &  $ 0.060\pm0.007$  \nl
 {\MgII}~2803    & 6199.493  &  $ 0.044\pm0.010$  \nl
 {\MgI}~2853     & 6308.805  &  $< 0.010$ \nl
\tablebreak
\cutinhead{S26 \hfil Q$0450-132$ \hfil $z_{\rm abs}=1.22948$}
 {\AlIII}~1855   & 4135.017  &  $ 0.042\pm0.009$  \nl 
 {\AlIII}~1863   & 4153.092  &  $ 0.010\pm0.005$  \nl 
 {\FeII}~2344    & 5226.367  &  $ 0.015\pm0.014$  \nl
 {\FeII}~2383    & 5312.315  &  $ 0.037\pm0.010$  \nl
 {\FeII}~2587    & 5766.872  &  $< 0.007$ \nl
 {\FeII}~2600    & 5797.020  &  $ 0.039\pm0.017$  \nl
 {\MgII}~2796    & 6234.397  &  $ 0.135\pm0.010$  \nl
 {\MgII}~2803    & 6250.402  &  $ 0.105\pm0.010$  \nl
 {\MgI}~2853     & 6360.612  &  $< 0.007$ \nl
\cutinhead{S27 \hfil Q$0450-132$ \hfil $z_{\rm abs}=1.23244$}
 {\FeII}~2344    & 5233.317  &  $< 0.008$ \nl
 {\FeII}~2374    & 5300.842  &  $< 0.008$ \nl
 {\FeII}~2383    & 5319.380  &  $ 0.011\pm0.007$  \nl
 {\FeII}~2587\tablenotemark{d}  & 5774.541  &  $ 0.008\pm0.002$  \nl
 {\MgII}~2796    & 6242.688  &  $ 0.101\pm0.009$  \nl
 {\MgII}~2803    & 6258.715  &  $ 0.070\pm0.008$  \nl
 {\MgI}~2853\tablenotemark{d}   & 6369.071  &  $ 0.005\pm0.002$  \nl
\cutinhead{S28 \hfil Q$0958+551$ \hfil $z_{\rm abs}=1.27238$}
 {\FeII}~2374    & 5395.671  &  $< 0.007$ \nl
 {\FeII}~2383\tablenotemark{d}  & 5414.540  &  $ 0.009\pm0.002$  \nl
 {\FeII}~2587    & 5877.844  &  $< 0.007$ \nl
 {\FeII}~2600\tablenotemark{d,f}  & 5908.573  &  $ 0.017\pm0.004$  \nl
 {\MgII}~2796    & 6354.366  &  $ 0.081\pm0.007$  \nl
 {\MgII}~2803    & 6370.679  &  $ 0.056\pm0.008$  \nl
 {\MgI}~2853\tablenotemark{d} & 6483.010  &  $0.007\pm0.002 $ \nl
\cutinhead{S29 \hfil Q$0117+212$ \hfil $z_{\rm abs}=1.32500$}
 {\AlIII}~1863   & 4331.013  &  $ 0.0162\pm0.004$  \nl
 {\FeII}~2344\tablenotemark{d}   & 5450.307  &  $ 0.006\pm0.002$  \nl
 {\FeII}~2374    & 5520.632  &  $< 0.017$ \nl
 {\FeII}~2383    & 5539.938  &  $ 0.030\pm0.010$  \nl
 {\FeII}~2587    & 6013.972  &  $ 0.010\pm0.005$  \nl
 {\FeII}~2600    & 6045.412  &  $ 0.026\pm0.009$  \nl
 {\MgII}~2796    & 6501.530  &  $ 0.291\pm0.009$  \nl
 {\MgII}~2803    & 6518.221  &  $ 0.180\pm0.011$  \nl
 {\MgI}~2853\tablenotemark{d}   & 6633.153  &  $ 0.005\pm0.002$  \nl
\tablebreak
\cutinhead{S30 \hfil Q$0117+212$ \hfil $z_{\rm abs}=1.34297$}
 {\AlIII}~1855   & 4345.600  &  $ 0.031\pm0.003$  \nl 
 {\AlIII}~1863   & 4364.461  &  $ 0.022\pm0.003$  \nl
 {\FeII}~2344    & 5492.421  &  $ 0.018\pm0.005$  \nl
 {\FeII}~2374    & 5563.289  &  $< 0.014$ \nl
 {\FeII}~2383    & 5582.745  &  $ 0.029\pm0.004$  \nl
 {\FeII}~2587    & 6060.441  &  $ 0.012\pm0.005$  \nl
 {\MgII}~2796    & 6501.530  &  $ \cdots$  \nl
 {\MgII}~2803    & 6568.586  &  $ 0.153\pm0.008$  \nl
 {\MgI}~2853     & 6684.406  &  $< 0.010$ \nl
\enddata
\tablenotetext{a}{When multiple components are present, $\lambda _{\rm
abs}$ is $\lambda_{\rm r}(1+z_{\rm abs})$.}
\tablenotetext{b}{Compromised by bad pixel (see \S\ref{sec:qsonotes}).}
\tablenotetext{c}{Confused by blending (see \S\ref{sec:qsonotes})}
\tablenotetext{d}{Not a $5\sigma$ detection, but at least a $>3\sigma$
detection.}
\tablenotetext{e}{Compromised by pen mark on the CCD.}
\tablenotetext{f}{Not captured by the CCD.}
\label{tab:ews}
\end{deluxetable}

\begin{deluxetable}{rcccccl}
\tablewidth{0pc}
\tablecaption{Ionization Conditions\tablenotemark{a}}
\tablehead
{
\colhead{ID} &
\colhead{$N_{\rm c}$} &
\colhead{W$_{\rm r}(\lambda 2796)$} &
\colhead{W$_{\rm r}(\lambda 2600)$} &
\colhead{W$_{\rm r}(\lambda 1548)$} &
\colhead{IC\tablenotemark{b}} &
\colhead{Other} \\
 & &
\colhead{[{\AA}]} &
\colhead{[{\AA}]} &
\colhead{[{\AA}]} &
 &
\colhead{Refs} 
}
\startdata 
S1  & 1 & $0.179\pm0.019$ & \nodata & \nodata                 & &  \nl  
S2  & 1 & $0.258\pm0.035$ & $<0.100 $       & $<0.393       $ & &  \nl  
S3  & 1 & $0.030\pm0.007$ & $<0.012 $       & $<0.247       $ & & 1 \nl  
S4\tablenotemark{c}  & 1 & $0.080\pm0.014$ & $<0.009 $       & $<0.645$ & & 2 \nl  
S5\tablenotemark{d}  & 1 & $0.135\pm0.014$ & $<0.012 $       & $0.175\pm0.055$ & H? & 1 \nl  
S6  & 1 & $0.103\pm0.008$ & $<0.012 $       & $<0.145       $ & & 1 \nl  
S7  & 1 & $0.118\pm0.008$ & $0.037\pm0.014$ & $<0.130       $ & L & 3 \nl  
S8  & 1 & $0.092\pm0.007$ & $<0.008 $       & $<0.130       $ & &  \nl  
S9  & 1 & $0.253\pm0.012$ & $0.017\pm0.005$ & \nodata         & L?,M?&   \nl  
S10 & 5 & $0.238\pm0.009$ & $0.075\pm0.016$ & $<0.088       $ & L & 1 \nl  
S11 & 1 & $0.234\pm0.024$ & $0.101\pm0.027$ & \nodata         & L?,M?&  \nl  
S12 & 1 & $0.030\pm0.018$ & $<0.008 $       & $<0.034       $ & L? & 4 \nl  
S13 & 1 & $0.086\pm0.008$ & $0.071\pm0.014$ & \nodata         & L?,M?&  \nl  
S14 & 6 & $0.235\pm0.014$ & $0.031\pm0.007$ & $0.718\pm0.600$ & M & 1 \nl  
S15 & 1 & $0.023\pm0.008$ & $<0.010 $       & $<0.110 $       & & 1 \nl  
S16 & 1 & $0.018\pm0.005$ & $<0.005 $       & $<0.093 $       & & 1 \nl  
S17 & 1 & $0.064\pm0.004$ & $<0.005 $       & $0.169\pm0.017$ & H & 4,5 \nl  
S18\tablenotemark{e} & 1 & $0.042\pm0.005$ & $0.022\pm0.008$ & $<0.111 $       &L &3 \nl  
S19 & 1 & $0.049\pm0.005$ & $<0.004 $       & $0.204\pm0.042$ & H & 1,6 \nl  
S20 & 1 & $0.052\pm0.007$ & $<0.005 $       & $0.479\pm0.040$ & H & 1 \nl  
S21\tablenotemark{d} & 1 & $0.181\pm0.035$ & $<0.028 $       & $<0.389 $       & &  \nl  
S22 & 1 & $0.142\pm0.010$ & $0.058\pm0.017$ & $<0.119 $       & L &  \nl  
S23 & 1 & $0.097\pm0.008$ & $<0.038 $       & $0.424\pm0.018$ & H? & 1,4,5 \nl  
S24 & 1 & $0.036\pm0.006$ & $<0.010 $       & \nodata         & &  \nl  
S25 & 1 & $0.060\pm0.007$ & $<0.006 $       & \nodata         & &  \nl  
S26 & 2 & $0.135\pm0.010$ & $0.039\pm0.017$ & \nodata         & L?,M?&  \nl  
S27\tablenotemark{e} & 2 & $0.101\pm0.009$ & $0.008\pm0.007$ & \nodata         & &  \nl  
S28 & 1 & $0.081\pm0.007$ & $0.017\pm0.004$ & $0.440\pm0.030$ & M & 7 \nl  
S29 & 6 & $0.291\pm0.011$ & $0.026\pm0.009$ & $0.890\pm0.060$ & M & 8 \nl  
S30 & 4 & $0.153\pm0.008$ & $0.022\pm0.004$ & $0.670\pm0.050$ & M & 8 \nl  
\enddata
\tablenotetext{a}{Limits are $3\sigma$.}
\tablenotetext{b}{``L'', ``H'', and ``M'' refer to ``low'', ``high'',
and ``multi--phase'' ionization conditions. See text for definitions.}
\tablenotetext{c}{Unrestrictive limit due to complex blend in {\Lya}
forest. Ambiguous case.}
\tablenotetext{d}{{\CIV}$\lambda 1548$ transitions could be {\Lya}
line.}
\tablenotetext{e}{$W_{\rm r}(2600)$ scaled from measured $W_{\rm
r}(2383)$ by ratio of oscillator strengths.}
\tablerefs{(1) Jannuzi \etal 1998; (2) Impey \etal 1996; (3) Churchill
\& Le Brun 1998; (4) Bachall \etal 1996; (5) Bergeron \etal 1994; (6)
Churchill \& Charlton 1988; (7) Sargent, Boksenberg, \& Steidel 1988;
(8) Steidel \& Sargent 1992}
\label{tab:civcomp}
\end{deluxetable}

\endgroup




\clearpage 

\begin{thebibliography}{ }
\begingroup
\footnotesize
\setlength{\parskip}{-3.0pt}

\bibitem[Bahcall \etal 1996]{kp7}
         Bahcall,~J.~N., \etal 1996, ApJ, 457, 19 (KP VII)

\bibitem[Bergeron \etal 1994]{bergeron94} 
         Bergeron,~J. \etal 1994, ApJ, 436, 33 

\bibitem[Bergeron \& Boiss\'{e} 1991]{bb91}
         Bergeron,~J., and Boiss\'{e},~P. 1991, A\&A, 243, 344

\bibitem[Bowen, Blades, \& Pettini 1995]{bbp95}
         Bowen,~D.~V., Blades,~J.~C., Pettini,~M. 1995, ApJ, 448, 634

\bibitem[Caulet 1989]{caulet89} 
         Caulet, A. 1989, ApJ, 340, 90  

\bibitem[Churchill 1995]{cwc95} 
         Churchill,~C.~W. 1995, Lick Observatory Technical Report, \#74

\bibitem[Churchill 1997a]{thesis} 
         Churchill,~C.~W. 1997a, Ph.~D. Thesis, University of
         California, Santa Cruz 

\bibitem[Paper II]{paper2}
         Churchill,~C.~W., Charlton,~J.~C., Rigby,~J.~R.,
         Januzzi,~B.~T., Kirhakos,~S., and Vogt,~S.~S. 1998, in
         progress (\cite{paper2})

\bibitem[Churchill \& Charlton 1998]{q1206}
         Churchill,~C.~W., and Charlton, J. C. 1998, ApJ, submitted

\bibitem[Churchill \& Le~Brun 1998]{cl98}
         Churchill,~C.~W., and Le~Brun,~V. 1998, ApJ, 499, 677

\bibitem[Churchill, Steidel, \& Vogt 1996]{csv96} 
         Churchill,~C.~W., Steidel,~C.~C., and Vogt,~S.~S, 1996, ApJ,
         471, 164 

\bibitem[Churchill, Vogt, \& Charlton 1998]{cvc98}
         Churchill,~C.~W., Vogt,~S.~S., and Charlton,~J.~C. 1998, ApJS,
         to be submitted (CVC98)

\bibitem[Dalcanton \etal 1997]{dalcanton}
         Dalcanton,~J.~J., Spergel,~D.~N., Gunn,~J.~E., Smith,~M.,
         and Schneider,~D.~P. 1997, AJ, 114, 635

\bibitem[Dobrzycki \etal 1998]{adam98}
         Dobrzycki,~A., Bechtold,~A., Wilden,~B., Morita,~M.,
         Dobrzycka,~D., Scott,~J., Tran,~K--V. 1998, ApJ, in
         preparation

\bibitem[de~Blok, McGaugh, \& van der Hulst 1996]{deblok}
         de~Blok,~W.~J.~G., McGaugh,~S.~S., and van der
         Hulst,~J.~M. 1996, MNRAS, 283, 18

\bibitem[Ferland 1996]{ferland}
         Ferland,~G. 1996, Hazy, University of Kentucky
         Internal Report

\bibitem[Freeman 1970]{freeman}
         Freeman,~K. 1970, ApJ, 160, 811 

\bibitem[Haardt \& Madau 1996]{hm96}
         Haardt,~F., and Madau,~P. 1996, ApJ, 461, 20

\bibitem[Holmberg 1975]{holmberg75}
         Holmberg,~E. 1975, in Galaxies and the Universe, ed.\
         A.~Sandage, M.~Sandage, \& J.~Christian (Chicago: University
         of Chicago Press), 123  

\bibitem[Impey \& Bothun 1989]{impey89}
         Impey,~C., and Bothun,~G. 1989, ApJ, 341, 89

\bibitem[Impey \etal 1996]{impey96}
         Impey,~C., Petry,~C.~E., Malkan,~M.~A., and Webb,~W. 1996,
         ApJ, 463, 473

\bibitem[Jannuzi \etal 1998]{kp13}
         Jannuzi,~B.~T., et~al., 1998, in press (KP XIII)

\bibitem[Kirhakos \etal 1992]{kirhakos92}
         Kirhakos, S., \etal 1992, PASP 646, 1994

\bibitem[Kulkarni \& Fall 1993]{kulkarni-fall}
         Kulkarni,~V.~P., and Fall,~S.~M. 1994, ApJ, 413, L63

\bibitem[Lanzetta, Turnshek, \& Wolfe 1987]{ltw87} 
         Lanzetta,~K.~M., Turnshek,~D.~A., and Wolfe,~A.~M. 1987, ApJ,
         322, 739 (LTW) 

\bibitem[Lanzetta 1988]{lanzetta88}
         Lanzetta,~K.~M. 1988, ApJ, 332, 96

\bibitem[Lanzetta \etal 1995]{lanzetta95}
         Lanzetta.~K.~M., Bowen,~D.~B., Tytler,~D., and
         Webb,~J.~K. 1995, ApJ, 442, 538

\bibitem[Lauroesch \etal 1996]{jtl}
         Lauroesch,~J.~T., Truran,~J.~W., Welty,~D.~E., and
         York,~D.~G. 1996, PASP, 108, 641

\bibitem[Le~Brun \etal 1997]{lebrundla}
         Le~Brun,~V., Bergeron,~J., Boiss\'{e},~P., and
         Deharving,~J.~M. 1997, A\&A, 321, 733

\bibitem[Le~Brun, Bergeron, \& Boiss\'{e} 1996]{lebrunlya} 
         Le~Brun,~V., Bergeron,~J., Boiss\'e,~P. 1996, A\&A, 306, 691

\bibitem[Lilly \etal 1995]{lilly95}
         Lilly,~S.~J., Tresse,~L., Hammer,~F., Crampton,~D., and
         Le~F\`{e}vre,~O. 1995, ApJ, 455, 108

\bibitem[Linder 1997]{linder}
         Linder,~S.~M. 1997, ApJ, in press

\bibitem[McGaugh 1994]{mcgaugh}
         McGaugh,~S. 1994, ApJ, 426, 135
 
\bibitem[Petitjean \& Bergeron 1990]{pb90} 
         Petitjean,~P., and Bergeron,~J. 1990, A\&A, 231, 309 (PB90)

\bibitem[Petitjean, Rauch, \& Carswell 1994]{prc94}
        Petitjean,~P., Rauch,~M., and Carswell,~R.~F. 1994, 
        A\&A, 291, 29

\bibitem[Salpeter 1993]{salpeter}
         Salpeter,~E.~E. 1993, AJ, 106, 1265

\bibitem[Sargent, Boksenberg, \& Steidel 1988]{sbs88} 
         Sargent,~W.~L.~W., Boksenberg,~A., and Steidel,~C.~C.  1988,
         ApJS, 68, 539

\bibitem[Sargent, Steidel, \& Boksenberg 1988]{ssb88} 
         Sargent,~W.~L.~W., Steidel,~C.~C., and Boksenberg,~A. 1988,
         ApJ, 334, 22 (SSB) 

\bibitem[Savage \& Sembach 1996]{savagearaa}
         Savage,~B.~D., and Sembach,~K.~R. 1996, ARAA, 34, 279

\bibitem[Savage \& Sembach 1991]{savage91}
         Savage,~B.~D., and Sembach,~K.~R. 1991, ApJ, 379, 245

\bibitem[Schechter 1976]{schechter76}
         Schechter,~P. 1976, ApJ, 203, 297

\bibitem[Schneider \etal 1993]{schneider93} 
         Schneider,~D.~P. {\etal} 1993, ApJS, 87, 45

\bibitem[Sembach \& Savage 1992]{sembach92}
         Sembach,~K.~R., and Savage,~B.~D. 1992, ApJS, 83, 147

\bibitem[Steidel \etal 1994]{steidel94}
         Steidel,~C.~C., Pettini,~M., Dickinson,~M., and
         Persson,~S.~E. 1994, AJ, 108, 2046

\bibitem[Steidel 1995]{steidel95} 
         Steidel,~C.~C. 1995, in ESO Workshop
         on Quasar Absorption Lines, ed.~G. Meylan, (Garching : 
         Springer--Verlag), 139

\bibitem[C.~Steidel, private communication]{chuckprivcomm}
         Steidel,~C.~C. 1996, private communication

\bibitem[Steidel, Dickinson, \& Persson 1994]{sdp94}
         Steidel,~C.~C., Dickinson,~M., and Persson,~S.~E. 1994, ApJ,
         L75 

\bibitem[Steidel \& Sargent 1992]{ss92} 
         Steidel,~C.~C., and Sargent,~W.~L.~W. 1992, ApJS, 80, 1
         (SS92) 

\bibitem[Stengler--Larrea \etal 1995]{stengler-larrea}
         Stengler--Larrea,~E.~A., \etal 1995, ApJ, 444, 64

\bibitem[Thimm 1995]{thimm95}
         Thimm, G. 1995,  in ESO Workshop on Quasar Absorption Lines,
         ed.\ G. Meylan, (Garching : Springer--Verlag), 169

\bibitem[Tripp, Lu, \& Savage 1997]{tripp97}
         Tripp,~T.~M., Lu,~L., and Savage,~B.~D. 1997, ApJS, 112, 1

\bibitem[Tytler \etal 1987]{tytler87} 
         Tytler,~D., Boksenberg,~A., Sargent,~W.~L.~W., Young,~P., and
         Kunth,~D. 1987, ApJS, 64, 667 (TBSYK)

\bibitem[Uomoto 1984]{uomoto84} 
         Uomoto,~A. 1984, ApJ, 284, 497 

\bibitem[van Gorkom et.~al.\ 1996]{vangorkom96}
        van Gorkom,~J.~H., Carilli,~C.~L., Stocke,~J.~T.,
        Perlman,~E.~S., Shull,~J.~M. 1996, AJ, 112, 1397
 
\bibitem[Vogt \etal 1994]{vogtspie}
         Vogt,~S.~S., \etal 1994, SPIE, 2198, 362

\bibitem[Womble 1995]{womble95} 
         Womble,~D. 1995 Workshop on Quasar Absorption Lines,
         ed.~G. Meylan, (Garching : Springer--Verlag), 158 

\bibitem[Yanny \& York 1992]{yanny-york}
         Yanny,~B., and York,~D.~G. 1992, ApJ, 391, 569

\endgroup

\end{thebibliography}
\end{document}